\documentclass[11pt]{article}
\usepackage[font=footnotesize]{caption}

\usepackage{natbib}
\setcitestyle{authoryear,round,notesep={; },aysep={},yysep={;}}

\usepackage{amsmath,amssymb,amsfonts,amsthm,mathrsfs}
\usepackage[dvipsnames]{xcolor}
\usepackage{hyperref}
\usepackage{tabularx}
\usepackage{graphicx}
\usepackage{multirow}
\usepackage{booktabs}
\usepackage{textcomp}
\usepackage{epstopdf}
\usepackage{lipsum}
\usepackage{algorithm}
\usepackage{algorithmicx}
\usepackage{algpseudocode}
\usepackage{listings}
\usepackage{subcaption}
\usepackage{makecell}

\newtheorem{theorem}{Theorem}

\newtheorem{lemma}{Lemma}

\numberwithin{equation}{section}
\numberwithin{table}{section}
\numberwithin{theorem}{section}
\numberwithin{lemma}{section}
\numberwithin{definition}{section}
\numberwithin{example}{section}

\newcommand*{\boldone}{\text{\usefont{U}{bbold}{m}{n}1}}

\makeatletter
\newcommand{\removeperiod}{\@ifnextchar.{\@gobble}\relax}
\makeatother
\hypersetup {
    colorlinks = true,
    linkcolor = blue,
    filecolor = magenta,
    urlcolor = NavyBlue,
    citecolor = NavyBlue
}

\begin{document}

\title{\LARGE \bf Shifts of dominant personality and spatial pattern formation due to spatially heterogeneous pollution}

\author{Tianxu Wang \thanks{Department of Mathematical and Statistical Sciences \& Interdisciplinary Lab for Mathematical Ecology and Epidemiology, University of Alberta, Edmonton, Alberta T6G 2G1, Canada} 
\and Jiwoon Sim \thanks{Department of Mathematical and Statistical Sciences, University of Alberta, Edmonton, Alberta T6G 2G1, Canada}
\and Hao Wang \thanks{The Corresponding Author (hao8@ualberta.ca). Department of Mathematical and Statistical Sciences \& Interdisciplinary Lab for Mathematical Ecology and Epidemiology, University of Alberta, Edmonton, Alberta T6G 2G1, Canada}}

\maketitle

\begin{abstract}
Personality traits, such as boldness and shyness, play a significant role in shaping the survival strategies of animals. Industrial pollution has long posed serious threats to ecosystems and is typically distributed heterogeneously. However, how animals with different personalities respond to spatially heterogeneous pollution remains largely unexplored. In this study, we introduce a prey-taxis model with nonlinear cross-diffusion to examine population dynamics in such environments. The global existence of classical solutions is established by deriving initial bounds through energy estimates and improving solution regularity via heat kernel properties and a bootstrap process. Our findings reveal that behavior, population structure, and spatial distribution are heavily influenced by pollution. Bold individuals maintain a competitive advantage in pollution-free or very low-toxin environments, whereas shy individuals become dominant in regions with low to moderate toxin levels. In highly polluted areas, no populations can survive. The spatial pattern of the population is also closely tied to the distribution of toxins. Grazers tend to move along toxin gradient and exhibit periodic behavior. As toxin concentrations rise, aggregation behavior becomes increasingly pronounced across all species. Interestingly, the total population in polluted areas may initially increase when toxin levels are low to moderate, but eventually declines, leading to extinction as toxin levels continue to rise.
\end{abstract}

{\bf Keywords} {\small Toxin, spatial heterogeneity, personality traits, global existence, cognitive movement}

{\bf MSC Classification} {\small 35K40, 35K55, 92C15, 92D25, 35A01}


\section{Introduction}
Animals often exhibit consistent behavioral patterns, referred to as personality traits, such as boldness and shyness, which play a significant role in shaping predator-prey interactions \citep{griffen2012role, pruitt2012behavioral, dirienzo2013combined, gebauer2023bold, gan2024boldness}. Bold individuals generally display shorter latencies to feed, attempt feeding over greater distances, and have faster feeding rates, whereas shy individuals typically explore smaller areas \citep{bourne2008boldness}. For instance, highly aggressive convict cichlids (amatitlania nigrofasciata) were observed to forage more frequently in open environments \citep{church2019ideal}. On the contrary, shy squids were found to take longer to feed and made fewer feeding attempts \citep{sinn2005personality}. Similarly, shy sheep were more likely to stop grazing and huddle together when disturbed \citep{sibbald2009individual}.

Personality differences not only influence foraging strategies but also shape behavioral responses to top predators \citep{belgrad2016predator}. While prey typically seek refuge or reduce activity in the presence of predators, individuals with different levels of boldness respond differently. Bold individuals are more likely to remain active in risky but energetically rewarding areas, whereas shy individuals prioritize safety by staying in more protected, resource-poor habitats \citep{church2019ideal} and often prioritize survival over reproductive investment \citep{cole2014shy}. For example, in a threat test, bold squids were observed confronting or even attacking perceived threats, while shy individuals exhibited more cautious behaviors \citep{sinn2005personality}. In general, prey personality and predator species can significantly interact to influence mortality rates \citep{belgrad2016predator} and it may be not easy to determine whether one personality type is more advantageous. Bold individuals sustain higher productivity, but this often comes at the cost of survival, whereas shy individuals tend to have lower productivity but higher survival rates \citep{wolf2007life,reale2010personality}.

Industrial activities such as crude oil extraction, metallurgy, chemical production, and paper manufacturing generate significant environmental pollution annually, including halogens, heavy metals, organic compounds, and microplastics \citep{hader2020anthropogenic, shan2023impact, singh2022environmental, amoatey2019effects, smith2013animal}. These pollutants are widespread and pose serious threats to aquatic ecosystems and wildlife \citep{baines2021linking, citterich2023plastic, mance2012pollution, jones1999persistent}. For instance, exposure to herbicides decrease activity and foraging in aquatic invertebrates \citep{steele2019express}. In another case, great tits (parus major) living in metal-polluted areas (with higher levels of lead and cadmium in their blood) exhibited slower exploration behaviors \citep{grunst2018variation}. Similarly, Trinidadian guppies (poecilia reticulata) exposed to crude oil showed reduced exploration tendency in a maze experiment \citep{jacquin2017evolutionary}. Pollutants can also disrupt predator avoidance behaviors \citep{weis2012pollutants} and reduce behavioral variability, such as prey capture success \citep{philibert2019persistent}.

Toxin-dependent models are commonly used to assess the negative effects of toxins on individual organisms and population dynamics. These models include discrete-time difference equation models \citep{hayashi2009population,spromberg2006relating,erickson2014daphnia,spromberg2005modeling} and continuous-time ordinary differential equation models \citep{freedman1991models,hallam1983effects1,hallam1984effects3,lan2019long,huang2015impact,wang2024stoichiometric}. However, these models are unable to capture toxicant-induced spatial behavioral changes, such as shifts in habitat preferences, body tremors, and altered migration patterns \citep{atchison1987effects,blaxter1992effect,scott2004effects}. More recently, a series of diffusive population-toxicant models have been developed to study polluted river systems \citep{deng2023global,deng2024toxicant,wang2023dynamics,zhou2022spatiotemporal,xing2024dynamical}. 

However, in larger water bodies, such as lakes, the dynamics differ significantly from those in rivers. Inputs from wastewater treatment facilities or other sources often result in uneven toxin distribution across the region, with this distribution remaining relatively stable over time. Near pollution sources, such as large industrial plants, toxin levels are much higher, while farther from these sources, concentrations may decrease rapidly or become negligible. In such cases, the behavioral responses of species, such as foraging behavior, predation risk, death rate, and cognitive movement, can be strongly influenced by the spatial variation of toxin levels \citep{wolfe2015movement}.

Moreover, it has been found that pollution-induced behavioral changes may potentially result in positive feedback \citep{jacquin2020effects}. For example, effluent exposure can trigger faster escape responses in aquatic species \citep{spath2022wastewater}. Perch (perca fluviatilis) exposed to psychiatric drugs became more active and bolder than control fish, with lower latency to feed \citep{brodin2013dilute}. One possible reason is that organisms exposed to pollutants tend to have higher metabolic rates and greater energetic needs, as detoxification and repair processes are energetically costly \citep{mckenzie2007complex}, which could lead to increased activity and foraging behaviors \citep{montiglio2014contaminants}.

Furthermore, animals with different personality traits may experience varying exposure risks. For example, bold zebrafish exposed to microplastics accumulated higher levels of microplastics than their shy counterparts \citep{chen2022fish}. Zebrafish embryos exposed to methylmercury (MeHg) showed significant variation in movement activity, and this effect was not proportional to the concentration of MeHg \citep{glazer2021developmental}. These observations raise important questions: Does this positive feedback always occur, and is it related to contaminant levels? Do animals with different personality traits experience the same feedback? If not, how might this influence individuals with different personalities, could it alter their foraging strategies or shift the population structure of personality types? Can this positive feedback, in some cases, mitigate the negative effects of pollutants, or does it amplify the negative impacts on fitness? Despite these critical questions, to the best of our knowledge, no models have yet addressed this gap in current research.

In this paper, we aim to tackle the aforementioned questions from a mathematical perspective. We propose a general personality-based three-species prey-taxis model with spatially dependent predation behavior and cognitive movement in arbitrary spatial dimensions. Our objective is to examine the impact of heterogeneous environmental pollutants on bold and shy individuals and assess how these personality traits influence species population dynamics and ecosystem structure in a stable aquatic environment.

The rest of this paper is organized as follows. We propose the model in section \ref{sec: models} and conduct analysis in Section \ref{sec: Analysis}. Section \ref{sec: Simulations} is devoted to the discussion of biological implication via numerical simulations. We end with conclusions in Section \ref{sec: Discussion}.

\section{Mathematical model}
\label{sec: models}
We extend the classic two-species Lotka-Volterra system \citep{kareiva1987swarms} to a general three-trophic level prey-taxis model that incorporates species' personality traits. In this model, \( X \), \( Y_1 \), \( Y_2 \), and \( Z \) represent the population densities of prey, bold middle predators, shy middle predators, and top predators, respectively. A typical example from marine ecosystems could involve a food chain consisting of phytoplankton, zooplankton, and fish species. Let \( \Omega \subset \mathbb{R}^n \) represent the living environment for all species, with pollutant distribution $T(x)$ varying spatially. The predation function, cognitive movement, and death rate are all influenced by the spatial distribution of toxins. The model is described by the following partial differential equations on $\Omega\times (0, \infty)$:
\begin{equation}
    \label{eq: general model}
    \begin{aligned}
        &\frac{\partial X}{\partial t}=r(X)-f_1(X,Y_1,T(x))-f_2(X,Y_2,T(x)) - d_1(T(x)) X+\delta_X \Delta X,\\
        &\frac{\partial Y_1}{\partial t}=e_1f_1(X,Y_1,T(x))-g_1(Y_1,Z,T(x))-d_{21}(T(x)) Y_1-\nabla \left(\alpha_1(T(x)) h(Y_1)\nabla X\right) +\delta_{Y_1}\Delta Y_1,\\
        &\frac{\partial Y_2}{\partial t}=e_1f_2(X,Y_2,T(x))-g_2(Y_2,Z,T(x))-d_{22}(T(x)) Y_2-\nabla \left(\alpha_2(T(x)) h(Y_2)\nabla X\right) + \delta_{Y_2}\Delta Y_2,\\
        &\frac{\partial Z}{\partial t}=e_2g_1(Y_1,Z,T(x))+ e_2g_2(Y_2,Z,T(x))-d_3(T(x)) Z - \nabla (\gamma_1(T(x)) h(Z)\nabla Y_1) \\
         & \qquad \quad - \nabla (\gamma_2(T(x)) h(Z)\nabla Y_2) +\delta_Z \Delta Z,
    \end{aligned}
\end{equation}
with boundary and initial conditions
\begin{equation}
    \label{eq: bc ic}
    \begin{aligned}
        &\frac{\partial X}{\partial \nu}(x,t)=\frac{\partial Y_1}{\partial \nu}(x,t)=\frac{\partial Y_2}{\partial \nu}(x,t)=\frac{\partial Z}{\partial \nu}(x,t)=0, \ (x, t) \in \partial\Omega \times (0, \infty), \\
        &(X(x, 0), Y_1(x, 0), Y_2(x, 0), Z(x, 0)) = (X_0(x), Y_{10}(x), Y_{20}(x), Z_0(x)), \ x \in \Omega.
    \end{aligned}
\end{equation}
Note that, this general model can be also applied to any heterogeneously distributed resources or environmental factors.

Here, \( r(X) \) denotes the growth function for prey, and $T(x)$ denotes the spatial distribution of pollution. The functions \( f_1(X,Y_1,T(x)) \) and \( f_2(X,Y_2,T(x)) \) represent predation on prey by bold and shy middle predators, respectively, while \( g_1(Y_1,Z,T(x)) \) and \( g_2(Y_2,Z,T(x)) \) denote predation on bold and shy middle predators by top predators. The taxis sensitivity function \( h(z) \) indicates the potential for density-dependent cognitive movement. The taxis coefficient functions \( \alpha_1(T(x)) \), \( \alpha_2(T(x)) \), \( \gamma_1(T(x)) \), and \( \gamma_2(T(x)) \) represent attraction or repulsion towards food sources or prey, with positive values indicating attraction and negative values indicating repulsion. The terms \( d_1(T(x)) \), \( d_{21}(T(x)) \), \( d_{22}(T(x)) \), and \( d_3(T(x)) \) represent death rates for prey, bold middle predators, shy middle predators, and top predators, respectively. The diffusion rates for the three species are represented by \( \delta \), and \( e_i \) denotes production efficiency.

For all the general functions in model \eqref{eq: general model}, we choose the following specific forms. 

Logistic growth is commonly used to model the growth of producers. The natural death rate of producers is typically quite low compared to the turnover driven by herbivory or environmental factors. Thus, the growth and death rates of producers are considered as
\begin{equation*}
    r(X) = rX\left(1-\frac{X}{K}\right), \quad d_1 (T(x)) = 0,
\end{equation*}
where  \( r \) is the intrinsic growth rate, and \( K \) is the carrying capacity. 

\begin{figure}[h!]
     \centering

    \begin{subfigure}[b]{0.44\textwidth}
         \centering        \includegraphics[width=\textwidth]{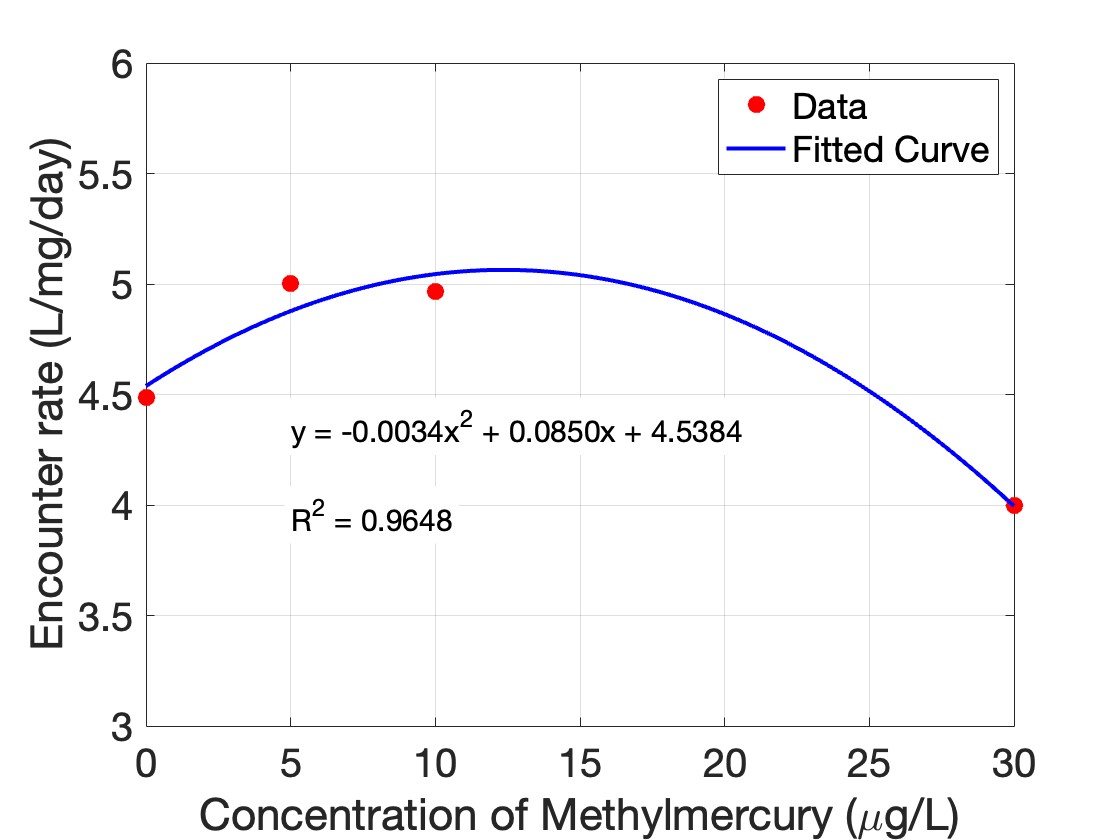}       
         \caption{}   
         \label{fig: Fit_abc1.jpg}
    \end{subfigure}
    \begin{subfigure}[b]{0.44\textwidth}
         \centering        \includegraphics[width=\textwidth]{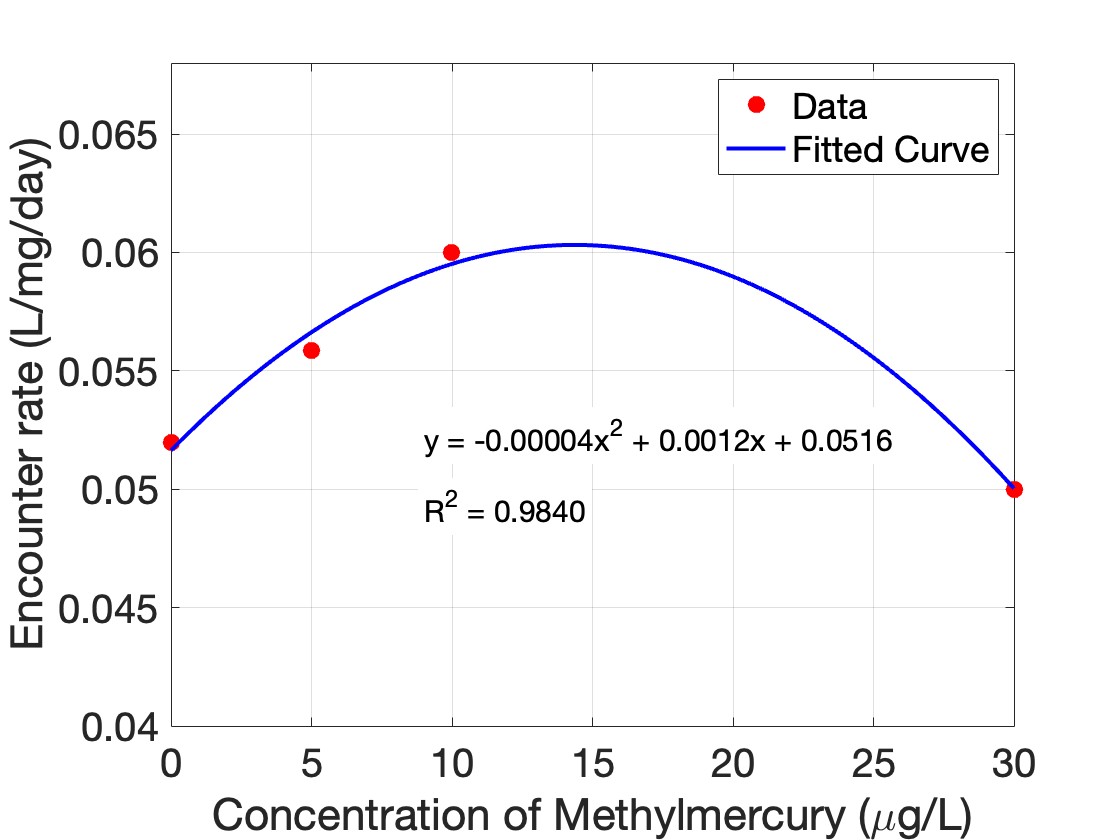}       
         \caption{}   
         \label{fig: Fit_abc2.jpg}
    \end{subfigure}   
\caption{\footnotesize Encounter rate fitted from \citep{glazer2021developmental}: (a) between producers and bold grazers, (b) between bold grazers and predators.}
     \label{fig: Fit encounter rate}
\end{figure}

The predation functions for grazers and predators, depending on the toxin level, are described by the following Holling type II form
\begin{equation*}
    \begin{aligned}
        f_1(X,Y_1,T(x)) &= \frac{\mu_{11}(T) X Y_1}{1 + \mu_{11}(T)\tau_1 X}, \quad f_2(X,Y_2,T(x)) = \frac{\mu_{12}(T) X Y_2}{1 + \mu_{12}(T)\tau_1 X}, \\
        g_1(Y_1,Z,T(x)) &= \frac{\mu_{21}(T) Y_1 Z}{1 + \mu_{21}(T)\tau_2 Y_1}, \quad g_2(Y_2,Z,T(x)) = \frac{\mu_{22}(T) Y_2 Z}{1 + \mu_{22}(T)\tau_2 Y_2},
    \end{aligned}
\end{equation*}
where \(\tau_1\) and \(\tau_2\) represent the handling times for grazers and predators, respectively. \(\mu_{i,j}\) \((i,j=1,2)\) represent the encounter rates. Data on movement, recorded over thirty-five minutes under light and dark settings (in Table S2 of the supplementary material \citep{glazer2021developmental}), is used to estimate the encounter rates. For each toxin concentration level, we calculate the mean value. The average encounter rates between grazers and producers, and between predators and grazers, used in \citep{chen2017global} and \citep{wang2024stoichiometric}, are 3.24 and 0.04, respectively. 
Since bold grazers are expected to have higher encounter rates, we consider a range of 4 to 5 for grazers-producers and 0.05 to 0.06 for predators-grazers. The data is linearized and fitted within these ranges for bold grazers, as shown in Figure \ref{fig: Fit encounter rate}. Thus, the encounter rates \(\mu_{11}(T)\) and \(\mu_{21}(T)\) are modeled as following quadratic functions
\begin{equation*}
    \mu_{11}(T(x)) = a_{1} T^2 + b_{1} T + c_{1}, \quad 
    \mu_{21}(T(x)) = a_{2} T^2 + b_{2} T + c_2.
\end{equation*}
For grazers, the encounter rate with producers reflects their foraging efficiency, while the encounter rate with predators represents the predation risk. Both rates are typically positively correlated with boldness. Grazers with a bolder personality tend to have higher foraging success but also face increased predation risk \citep{church2019ideal,wolf2007life,reale2010personality}. Hence, the encounter rates with both producers and predators for shy grazers are assumed to be proportional to those of bold grazers by a factor of \(\xi\) (shyness index), as follows
\begin{equation*}
    \mu_{12}(T(x)) = \xi \mu_{11}(T), \quad \mu_{22}(T(x)) = \xi \mu_{21}(T).
\end{equation*}
\begin{figure}[h!]
    \centering
    \begin{subfigure}[b]{0.44\textwidth}
         \centering        \includegraphics[width=\textwidth]{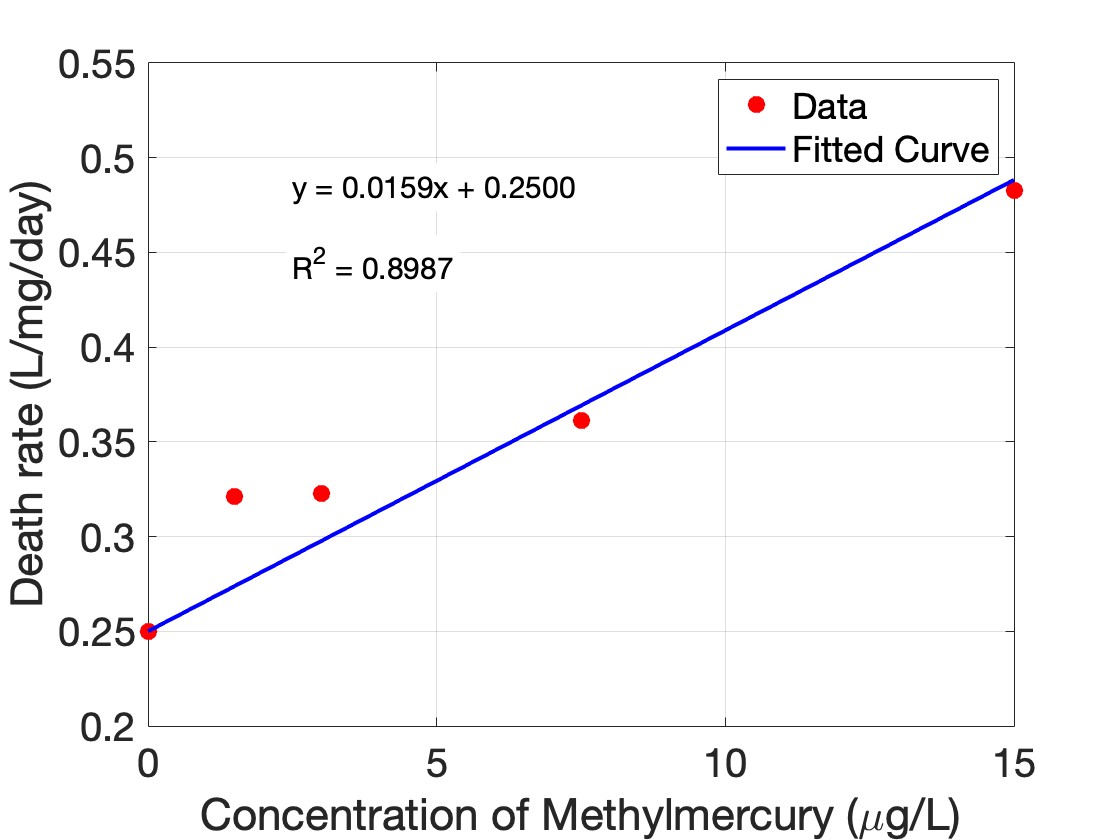}       
         \caption{}   
         \label{fig: Fit_d2.jpg}
    \end{subfigure}
    \begin{subfigure}[b]{0.44\textwidth}
         \centering        \includegraphics[width=\textwidth]{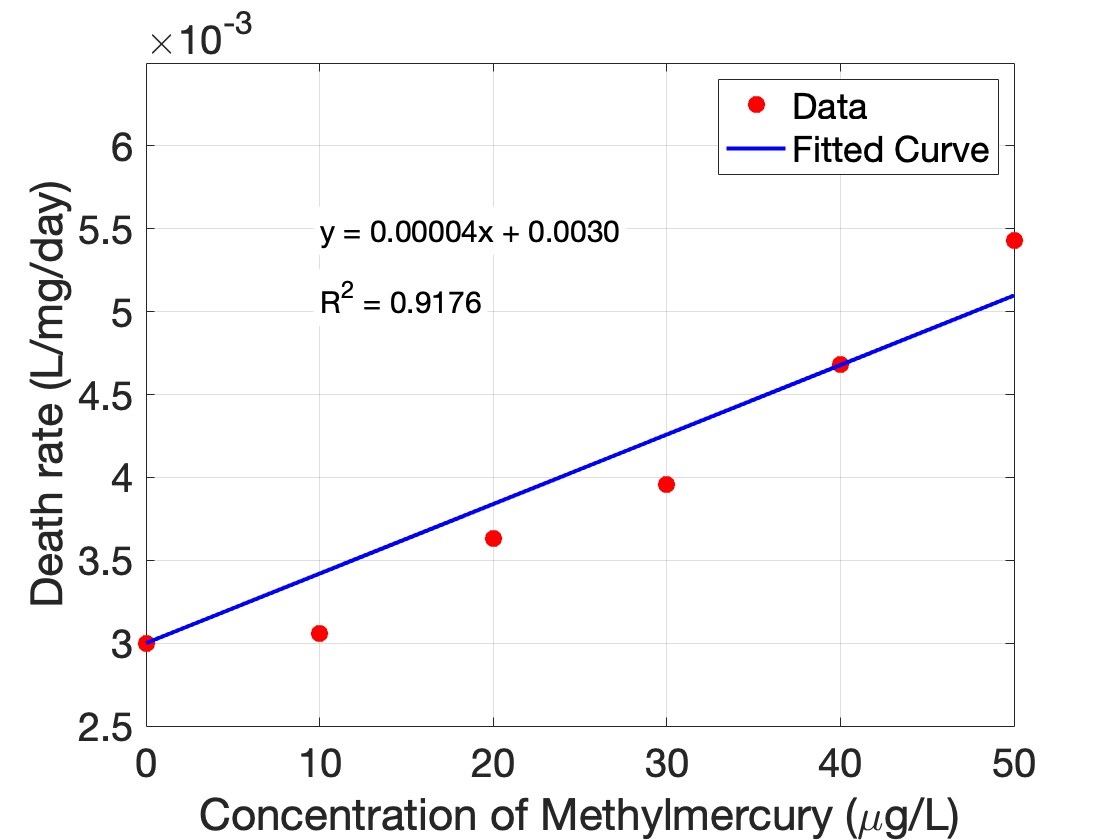}       
         \caption{}   
         \label{fig: Fit_d3.jpg}
    \end{subfigure}   
\caption{\footnotesize Fitted death rates: (a) Bold grazers from \citep{jensen2007lethal}, $d_2 = 0.0159$, (b) Predators from \citep{huang2011toxicity}, $d_{31} = 0.00004$.}
     \label{fig: Fit death rate}
\end{figure}
Since bolder animals have higher encounter rates and shyer animals have lower ones, the encounter rate can be interpreted as a measure of boldness. The taxis coefficient terms are considered to be positively proportional to the boldness level and are given by
\begin{equation*}
    \alpha_1(T(x)) = \eta \mu_{11}(T), \quad
    \alpha_2(T(x)) = \eta \mu_{12}(T), \quad 
    \gamma_1(T(x)) = \eta \mu_{21}(T), \quad
    \gamma_2(T(x)) = \eta \mu_{22}(T).
\end{equation*}
The taxis sensitivity function is
\begin{equation*}
    h(z)=\frac{z}{1 + \kappa z}.
\end{equation*}
The death rates for grazers and predators are influenced by toxin levels. The natural death rates for grazers and predators, as used in \citep{chen2017global} and \citep{wang2024stoichiometric}, are 0.25 and 0.003, respectively. To estimate the effect of toxins on these rates, we use death proportion data from \citep{jensen2007lethal} and \citep{huang2011toxicity}. This data is linearized to match the natural death rates, and the remaining data is fitted accordingly, as shown in Figures \ref{fig: Fit_d2.jpg} and \ref{fig: Fit_d3.jpg}. The death rates for shy grazers are assumed to scale with their boldness level. Thus, these death rate functions are given by following linear functions 
\begin{equation}
\label{eq: d functions}
    d_{21}(T(x)) = d_{20} + d_{2} T, \quad d_{22}(T(x)) = d_{20} + \xi d_{2} T, \quad
    d_3(T(x)) = d_{30} + d_{31} T.
\end{equation}

\section{Mathematical analysis}
\label{sec: Analysis}

We now perform mathematical analysis to the general model \eqref{eq: general model} under the following assumptions. For the following analysis, to simplify notation, we abbreviate \( f_i(X,Y_i,T(x)) \), \( g_i(Y_i,Z,T(x)) \), \( \alpha_i(T(x)) \), and \( \gamma_i(T(x)) \) as \( f_i(X,Y_i,x) \), \( g_i(Y_i,Z,x) \), \( \alpha_i(x) \), and \( \gamma_i(x) \), where \( i = 1,2 \). Similarly, we simplify \( d_1(T(x)) \), \( d_{21}(T(x)) \), \( d_{22}(T(x)) \), and \( d_3(T(x)) \) to \( d_1(x) \), \( d_{21}(x) \), \( d_{22}(x) \), and \( d_3(x) \), respectively. 

\subsection{Assumptions}
We always assume that the following properties hold when discussing equation \eqref{eq: general model}:

\begin{itemize}
    \item Each one (denoted as $u_0$) of the initial values $X_0$, $Y_{10}$, $Y_{20}$, $Z_0$ satisfy 
    \begin{equation}
        \label{eq: assumption initial}
        u_0 \in C^{\infty} (\Omega) \backslash \{\mathbf{0}\} \quad \text{and} \quad u_0(x) \geq 0, \ \forall x \in \overline{\Omega}.
    \end{equation}
    \item The parameters $e_1$, $e_2$, $\delta_X$, $\delta_{Y_1}$, $\delta_{Y_2}$, $\delta_Z$ are positive constants.

    \item \label{eq: assumption functions} All functions on the right-hand side of equation \eqref{eq: general model} are smooth functions on $\overline{\Omega}$. For $i = 1$, 2, the funtions \(\alpha_i\), \(\gamma_i\), \(d_{1}\), and $T$ are nonnegative and \(d_{2i}\) and \(d_3\) are strictly positive. The minimum and maximum of these functions are indicated with a subscript of $m$ and $M$ respectively.

    \item The growth function $r(X)$ satisfies
    \begin{equation}
        \label{eq: assumption r}
        r(X) \geq 0, \ X \in [0, K] \quad \text{and} \quad r(X) < 0, \ X > K.
    \end{equation}

    \item The predation functions satisfy
    \begin{equation}
        \label{eq: assumption f}
        0 \leq f_i(X, Y_i, x) \leq C_{f, i} Y_i, \ \forall X, Y_i \geq 0, \ i = 1, 2,
    \end{equation}
    and
    \begin{equation}
        \label{eq: assumption g}
        0 \leq g_i(Y_i, Z, x) \leq C_{g, i} Z, \ \ \forall Y_i, Z \geq 0, \ i = 1, 2.
    \end{equation}

    \item The taxis sensitivity function $h(z)$ satisfies
    \begin{equation}
        \label{eq: assumption h}
        0 \leq h(z) \leq C_h, \ \forall z \geq 0.
    \end{equation}
\end{itemize}

These assumptions are sufficiently general, and all the functions outlined in Section \ref{sec: models} satisfy them. Biologically, the first two assumptions ensure that all initial population densities are nonnegative, and that all behavior-related functions and rates, such as death rate, toxin concentration, and taxis coefficient, are also nonnegative. Assumption \ref{eq: assumption r} guarantees that prey population density will not grow indefinitely but will instead saturate at a point \(K\), which represents the maximum carrying capacity. Before reaching this capacity, the growth rate is positive, but once surpassed, it becomes negative, as seen in classic logistic growth. Assumptions \ref{eq: assumption f} and \ref{eq: assumption g} specify that the predation functions for grazers and predators are sublinear, which is a common modeling assumption. For example, \(f\) and \(g\) could represent Holling type II, or III functional responses. Assumption \ref{eq: assumption h} implies that taxis sensitivity is bounded, which is relatively realistic. Typically, as predator populations increase, their tendency to move and hunt also increases, but only up to a certain threshold. Beyond this point, even with more predators, their inclination to move and hunt does not increase further.

\subsection{Main results}

Under the general assumptions outlined above, we now proceed with further analysis. Below, we abbreviate the norm $||\cdot||_{L^p(\Omega)}$ and $||\cdot||_{W^{k, p}(\Omega)}$ into $||\cdot||_{L^p}$ and $||\cdot||_{W^{k, p}}$. Besides, we denote $C_k (\ldots)$ to be various constants that are independent of the solution to the differential equation. These constants depend continuously on the parameters in the bracket, and remain finite when the parameters are bounded. When the constant is dependent on time $t$, then we will abbreviate it into $C_{k, t}$. We omit the integration variable when performing spatial integration.

The following theorem on the local existence and positivity of solutions ensures that the model \eqref{eq: general model} is biologically meaningful. It guarantees that, for a short time, a positive solution always exists.

\begin{theorem}[Local existence]
    \label{th: local existence}
    Equation \eqref{eq: general model}-\eqref{eq: bc ic} has a unique maximal solution $U = (X, Y_1, Y_2, Z)$ satisfying
    \begin{equation}
        \label{eq: smoothness}
        U \in C \big( \overline{\Omega} \times [0, T_{\max}); \mathbb R_{\geq 0}^4 \big) \cap C^\infty \big( \overline{\Omega} \times (0, T_{\max}); \mathbb R_+^4 \big),
    \end{equation}
    where $T_{\max} \in (0, \infty]$. If $T_{\max} < \infty$, then
    \begin{equation}
        \label{eq: blow up condition simplified}
        \lim_{t \to T_{\max}} \big( ||X (\cdot, t)||_{L^\infty} + ||Y_1 (\cdot, t)||_{L^\infty} + ||Y_2 (\cdot, t)||_{L^\infty} + ||Z (\cdot, t)||_{L^\infty} \big) = \infty.
    \end{equation}
\end{theorem}
The analysis and proof can be found in Appendix \ref{app: proof for local existence}.

For the following, we establish \(L^1\) boundedness of the solution. These estimates ensure that the total population of all species remains bounded in all forward time. All proofs can be found in Appendix \ref{App: proof for L1 L2 estimates}.
\begin{lemma}
    \label{lem: X bounded}
    The solution to equation \eqref{eq: general model}-\eqref{eq: bc ic} satisfies 
    \begin{equation}
        \label{eq: X bounded}
        X(x, t) \leq \max \left\{ ||X_0||_{L^\infty}, K \right\} =: \overline{X}, \ \forall (x, t) \in \overline{\Omega} \times [0, T_{\max}).
    \end{equation}
\end{lemma}

Lemma \ref{lem: X bounded} indicates that the population density of producers cannot exceed either its initial state or the maximum carrying capacity. This is further demonstrated in Section \ref{sec: Simulations}, where we set the initial condition to be less than \(K\), and in all simulations, the population densities of prey remain below the maximum carrying capacity at all times.
Additionally, the following two lemmas establish the \( L^1 \)-boundedness of \( Y_1 \) and \( Y_2 \), indicating that the total population densities of bold and shy grazers are also bounded.
\begin{lemma}
    \label{lemma: boundedness of integral of Y_1}
    The solution for equation \eqref{eq: general model}-\eqref{eq: bc ic} satisfies
    \begin{equation}
        \label{eq: lemma of boundedness of integral of Y_1}
        \left\|Y_1\right\|_{L^1} 
        \leq C_1\big( ||X_0||_{L^\infty}, ||Y_{10}||_{L^1} \big), \quad \forall t \in [0, T_{\max}).
    \end{equation}
    Moreover, if $e_1 C_{f,1} < d_{21m}$, then $\lim\limits_{t\to \infty} \|Y_1\|_{L^1} =0$.
\end{lemma}

This implies that when the predation rate is lower than the death rate, bold grazers will eventually become extinct. Similarly, shy grazers will also face extinction when the predation rate is lower than the death rate, as stated in the following lemma.

\begin{lemma}
    \label{lemma: boundedness of integral of Y_2}
    The solution for equation \eqref{eq: general model}-\eqref{eq: bc ic} satisfies
    \begin{equation}
        \label{eq: lemma of boundedness of integral of Y_2}
        \left\|Y_2\right\|_{L^1} \leq C_2 \big( ||X_0||_{L^\infty}, ||Y_{20}||_{L^1} \big), \quad \forall t \in [0, T_{\max}).
    \end{equation}
    Moreover, if $e_1 C_{f,2} < d_{22m}$, then $\lim\limits_{t\to \infty} ||Y_2||_{L^1} = 0$.
\end{lemma}

\begin{lemma}
    \label{lemma: boundedness of integral of Z}
    The solution for equation \eqref{eq: general model}-\eqref{eq: bc ic} satisfies
    \begin{equation}
        \label{eq: lemma of boundedness of integral of Z}
        \left\|Z\right\|_{L^1}\leq C_3 \big( ||U_0||_{L^1}, ||X_0||_{L^\infty} \big), \quad \forall t \in [0, T_{\max}).
    \end{equation}
    Moreover, $\lim\limits_{T_m \to \infty} \limsup\limits_{t\to \infty}\int_\Omega \left(e_1 e_2 X + e_2 Y_1 + e_2 Y_2 + Z \right) \leq e_1 e_2 \overline{X} |\Omega|$. Additionally, if $e_2\left( C_{g,1} + C_{g,2}\right) < d_{3m}$, then $\lim\limits_{t\to \infty} ||Z||_{L^1} = 0$.
\end{lemma}

Lemma \ref{lemma: boundedness of integral of Z} implies that for any $A > e_1 e_2 \overline{X} |\Omega|$, as the toxin concentration increases sufficiently, the system admits a global attractor 
\begin{equation*}
    \left\{X\geq 0, \; Y_1 \geq 0, \;  Y_2\geq 0, \; Z\geq 0, \; \int_\Omega \left(e_1 e_2 X + e_2 Y_1 + e_2 Y_2 + Z \right) < A \right\}.
\end{equation*}
Moreover, since high toxin levels dramatically raise the death rate to a level exceeding the predation rate, both grazers and predators are unable to survive under such conditions. As a result, only the producers persist. This aligns with the findings discussed in Section \ref{sec: Influence spatial distribution of toxin}. For instance, in regions where toxin levels are highest, such as the edges in a rectangular toxin distribution or the left edge in a linear gradient distribution, only producers survive. These patterns are illustrated in Figures \ref{fig: Multiple_toxin_distribution_2D_T0=30}.

The following theorem demonstrates that the solution not only exists for a short time, but also persists for all future times without blowing up. Therefore, the model is well-defined.

\begin{theorem}[Global existence]
    \label{th: global existence}
    Under assumption \eqref{eq: condition A positive definite}, equation \eqref{eq: general model}-\eqref{eq: bc ic} has a unique maximal solution $U = (X, Y_1, Y_2, Z)$ satisfying
    \begin{equation*}
        U \in C \big( \overline{\Omega} \times [0, \infty); \mathbb R_{\geq 0}^4 \big) \cap C^\infty \big( \overline{\Omega} \times (0, \infty); \mathbb R_+^4 \big).
    \end{equation*}
\end{theorem}
All proofs are provided in Appendix \ref{App: proof for global existence}. Our proof relies on coupled energy estimates to establish initial bounds on \(X\), \(Y_1\), \(Y_2\), and \(Z\). We then utilize various properties of the heat kernel, combined with a bootstrap argument, to enhance the regularity and integrability of the solution. The primary difficulty arises from the fact that, to establish an \(L^\infty\) bound on \(Z\), it is necessary to first derive an a priori bound for \(||\nabla Y_i||_{L^\infty}\), which depends on both \(Y_i\) and \(X\). 

The main idea of the proof is as follows: Lemmas \ref{lem: Ln+1 Boundedness of Y_1} and \ref{lemma: Ln+1 Boundedness of $Y_2$} provide estimates for $||Y_i||_{L^{n + 1}}$ through estimates of \( ||\nabla X||_{L^2} \) and \( ||Y_i||_{L^2} \), $i = 1$, 2. Similarly, Lemma \ref{lemma: L^n+1 Boundedness of nabla X} gives an estimate for \( ||\nabla X||_{L^{n + 1}} \) using \( ||Y_1||_{L^2} \) and \( ||Y_2||_{L^2} \). Replacing the index pair \( (2, n + 1) \) with \( (n + 1, \infty) \) in the above proofs, we deduce \( L^\infty \) estimates for \( Y_1 \), \( Y_2 \), and \( \nabla X \). Using Lemma \ref{lem: Young's convolution inequality on T}, along with the \( L^\infty \) boundedness of \( X \), \( Y_1 \), \( Y_2 \), and \( \left\|\nabla U\right\|_{L^2(\Omega)L^2(0,t)} \), we establish the boundedness of \( \|\Delta X\|_{L^2(\Omega)L^{p}(0,t)} \), followed by \( \|\nabla Y_1 \|_{L^2(\Omega)L^{p}(0,t)} \). Then, \( \|\nabla Y_1 \|_{L^{n+1}} \) and \( \|\nabla Y_2 \|_{L^{n+1}} \) can be obtained, and by applying the same process with the index pair \( (2, n + 1) \) replaced by \( (n + 1, \infty) \), we ultimately obtain \( \| Z \|_{L^\infty} \). These boundedness results, together with the local existence from Theorem \ref{th: local existence}, establish the global existence of the solution.

\section{Numerical results}
\label{sec: Simulations}
In this section, we conduct numerical simulations to further explore the influence of toxins on population dynamics. We performed simulations in both one and two dimensional space. All parameter details can be found in Table \ref{Table: Parameters}.

\begin{table}[h!]
    \centering
    \caption{\footnotesize Parameters in the system \eqref{eq: general model}}
    \label{Table: Parameters}
    \begin{tabular}{lllll}
    \hline
        \toprule
Para. & Description & Value & Unit & Ref. \\
\midrule
$r$ & Growth rate of producers & 1.2 & 1/day & $\bullet$ \\
$K$ & Carrying capacity for the producers & 1.5 & mg/(L$\cdot$m) & $\bullet$ \\
$e_1$ & Production efficiency of zooplankton & 0.8 & no unit & $\bullet$ \\
$e_2$ & Production efficiency of fish & 0.5 & no unit & $\bullet$ \\
$a_1$ & Coefficient of encounter rate & -0.0034 & L$^3$/(day$\cdot$mgC$\cdot $µgT$^2$) & $\bullet\vartriangle$ \\
$b_1$ & Coefficient of encounter rate & 0.085 & L$^2$/(day$\cdot$mgC$\cdot$µgT) & $\bullet\vartriangle$ \\
$c_1$ & Coefficient of encounter rate & 4.5384 & L/(day$\cdot$mgC) & $\bullet\vartriangle$ \\
$\xi$ & Shyness index & 0.6 & no unit & $\circ$ \\
$a_2$ & Coefficient of encounter rate & -0.0004 & L$^3$/(day$\cdot$mgC$\cdot $µgT$^2$) & $\bullet\vartriangle$ \\
$b_2$ & Coefficient of encounter rate & 0.0012 & L$^2$/(day$\cdot$mgC$\cdot $µgT) & $\bullet\vartriangle$ \\
$c_2$ & Coefficient of encounter rate & 0.0516 & L/(day$\cdot$mgC) & $\bullet\vartriangle$ \\
$\eta$ & Taxis coefficient ratio based on boldness & 1 & no unit & $\circ$ \\
$d_{20}$ & Natural death rate of grazers & 0.25 & day$^{-1}$ & $\bullet$ \\
$d_{30}$ & Natural death rate of predators & 0.003 & day$^{-1}$ & $\bullet$ \\
$d_{2}$ & Effect of toxin on bold grazer death rate & 0.0159 & L$\cdot$m/(µg$\cdot$day) & $\blacktriangledown$ \\
$d_{31}$ & Effect of toxin on predator death rate & 0.00004 & L$\cdot$m/(µg$\cdot$day) & $*$ \\
$\tau_{1}$ & Handling time for zooplankton & 1.235 & day & $\bullet$ \\
$\tau_{2}$ & Handling time for fish & 33.33 & day & $\bullet$ \\
$\delta_X$ & Diffusion coefficient of phytoplankton & 0.05 & m$^2$/day & $\blacklozenge$ \\
$\delta_{Y_1}$ & Diffusion coefficient of bold zooplankton & 0.05 & m$^2$/day & $\blacklozenge$ \\
$\delta_{Y_2}$ & Diffusion coefficient of shy zooplankton & 0.05 & m$^2$/day & $\blacklozenge$ \\
$\delta_Z$ & Diffusion coefficient of fish & 0.05 & m$^2$/day & $\blacklozenge$ \\
$\kappa$ & Coefficient for taxis function & 0.1 & L$\cdot$m/mg & $\circ$ \\
$T_0$ & Maximal toxin concentration & 0 - 1000 & µg/(L$\cdot$m) & $\vartriangle \blacktriangledown * \circ$ \\
        \bottomrule
        \hline
    \end{tabular} 
    \caption*{\footnotesize 
    Note: $\bullet$ Parameters related to producers (e.g., phytoplankton) and consumers (e.g., zooplankton) are selected from \citep{andersen2013pelagic}, \citep{urabe1996regulation}. Parameters related to predators (e.g., fish) are chosen from \citep{jorgensen1991handbook}, \citep{magnea2013model}. These parameters have been widely used in models such as \citep{loladze2000stoichiometry,chen2017global,wang2024stoichiometric}. The maximal production efficiency of fish is 0.75, but here we take 0.5. $\vartriangle$ Fitted from data in \citep{glazer2021developmental}. $\blacktriangledown$ Fitted from data in \citep{jensen2007lethal}. $*$ Fitted from data in \citep{huang2011toxicity}. $\blacklozenge$ Followed from \citep{demir2021spatial}. $\circ$ Assumed values. The maximal toxin concentration in experiments from \citep{glazer2021developmental}, \citep{huang2011toxicity}, and \citep{jensen2007lethal} is below 50 µg/L. Here, we consider a wider range of 0–1 mg/L.
    }
\end{table}

We aim to address the following key issues. In one-dimensional space, we first explore the long-term behavioral responses of grazers with different personality traits in both toxin-free and toxin-present environments. Additionally, we investigate how variations in boldness disparity between the two grazer types affect their behavior. Finally, from a broader ecosystem perspective, we examine the overall impact of toxins on the total grazer population. In a two-dimensional region, we analyze how toxin levels and spatial distributions of toxins influence the spatial patterns of populations.

\subsection{One dimensional space}

For the one-dimensional simulation, the domain is the line segment from \(-100\) to \(100\) meters, with initial values given by $$(X_0, Y_{10}, Y_{20}, Z_0) = (0.5 + 0.2 \cos\left(0.07 x\right), 0.5 - 0.2 \cos\left(0.05 x\right), 0.5 - 0.2 \cos\left(0.05 x\right), 0.5 + 0.3 \cos\left(0.09 x\right) ).$$ The time span is set from 0 to 500 days. 

In large aquatic environments, toxins are typically distributed in a highly spatially heterogeneous manner. When a pollutant, such as oil, chemicals, or heavy metals, is released into a large lake or ocean, its concentration decreases rapidly as it moves away from the source due to processes like dilution, dispersion, and degradation. The exponential decay function effectively models this behavior by representing a high concentration near the source that diminishes quickly with increasing distance. In cases where the pollution source is centrally located, the distribution tends to be symmetric. Here, we model the toxin concentration as
\[
T(x) = T_0 e^{-(x/15)^2},
\]
where \(T_0\) represents the maximum concentration at the origin (\(x = 0\)), and the concentration decays exponentially as distance from the source increases.

\subsubsection{Behavioral responses to toxins based on personality traits} 

\begin{figure}[h!]
    \centering
    \begin{subfigure}[b]{0.44\textwidth}
         \centering        \includegraphics[width=\textwidth]{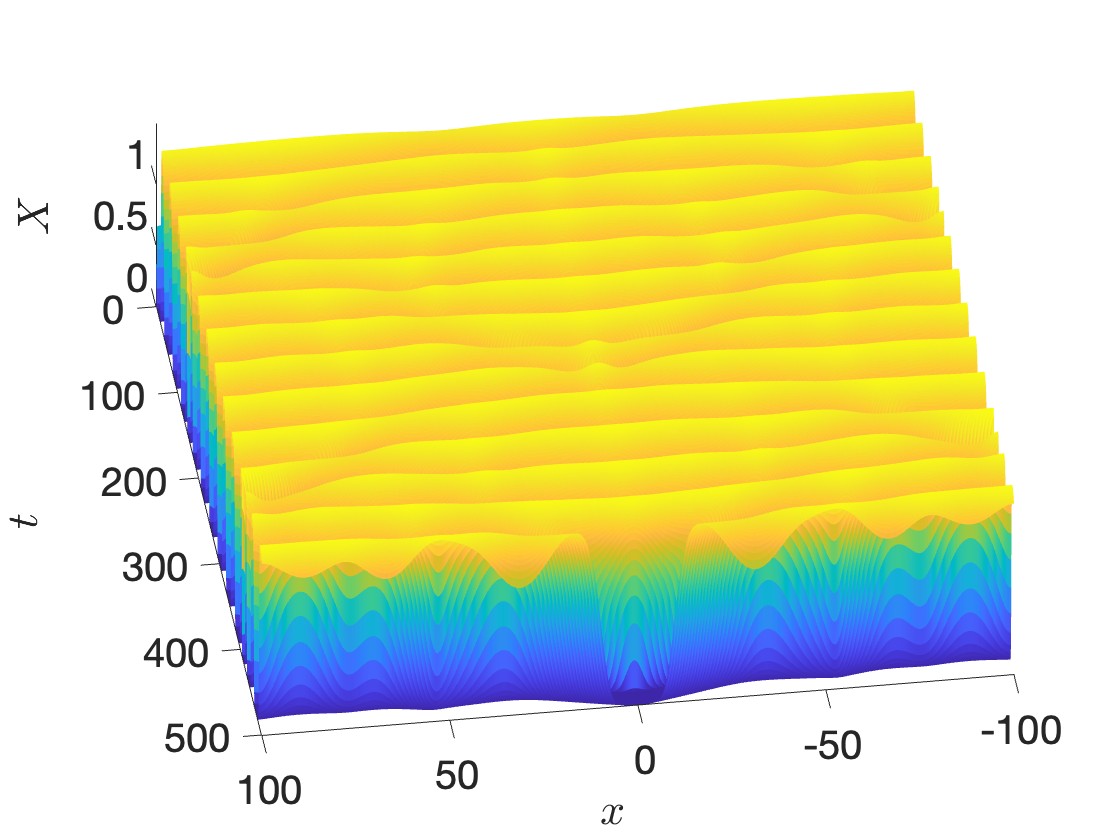}       \caption{}   
         \label{fig: X_t=500_T0=0.jpg}
    \end{subfigure}
          \begin{subfigure}[b]{0.44\textwidth}
         \centering        \includegraphics[width=\textwidth]{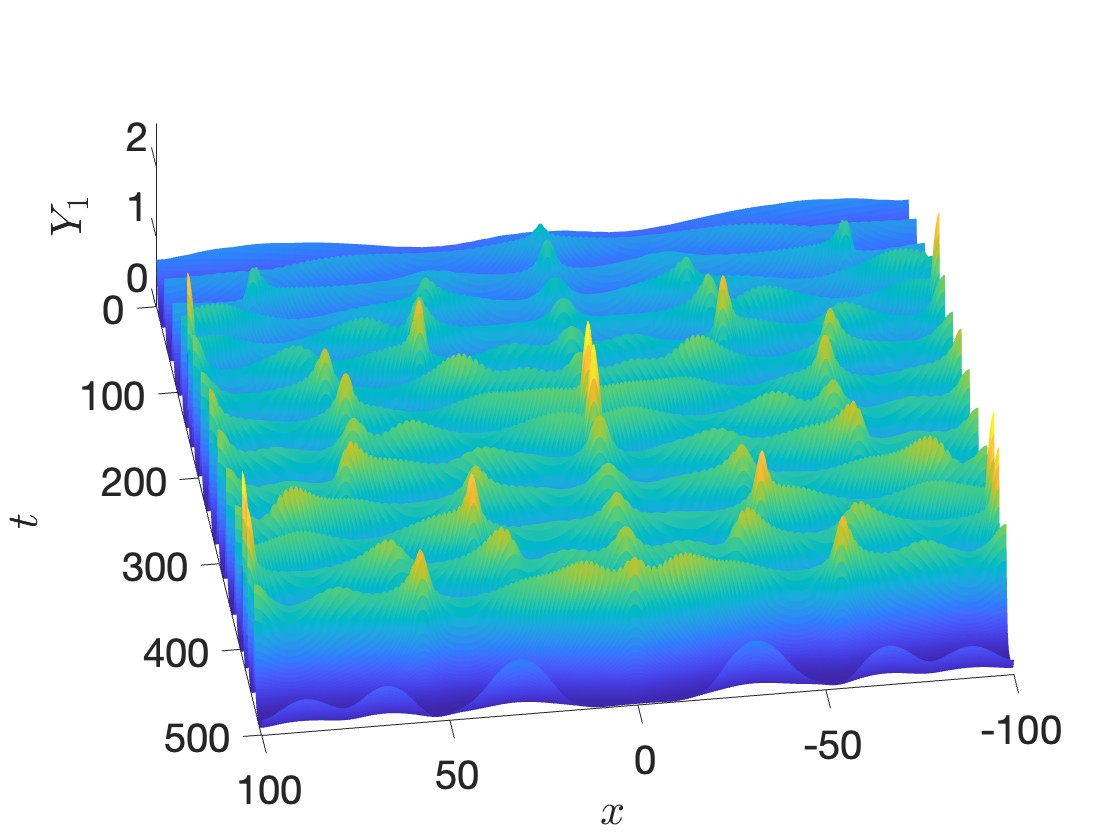}       \caption{}   \label{fig: Y1_t=500_T0=0.jpg}
    \end{subfigure}   
    \begin{subfigure}[b]{0.44\textwidth}
         \centering        \includegraphics[width=\textwidth]{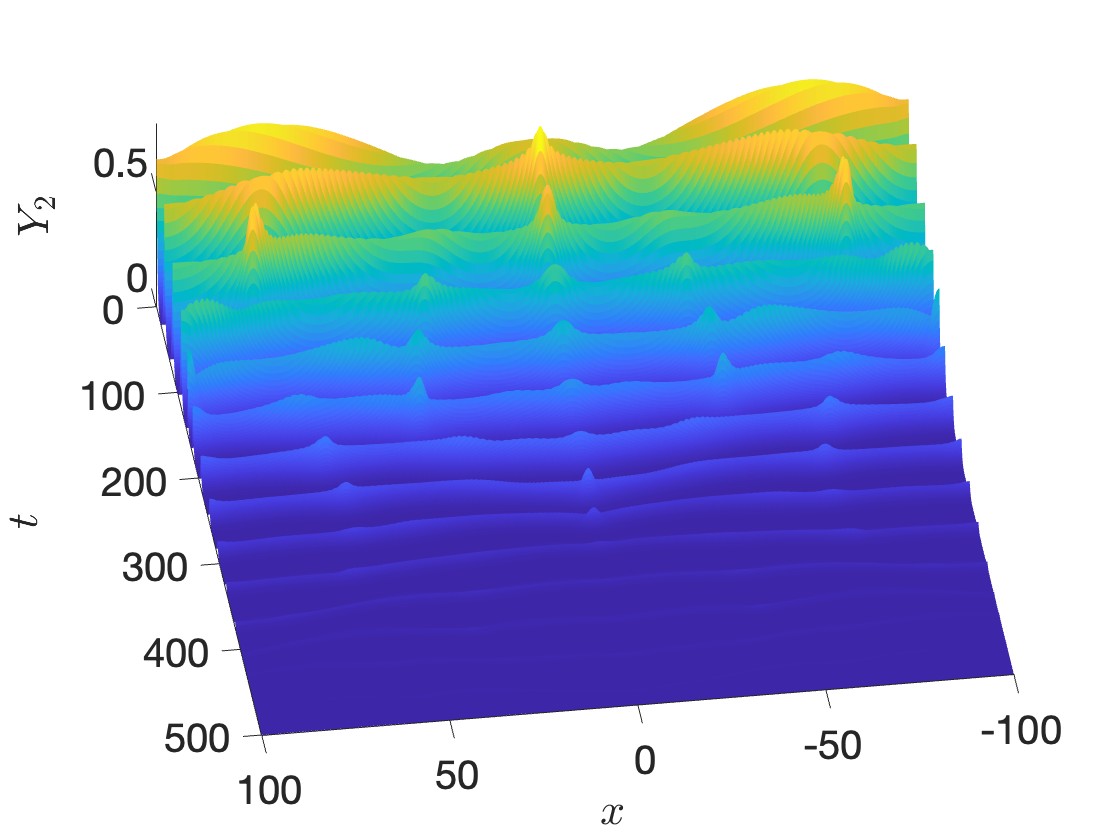}       \caption{}   \label{fig: Y2_t=500_T0=0.jpg}
     \end{subfigure}
    \begin{subfigure}[b]{0.44\textwidth}
         \centering        \includegraphics[width=\textwidth]{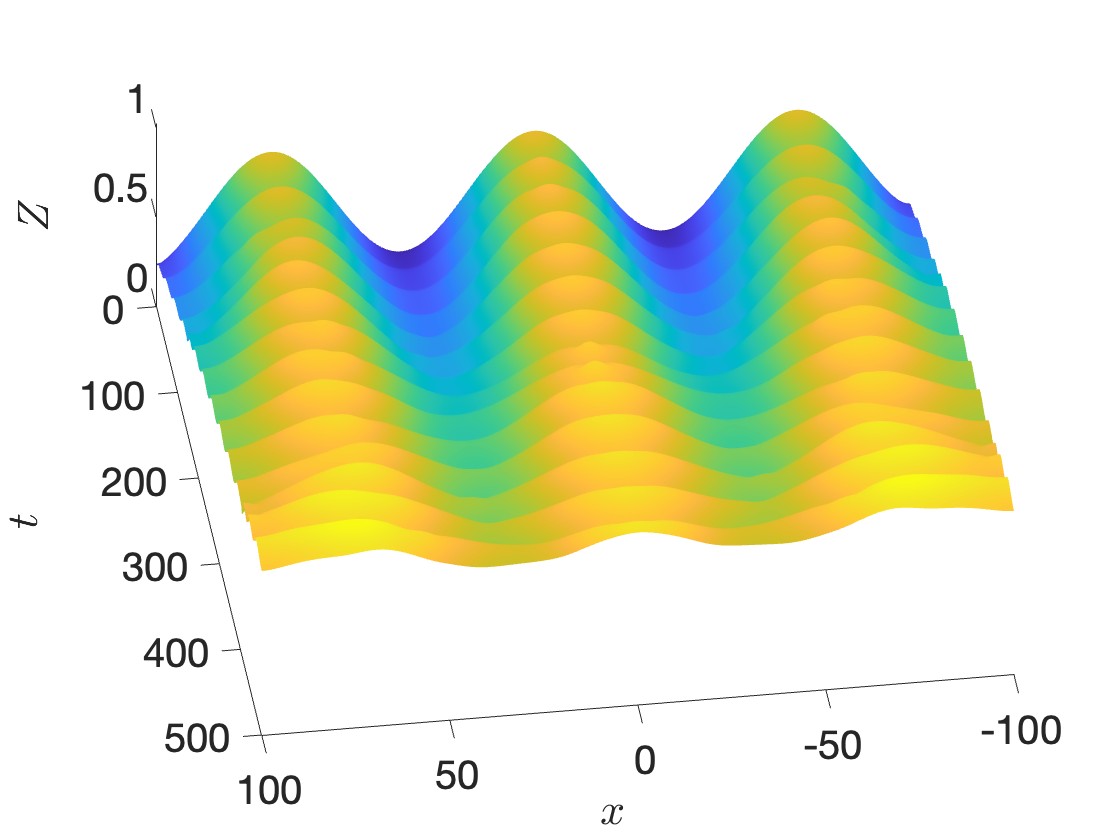}       \caption{}   
         \label{fig: Z_t=500_T0=0.jpg}
    \end{subfigure} 
\caption{\footnotesize Spatial-temporal population densities in a toxin-free environment: (a) producers, (b) bold grazers, (c) shy grazers, (d) predators.}
     \label{fig: T0=0}
\end{figure}

We first investigate the differences between grazers with different personality traits in a toxin-free environment. The result can be seen in Figure \ref{fig: T0=0}. Over time, the producer population oscillates periodically with an almost uniform spatial distribution, as seen in Figure \ref{fig: X_t=500_T0=0.jpg}. Bold grazers outperform shy grazers, eventually leading to the extinction of the shy population (\ref{fig: Y1_t=500_T0=0.jpg}, \ref{fig: Y2_t=500_T0=0.jpg}). The total predators population increases, as shown in Figure \ref{fig: Z_t=500_T0=0.jpg}. This suggests that, in a toxin-free environment, bold personalities are more competitive. Note that although bold grazers have a greater likelihood of successful foraging and predation, they also face a higher risk of predation. Our results show that despite this trade-off, bold grazers still tend to maintain a competitive advantage in most cases within a toxin-free ecosystem.

\begin{figure}[h!]
    \centering
    \begin{subfigure}[b]{0.32\textwidth}
         \centering        \includegraphics[width=\textwidth]{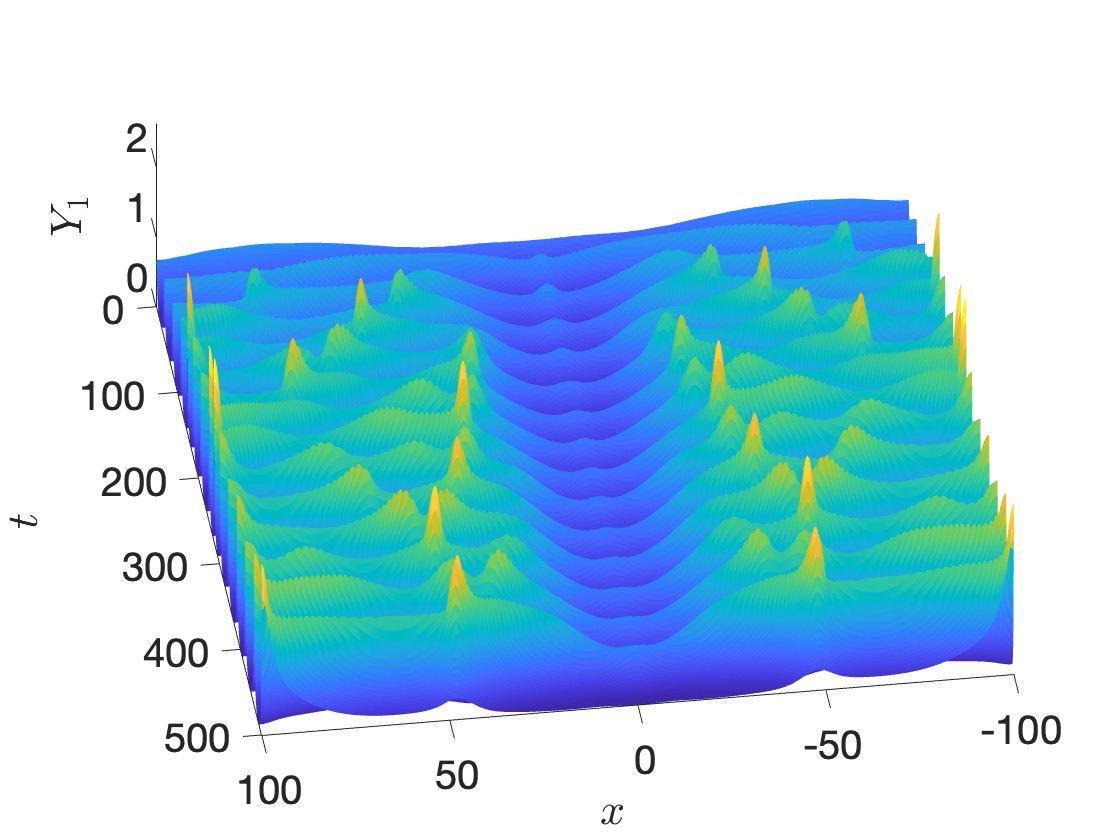}       \caption{}   \label{fig: Y1_t=500_T0=6.jpg}
    \end{subfigure}   
    \begin{subfigure}[b]{0.32\textwidth}
         \centering        \includegraphics[width=\textwidth]{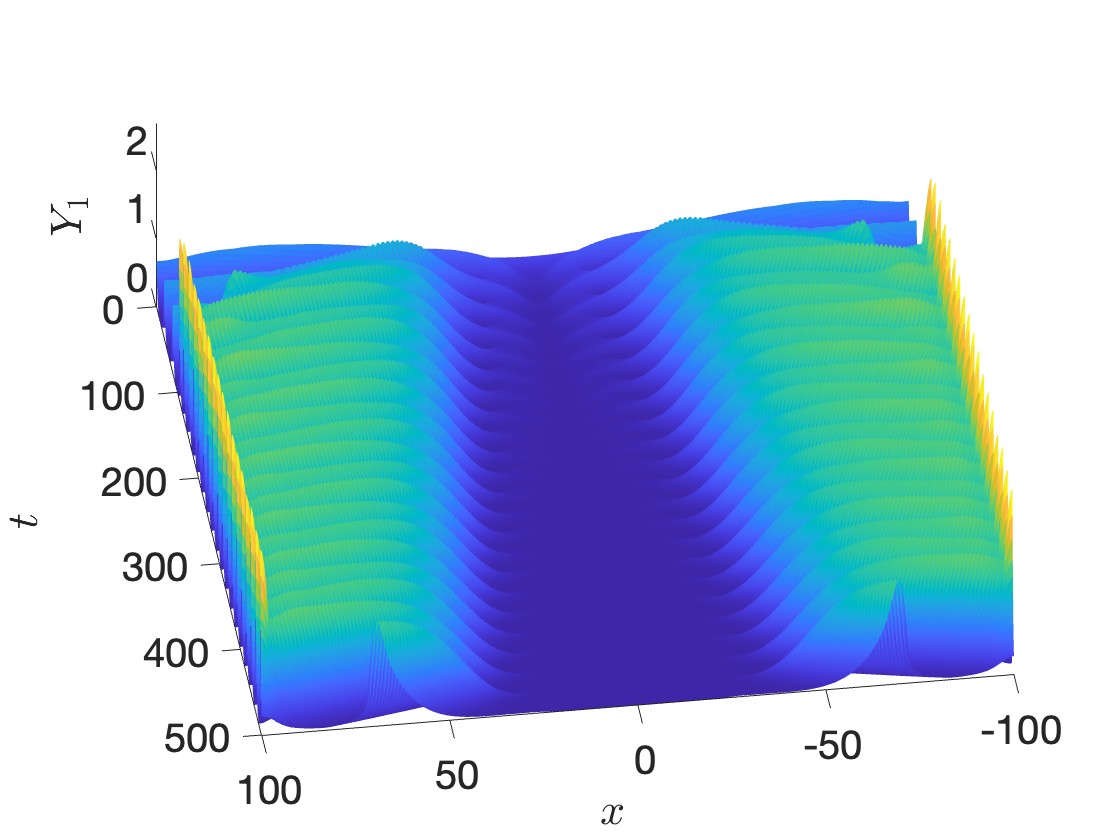}       \caption{}   \label{fig: Y1_t=500_T0=30.jpg}
    \end{subfigure}
    \begin{subfigure}[b]{0.32\textwidth}
         \centering        \includegraphics[width=\textwidth]{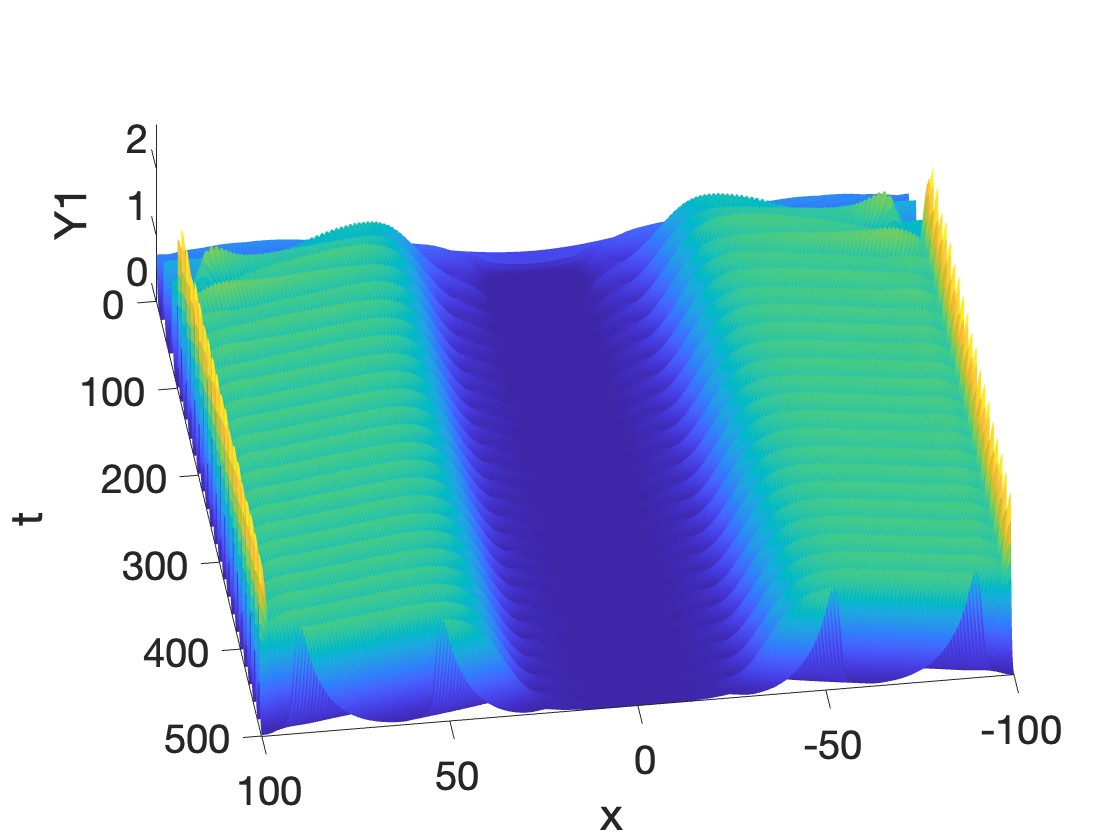}       \caption{}   \label{fig: Y1_t=500_T0=100.jpg}
           \end{subfigure} 
    \begin{subfigure}[b]{0.32\textwidth}
         \centering        \includegraphics[width=\textwidth]{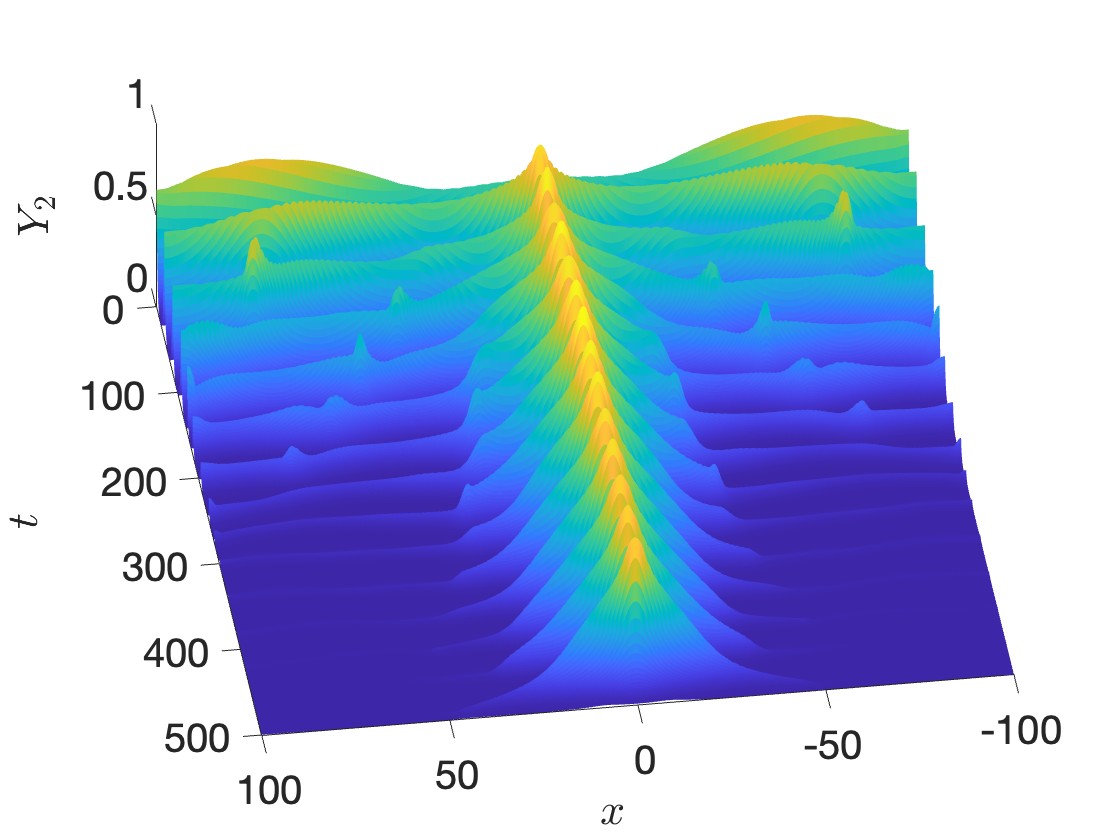}       \caption{}   \label{fig: Y2_t=500_T0=6.jpg}
     \end{subfigure}   
    \begin{subfigure}[b]{0.32\textwidth}
         \centering        \includegraphics[width=\textwidth]{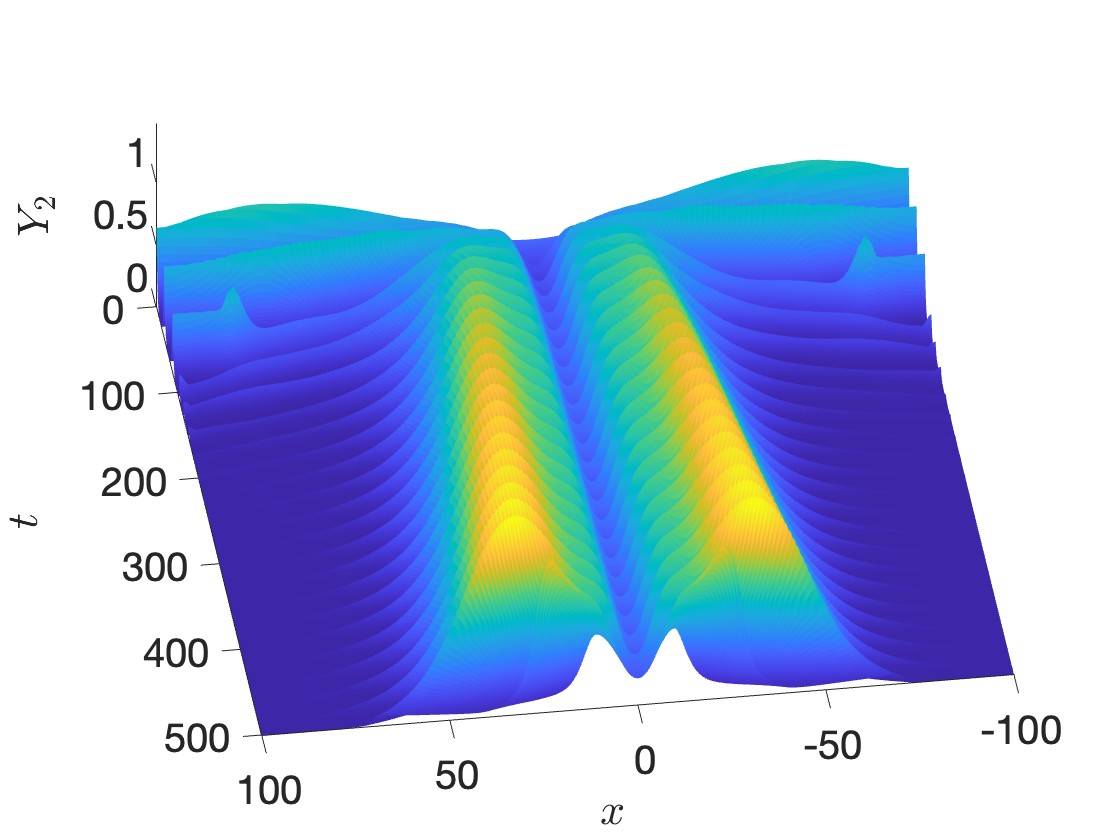}       \caption{}   \label{fig: Y2_t=500_T0=30.jpg}
    \end{subfigure}
    \begin{subfigure}[b]{0.32\textwidth}
         \centering        \includegraphics[width=\textwidth]{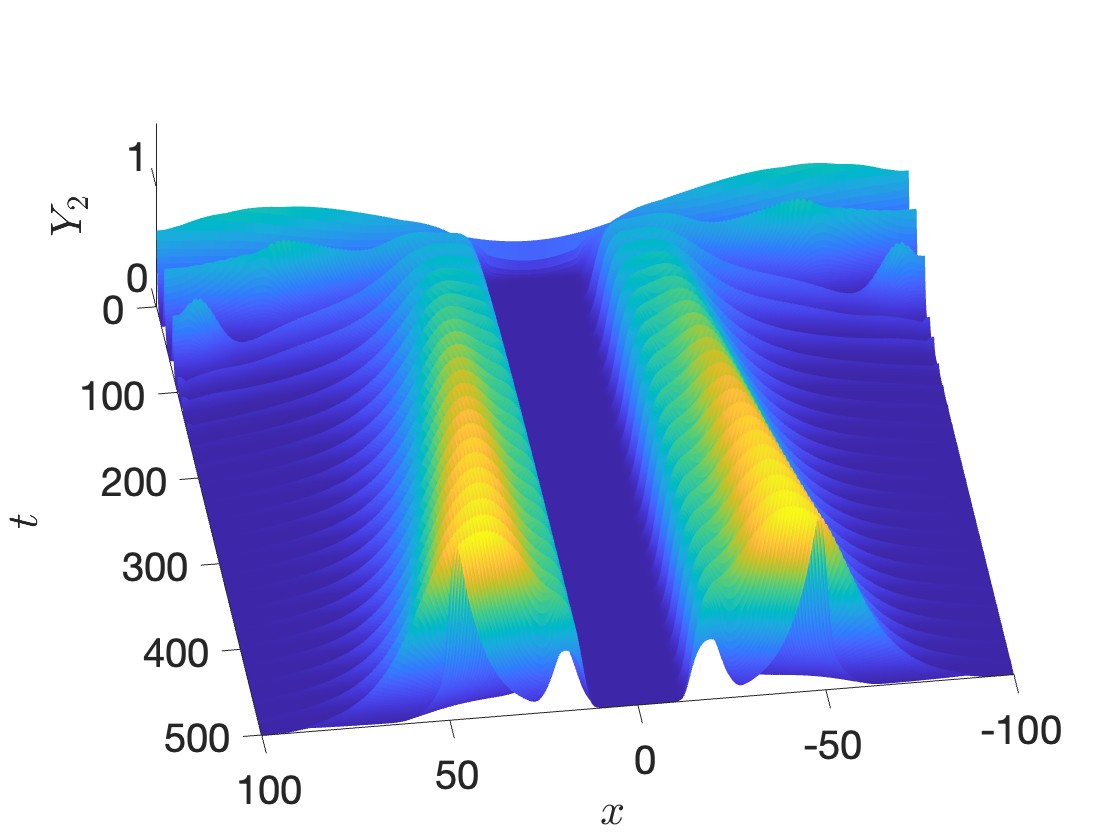}       \caption{}   \label{fig: Y2_t=500_T0=100.jpg}
    \end{subfigure}
 \caption{\footnotesize Spatial-temporal population densities of bold (a–c) and shy (d–f) grazers under varying toxin concentrations. (a)(d) $T_0 = 6$, (b)(e) $T_0 = 30$, (c)(f) $T_0 = 100$.}
     \label{fig: Y1Y2}
\end{figure}

Now we consider how the behavioral responses of grazers with different personalities change under varying toxin concentrations. This is illustrated in Figure \ref{fig: Y1Y2}. Panels (a)–(c) display the population densities of bold grazers, while panels (d)–(f) show those of shy grazers. Panels (a) and (d) represent low toxin levels (\(T_0 = 6\)), (b) and (e) show moderate toxin levels (\(T_0 = 30\)), and (c) and (f) illustrate high toxin levels (\(T_0 = 100\)).

The differences in behavioral responses to toxins based on personality traits are significant. Compared to the toxin-free scenario, where bold grazers are always more competitive, surprisingly, shyness becomes an advantage in the presence of toxins. At low toxin levels ($T_0 = 6$), shy grazers quickly aggregate near the toxic spot, and both bold and shy grazers coexist within 500 days. At moderate toxin levels ($T_0 = 30$), bold grazers cannot survive near the highly toxic region and spread away from the polluted area, while shy grazers aggregate closer to the pollution source. When the toxin concentration increases to $T_0 = 100$, neither personality type can survive in the highest polluted region near origin. Bold grazers remain far from the polluted area, while shy grazers cluster in neighboring regions with low to moderate toxin levels.

\begin{figure}[ht!]
         \centering        
         \includegraphics[width=0.45\textwidth]{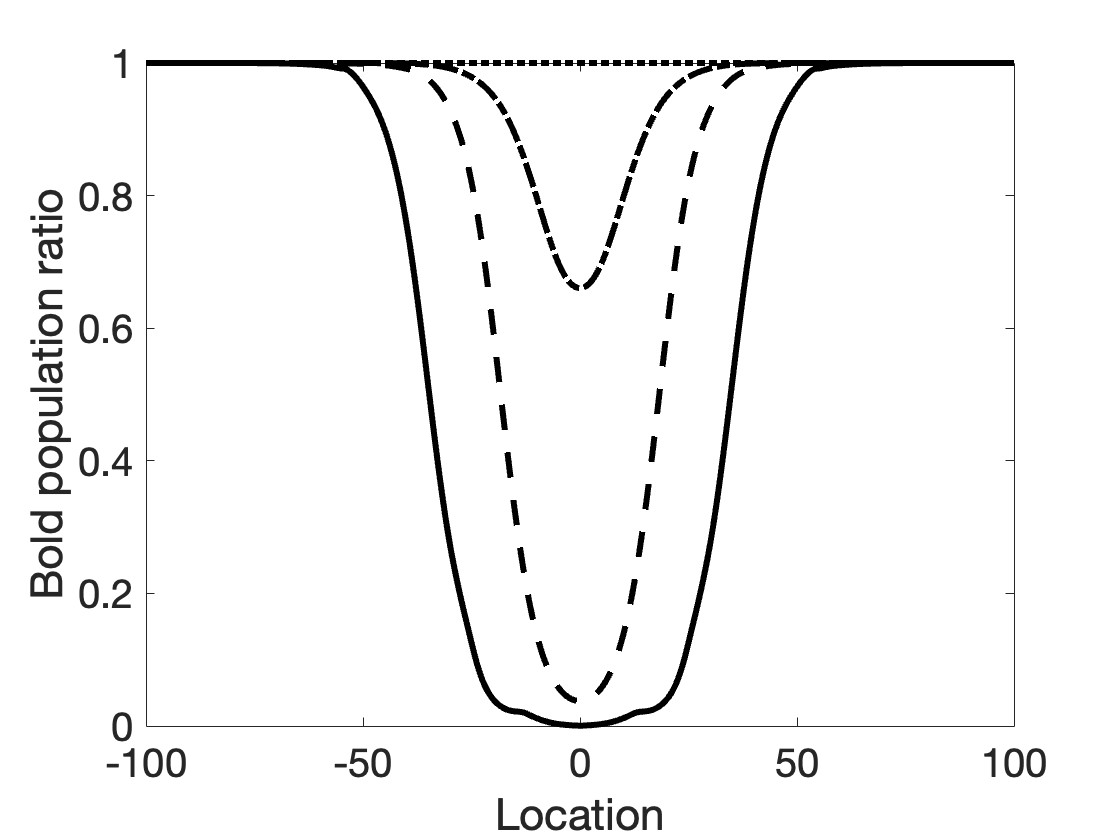} 
\caption{\footnotesize Bold population ratio. The dotted curve represents $T_0 = 0$, dash-dotted curve $T_0 = 6$, dashed curve $T_0 = 30$, and solid curve $T_0 = 100$. }
     \label{fig: ratio.jpg}
\end{figure}

The reshaping of population structure due to toxin levels and spatial distribution can further be analyzed by examining the bold population ratio (the proportion of bold grazers relative to the total grazer population). Figure \ref{fig: ratio.jpg} shows the bold population ratio, ranging from 0 to 1, at time \(t = 500\). As the maximum toxin concentration \(T_0\) increases, the polluted area expands, and the bold population ratio decreases accordingly. Near the origin, where toxin levels are highest, the bold population ratio becomes very low as \(T_0\) increases. In toxin-free regions farther from the polluted area, the bold population ratio remains at 1, indicating that only bold grazers are present. Closer to the polluted center, where toxin levels are low to moderate, the bold population ratio decreases significantly. This further confirms that shy grazers tend to cluster in these low to moderate pollution zones. As toxin levels rise, these regions expand, providing shy grazers with more area to survive.

\begin{figure}[ht!]
     \centering
    \begin{subfigure}[b]{0.45\textwidth}
         \centering        \includegraphics[width=\textwidth]{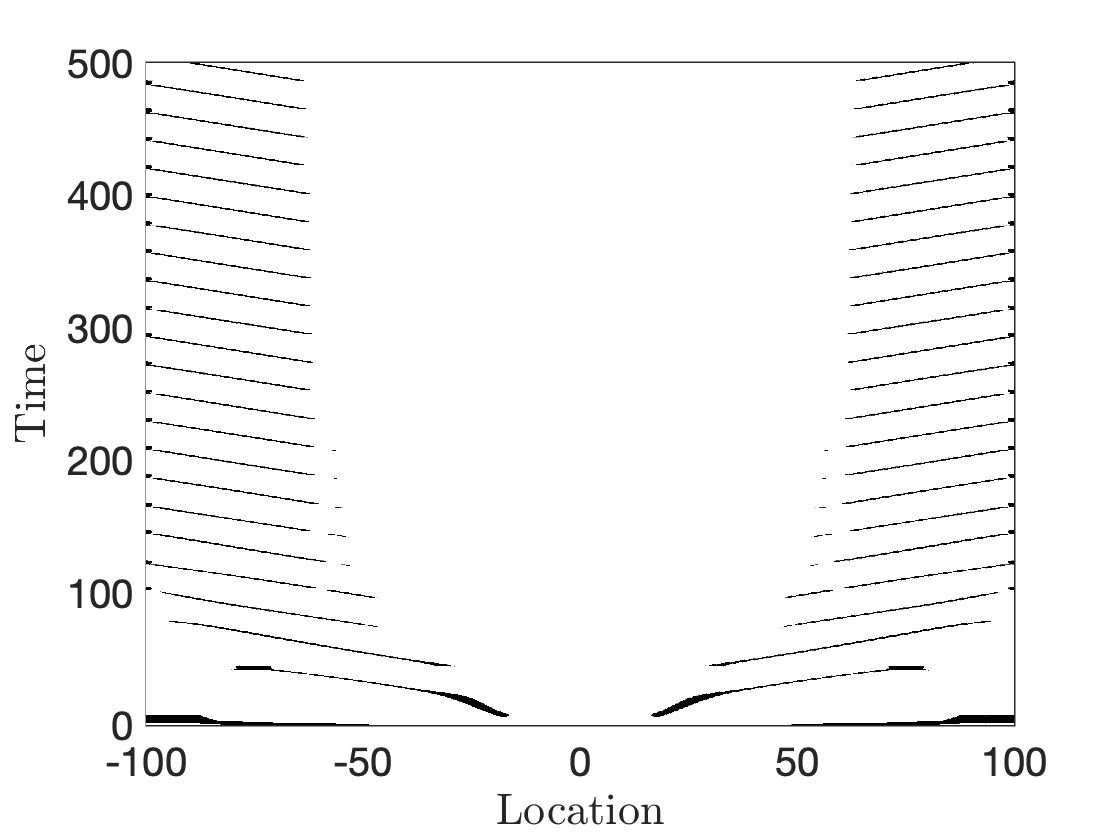}       
         \caption{}   
         \label{fig: trace_Y1.jpg}
    \end{subfigure} 
    \begin{subfigure}[b]{0.45\textwidth}
         \centering        \includegraphics[width=\textwidth]{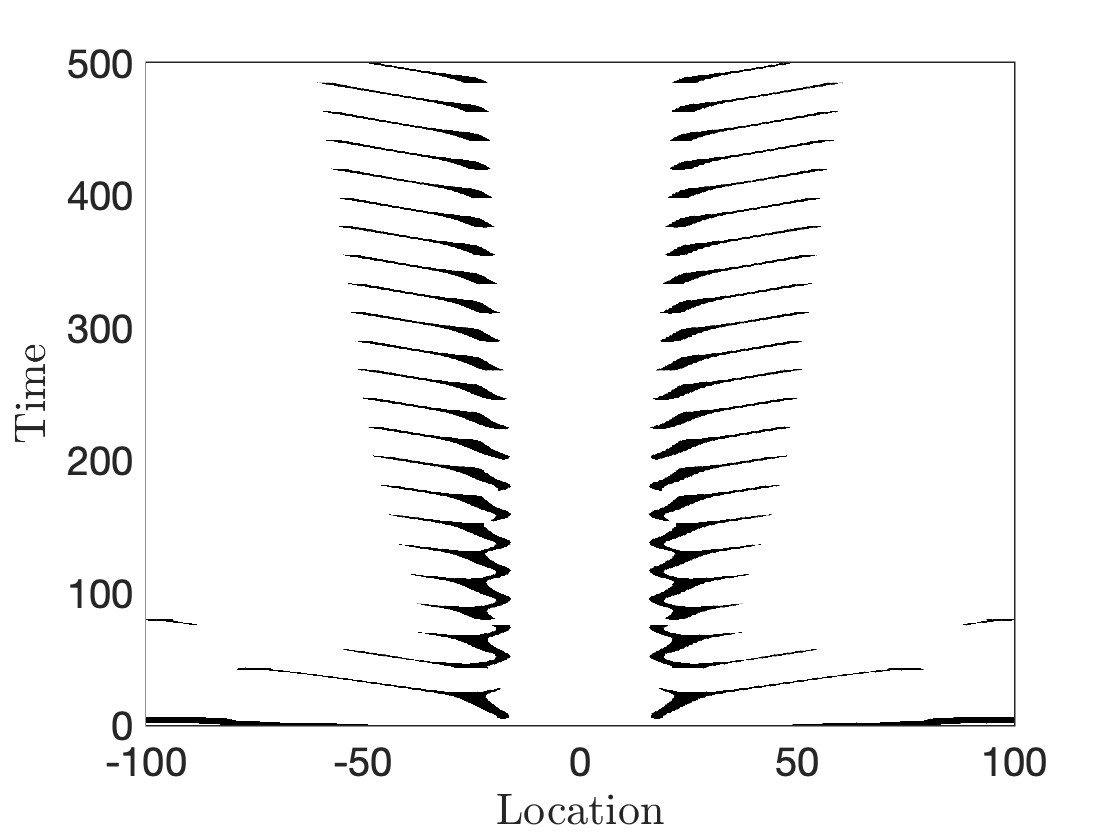}       
         \caption{}   
         \label{fig: trace_Y2.jpg}
    \end{subfigure}
\caption{\footnotesize Movement trace of the majority population at $T_0 = 100$: (a) bold grazers, (b) shy grazers.}
     \label{fig: trace.jpg}
\end{figure}

Finally, we compare the movement patterns between grazers with different personalities. Figure \ref{fig: trace.jpg} shows the movement trace of the majority of the grazer population at \(T_0 = 100\). At each time point, we record the locations where the population density exceeds 90\% of the maximum. Figure \ref{fig: trace_Y1.jpg} represents the trace for bold grazers. Over time, this population follows the toxin gradient, gradually aggregating in toxin-free regions. Figure \ref{fig: trace_Y2.jpg} shows the movement trace of shy grazers. While most shy grazers also tend to move away from the polluted center, their trace remains closer to low-pollution areas.

\subsubsection{Personality discrepancy influence behavioral responses to toxins}

\begin{figure}[h!]
    \centering
    \begin{subfigure}[b]{0.44\textwidth}
         \centering        \includegraphics[width=\textwidth]{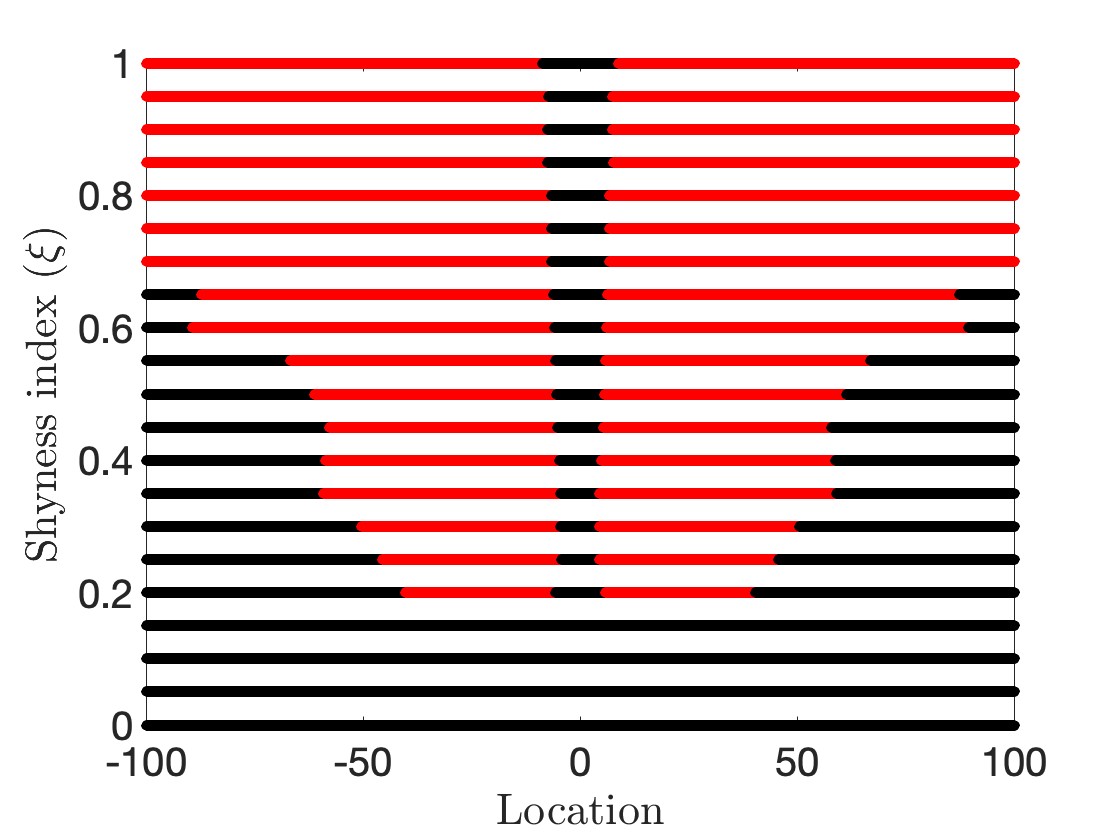}       \caption{}   
         \label{fig: xi.jpg}
    \end{subfigure}
          \begin{subfigure}[b]{0.44\textwidth}
         \centering        \includegraphics[width=\textwidth]{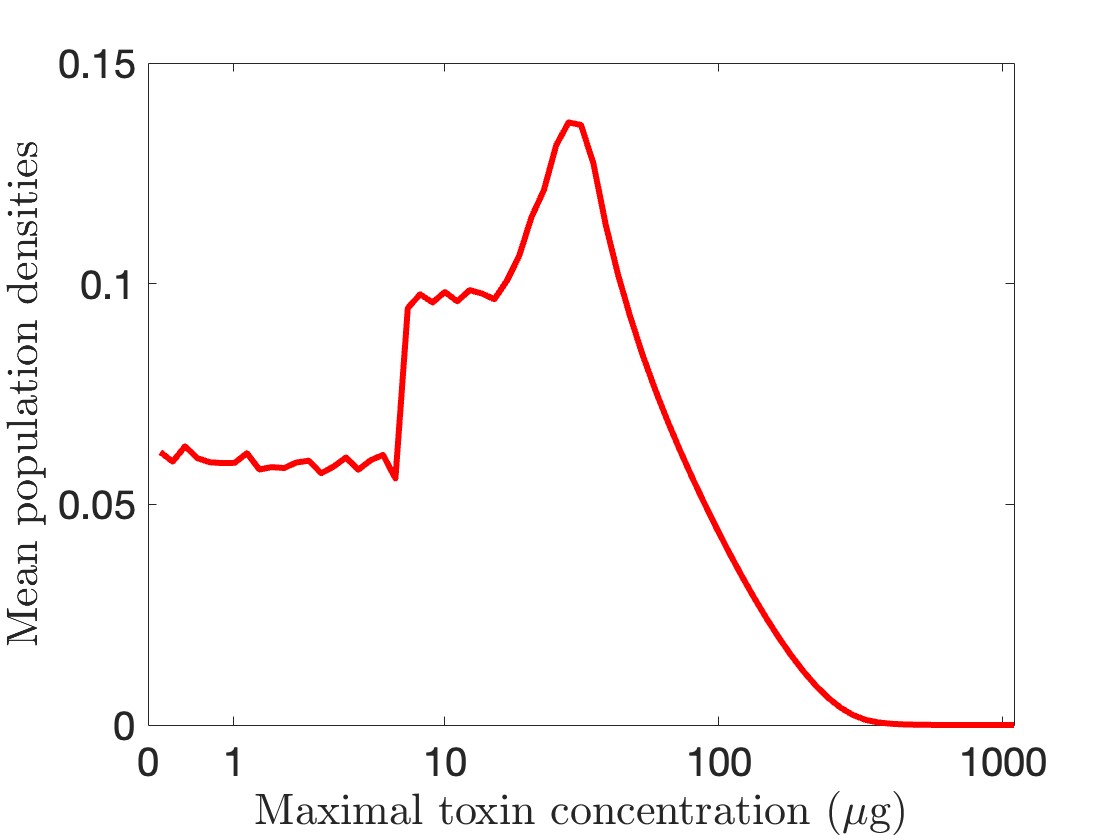}       
     \caption{}   
     \label{fig: mean grazer population.jpg}
    \end{subfigure}   
\caption{\footnotesize (a) Survival state of shy grazers at different locations with varying shyness indices. $T_0 = 100$. (b) Mean population density of all grazers under varying toxin concentrations near a polluted spot.}
     \label{}
\end{figure}

The shyness index represents the personality discrepancy between shy and bold animals. When \(\xi = 1\), shy grazers share the same boldness as bold grazers, whereas \(\xi = 0\) indicates an extreme difference between the two personalities, with shy grazers being highly timid. This raises an important question: how would population dynamics shift with varying degrees of personality discrepancy between the two groups.

Figure \ref{fig: xi.jpg} shows the spatial distribution of shy grazers at \(T_0 = 100\), at time \(t = 500\). The red regions indicate the presence of shy populations, while black regions indicate their absence. When the shyness index is extremely low, i.e. the shy population is very timid, then shy grazers cannot survive due to their low predation activity. However, when the shyness index is in a moderate range, shy grazers tend to stay near the polluted center, specifically in regions with low to moderate pollution levels. On the other hand, as the shyness index increases, the personality differences between bold and shy populations diminish. Consequently, the survival region of the shy population expands, and their behavior becomes more similar to that of the bold population.

\subsubsection{Influence of toxins on total population}
As previously discussed, individuals with different personalities exhibit different responses to toxins. Now, we will take a broader ecosystem perspective to explore how toxins affect the overall population dynamics of grazers.

Figure \ref{fig: mean grazer population.jpg} illustrates the mean population density of all grazers within a polluted region from -20 to 20. When toxin concentration is relatively low, the mean population density remains almost unchanged or exhibits slight oscillations. This suggests that the system has some resistance to toxins and can maintain a relatively stable state in a mildly polluted environment. When the toxin concentration is in the lower range (\(6.4 < T_0 < 32.1\)), surprisingly, the mean population increases as toxin concentration rises, with certain phases exhibiting steady growth. Moreover, this process go through multiple important thresholds. However, as toxin levels accumulate and exceeds \(T_0 = 32.1\), the mean population density starts to decrease as \(T_0\) increases. This implies that the population can initially benefit from low toxin levels, possibly due to adaptive behaviors like adjusting predation strategies, increasing predation activity, or enhancing agility to cope with the polluted environment. But once the critical toxin threshold are surpassed, higher toxin concentrations lead to a continuous decline in population. Eventually, at very high toxin levels, the population faces extinction.

\subsection{Two dimensional space}

For two-dimensional simulations, the region is \([-100, 100]^2\), the initial values are uniformly distributed random variables in $[0, 1]$, and the time span is 200 days. 
\subsubsection{Influence of toxin intensity}
\begin{figure}[h!]
    \centering
    \begin{subfigure}[b]{0.24\textwidth}
         \centering        \includegraphics[width=\textwidth]{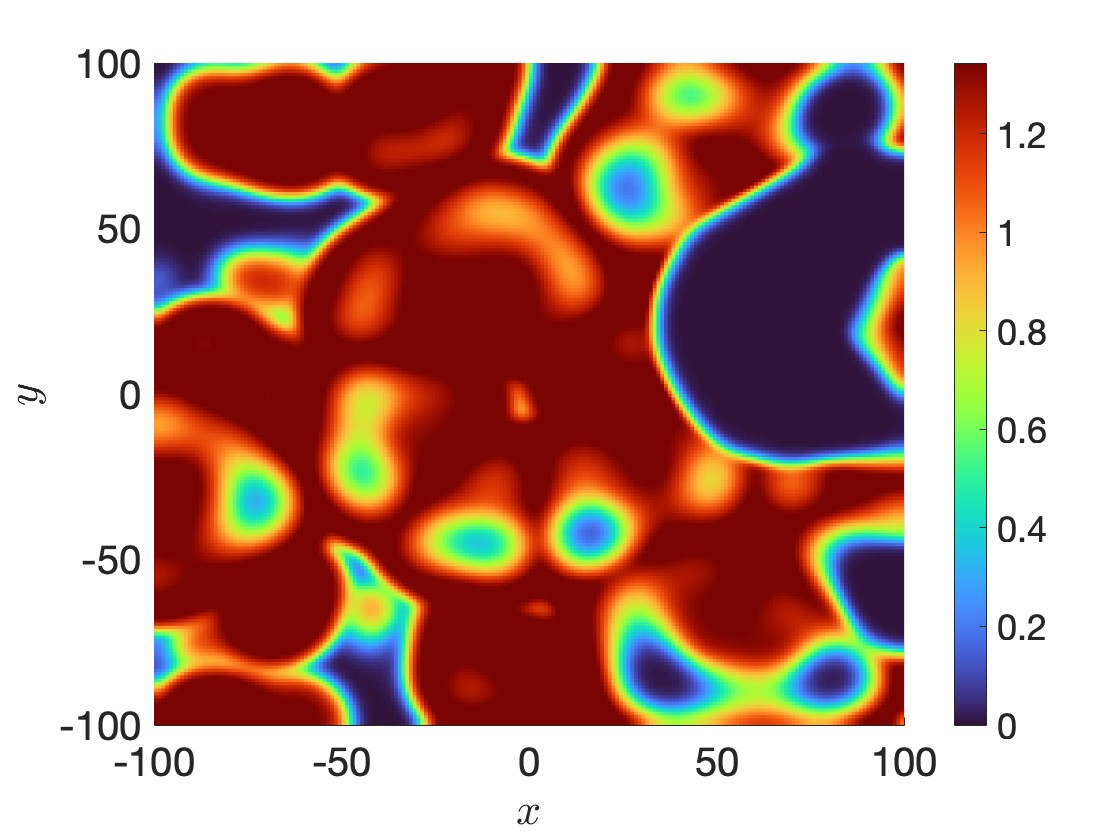}       \caption*{Prey - \(T_0 = 0\)}   \label{fig: colormap_X_T0=0_t=201.jpg}
    \end{subfigure}   
    \begin{subfigure}[b]{0.24\textwidth}
         \centering        \includegraphics[width=\textwidth]{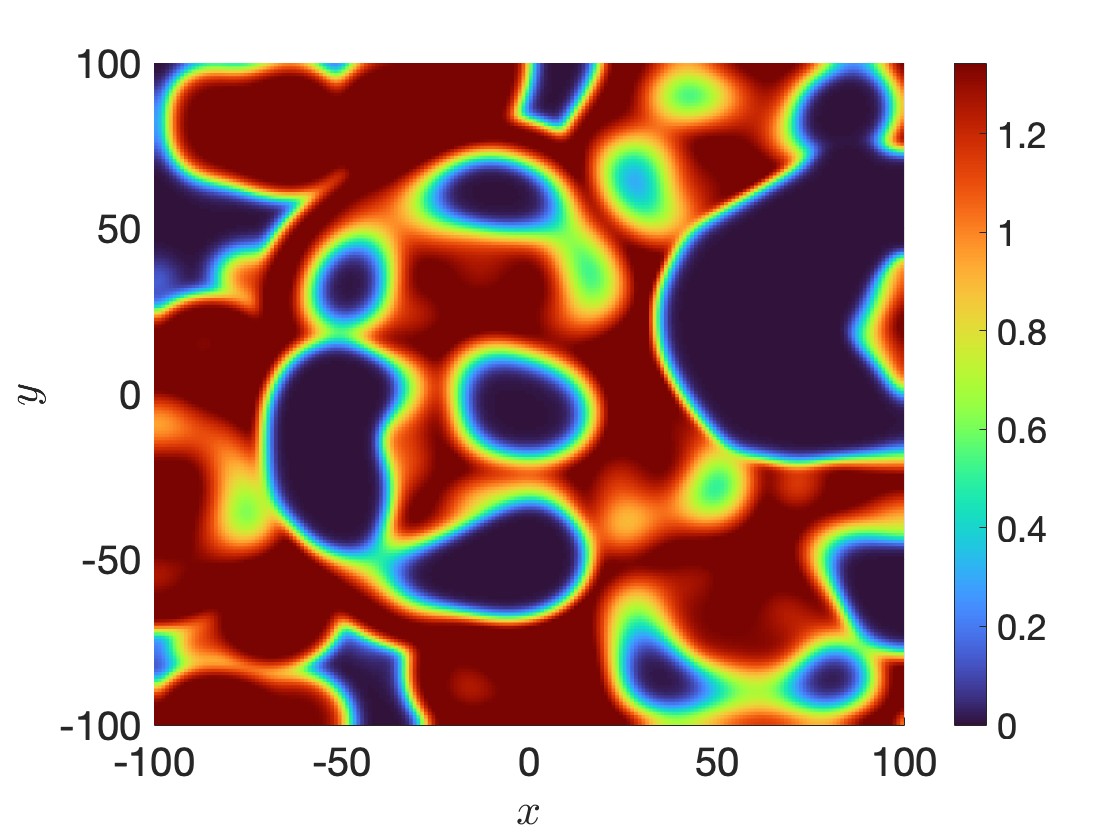}       \caption*{Prey - \(T_0 = 6\)}   \label{fig: colormap_X_T0=6_t=201.jpg}
    \end{subfigure}
    \begin{subfigure}[b]{0.24\textwidth}
         \centering        \includegraphics[width=\textwidth]{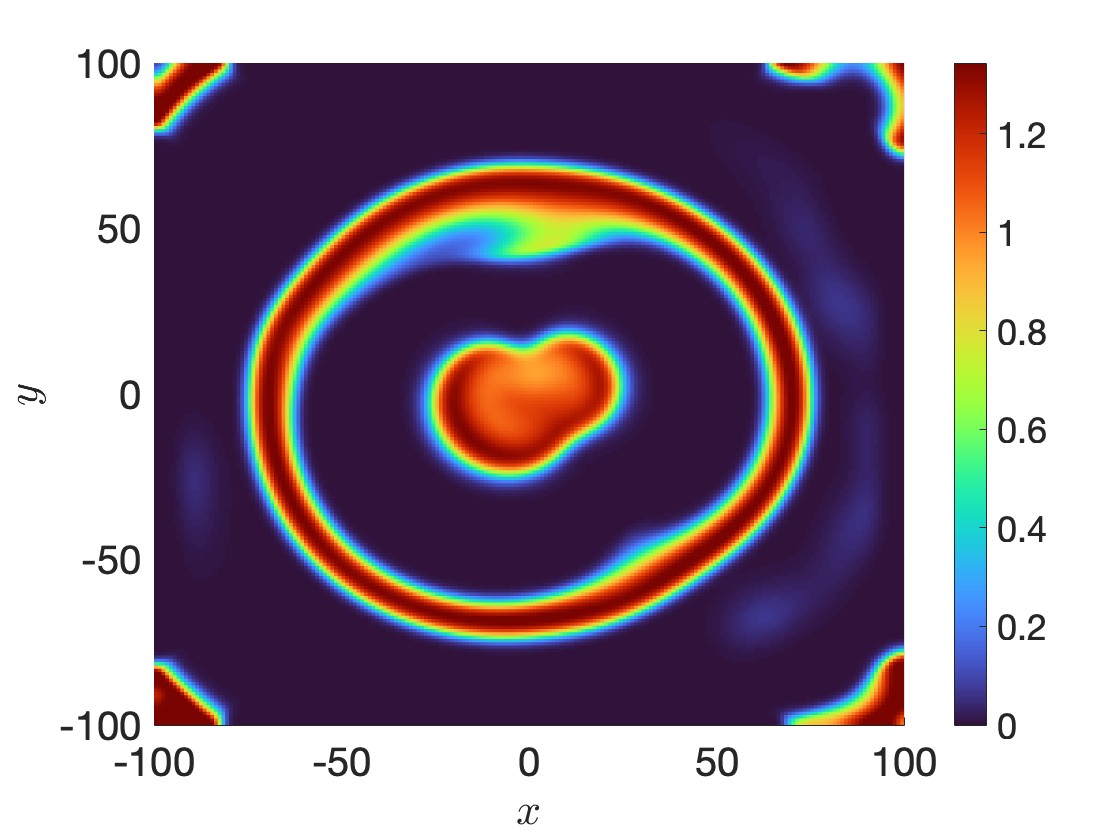}       \caption*{Prey - \(T_0 = 30\)}   \label{fig: colormap_X_T0=30_t=201.jpg}
           \end{subfigure} 
    \begin{subfigure}[b]{0.24\textwidth}
         \centering        \includegraphics[width=\textwidth]{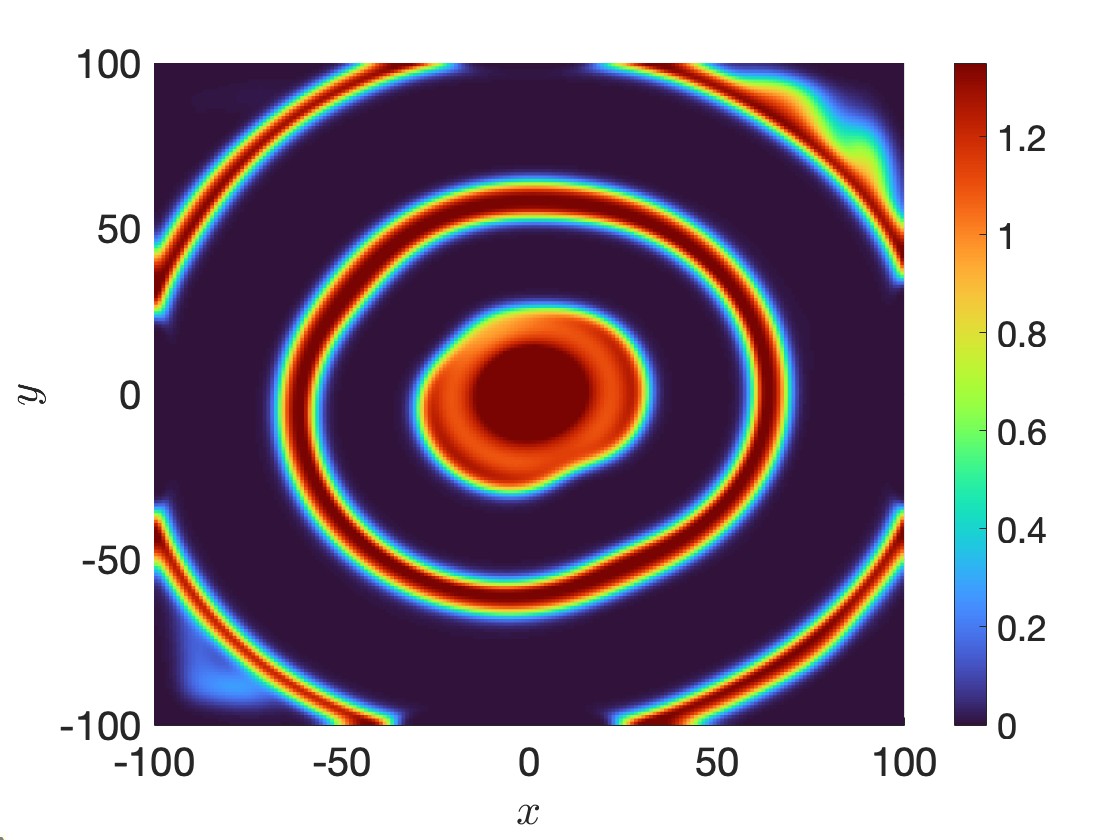}       \caption*{Prey - \(T_0 = 100\)}   \label{fig: colormap_X_T0=100_t=201.jpg}
     \end{subfigure} 
     \begin{subfigure}[b]{0.24\textwidth}
         \centering        \includegraphics[width=\textwidth]{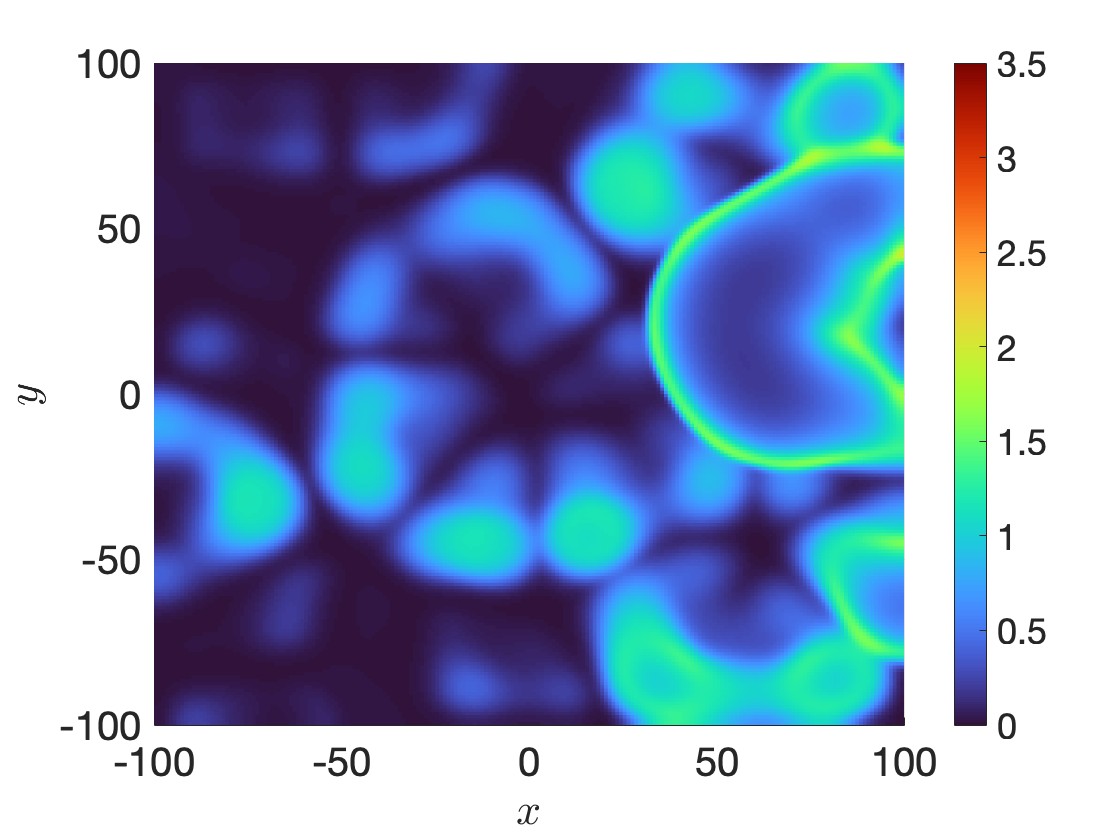}       \caption*{Bold Grazers - \(T_0 = 0\)}   \label{fig: colormap_Y1_T0=0_t=201.jpg}
    \end{subfigure}   
    \begin{subfigure}[b]{0.24\textwidth}
         \centering        \includegraphics[width=\textwidth]{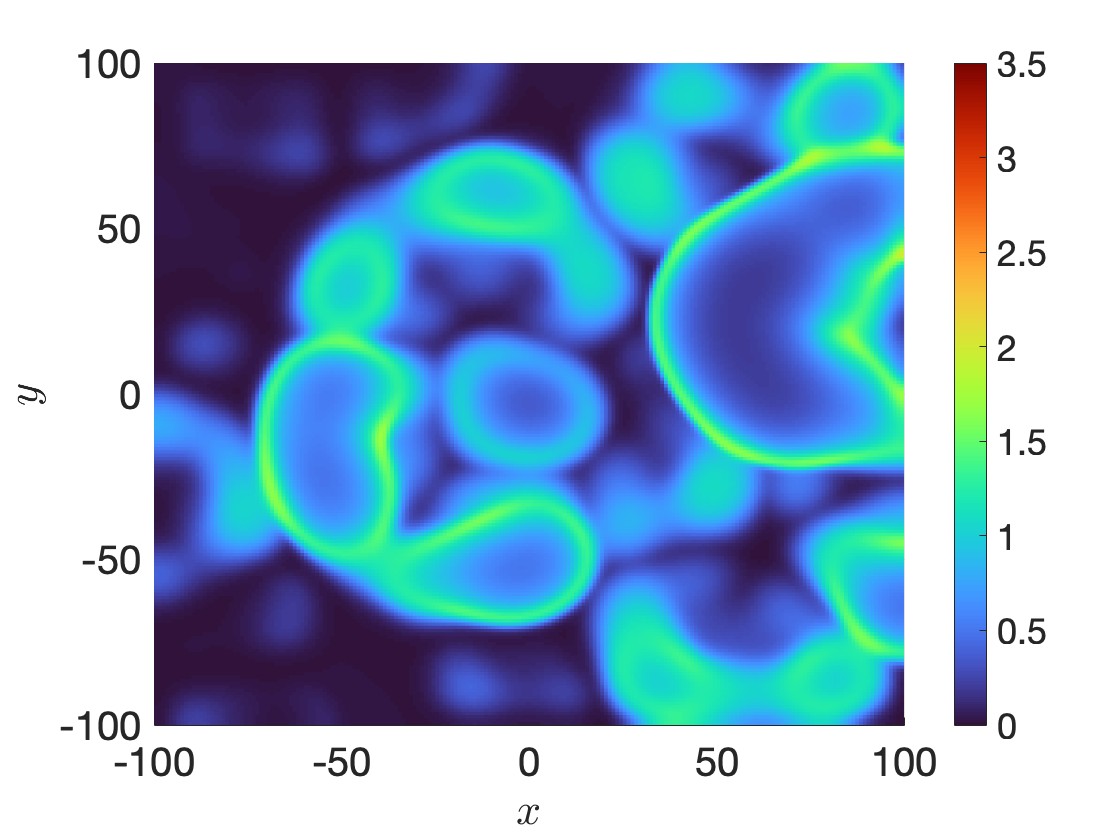}       \caption*{Bold Grazers - \(T_0 = 6\)}   \label{fig: colormap_Y1_T0=6_t=201.jpg}
    \end{subfigure}
    \begin{subfigure}[b]{0.24\textwidth}
         \centering        \includegraphics[width=\textwidth]{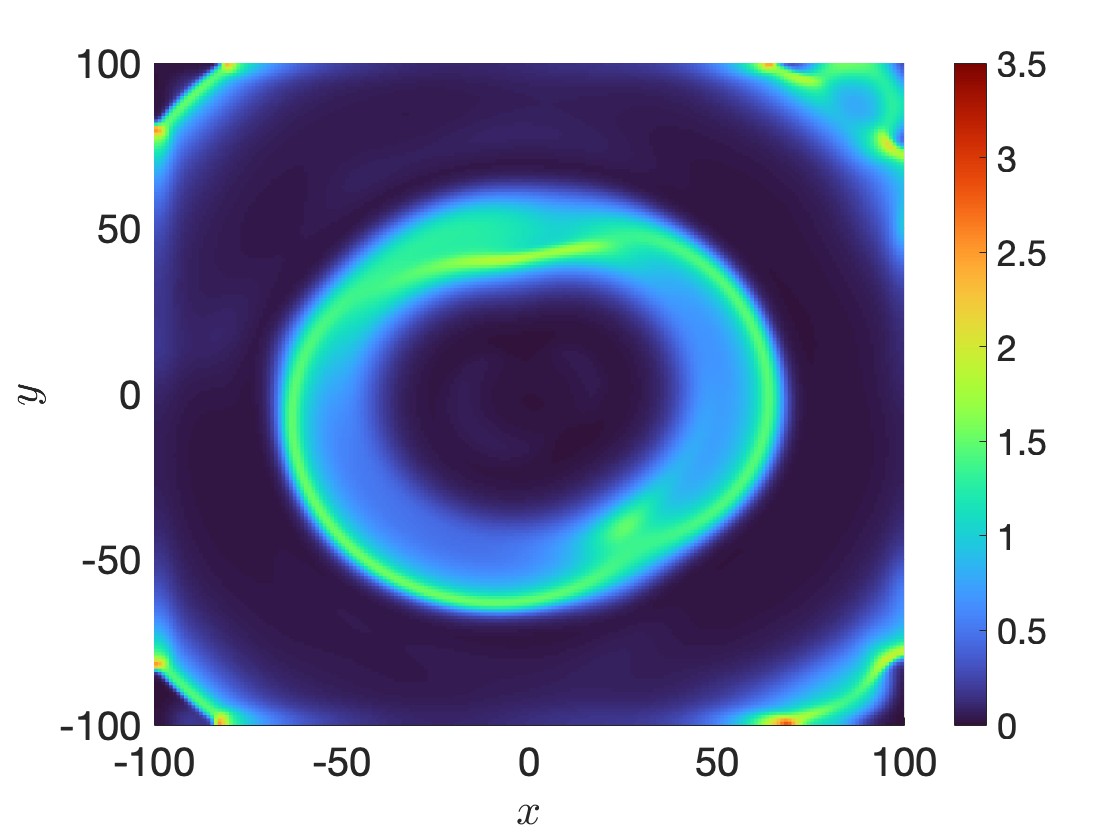}       \caption*{Bold Grazers - \(T_0 = 30\)}   \label{fig: colormap_Y1_T0=30_t=201.jpg}
           \end{subfigure} 
    \begin{subfigure}[b]{0.24\textwidth}
         \centering        \includegraphics[width=\textwidth]{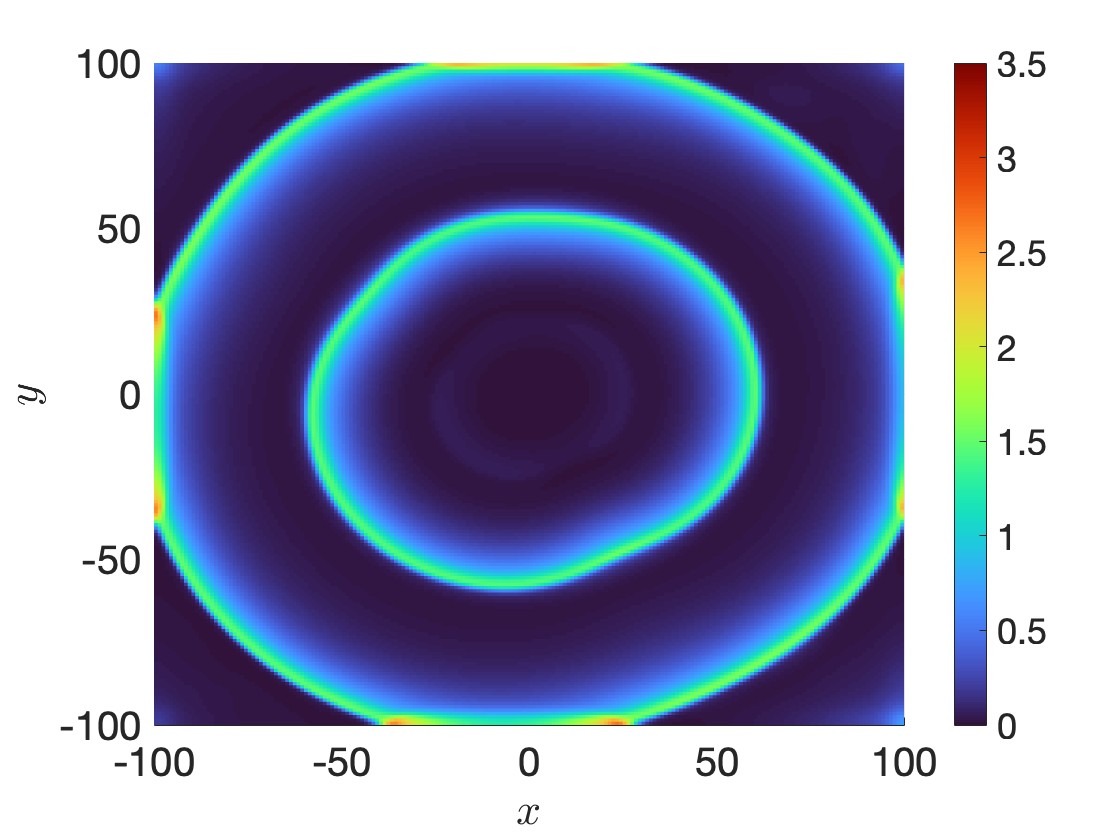}       \caption*{Bold Grazers - \(T_0 = 100\)}   \label{fig: colormap_Y1_T0=100_t=201.jpg}       
     \end{subfigure}
     \begin{subfigure}[b]{0.24\textwidth}
         \centering        \includegraphics[width=\textwidth]{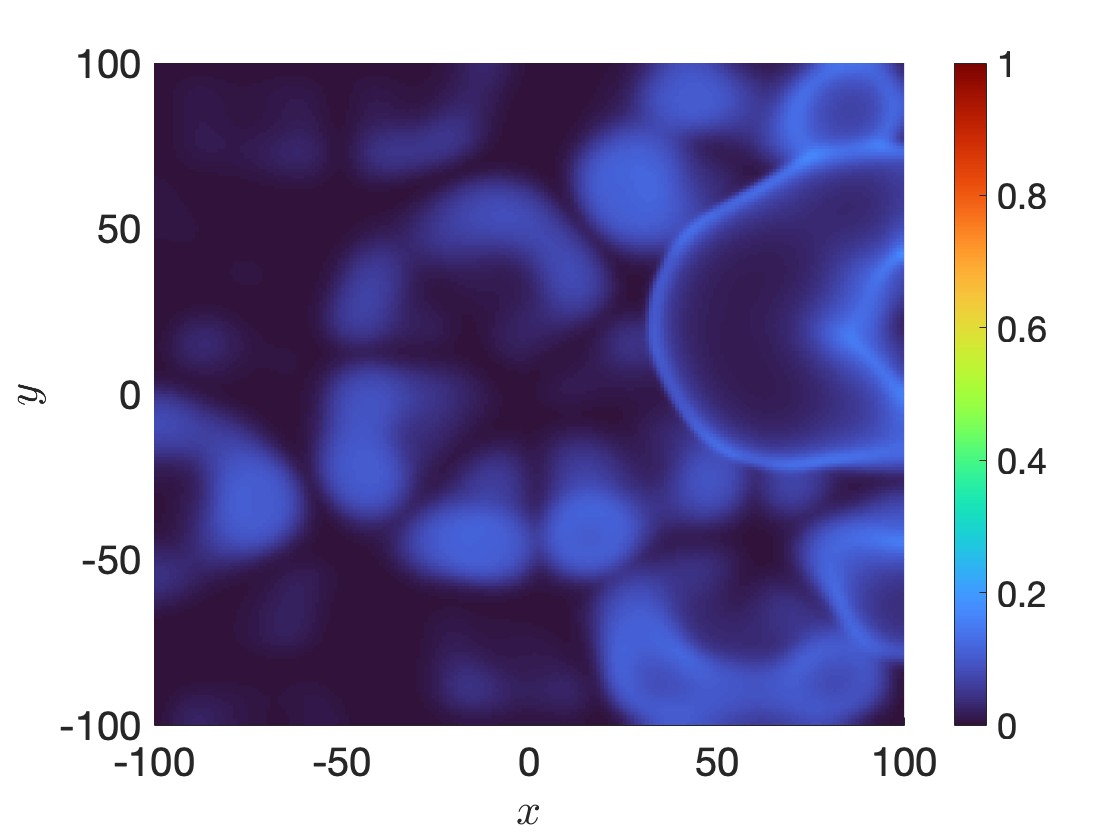}       \caption*{Shy Grazers - \(T_0 = 0\)}   \label{fig: colormap_Y2_T0=0_t=201.jpg}
    \end{subfigure}   
    \begin{subfigure}[b]{0.24\textwidth}
         \centering        \includegraphics[width=\textwidth]{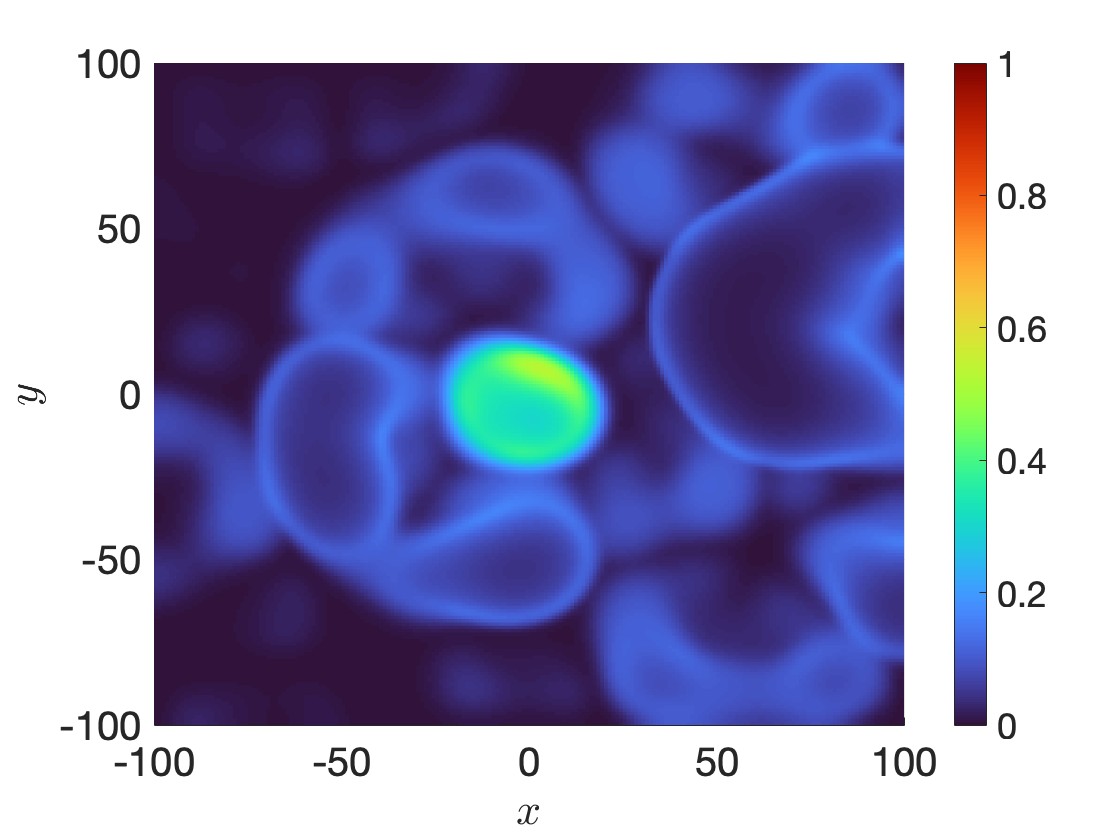}       \caption*{Shy Grazers - \(T_0 = 6\)}   \label{fig: colormap_Y2_T0=6_t=201.jpg}
    \end{subfigure}
    \begin{subfigure}[b]{0.24\textwidth}
         \centering        \includegraphics[width=\textwidth]{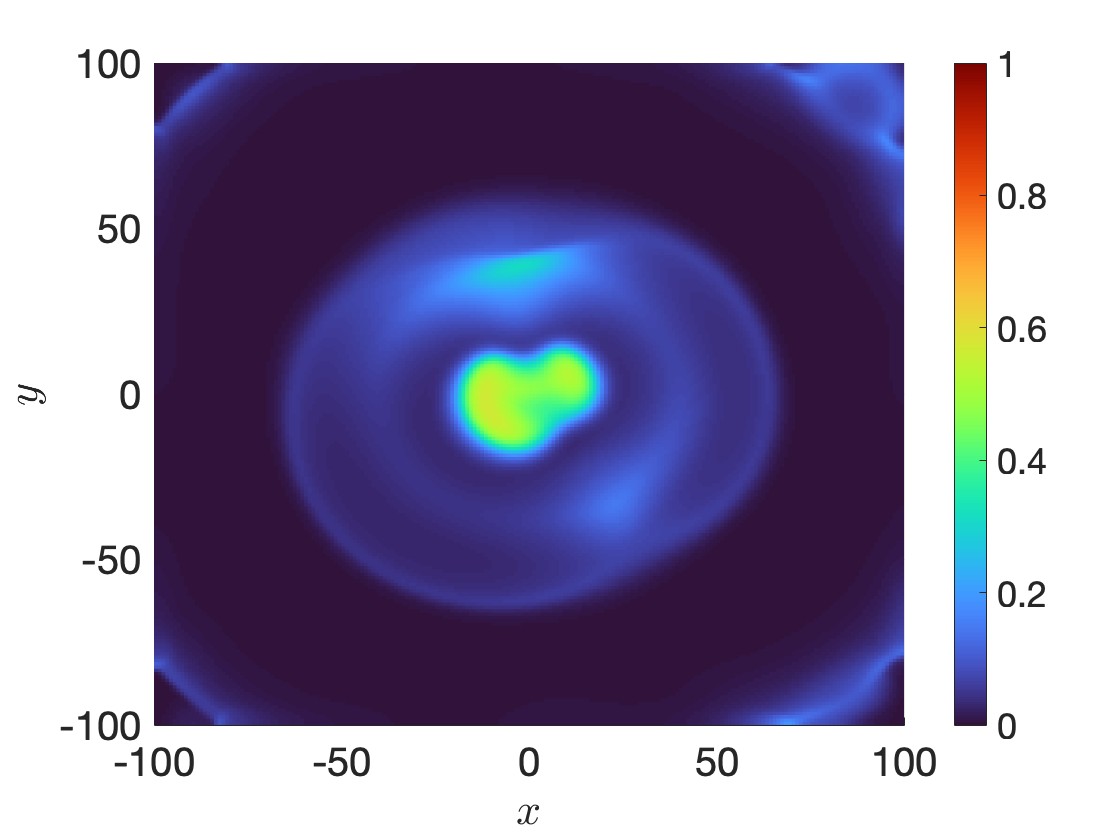}       \caption*{Shy Grazers - \(T_0 = 30\)}   \label{fig: colormap_Y2_T0=30_t=201.jpg}
           \end{subfigure} 
    \begin{subfigure}[b]{0.24\textwidth}
         \centering        \includegraphics[width=\textwidth]{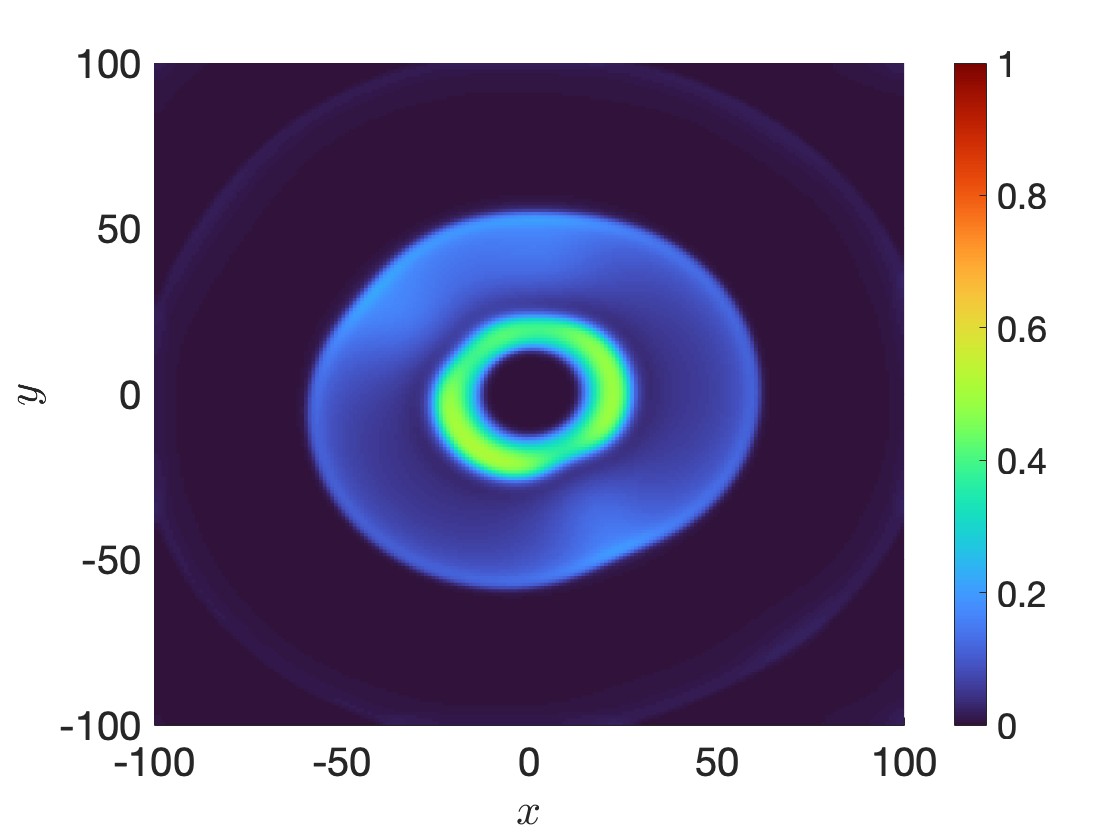}       \caption*{Shy Grazers - \(T_0 = 100\)}   \label{fig: colormap_Y2_T0=100_t=201.jpg}       
     \end{subfigure}
     \begin{subfigure}[b]{0.24\textwidth}
         \centering        \includegraphics[width=\textwidth]{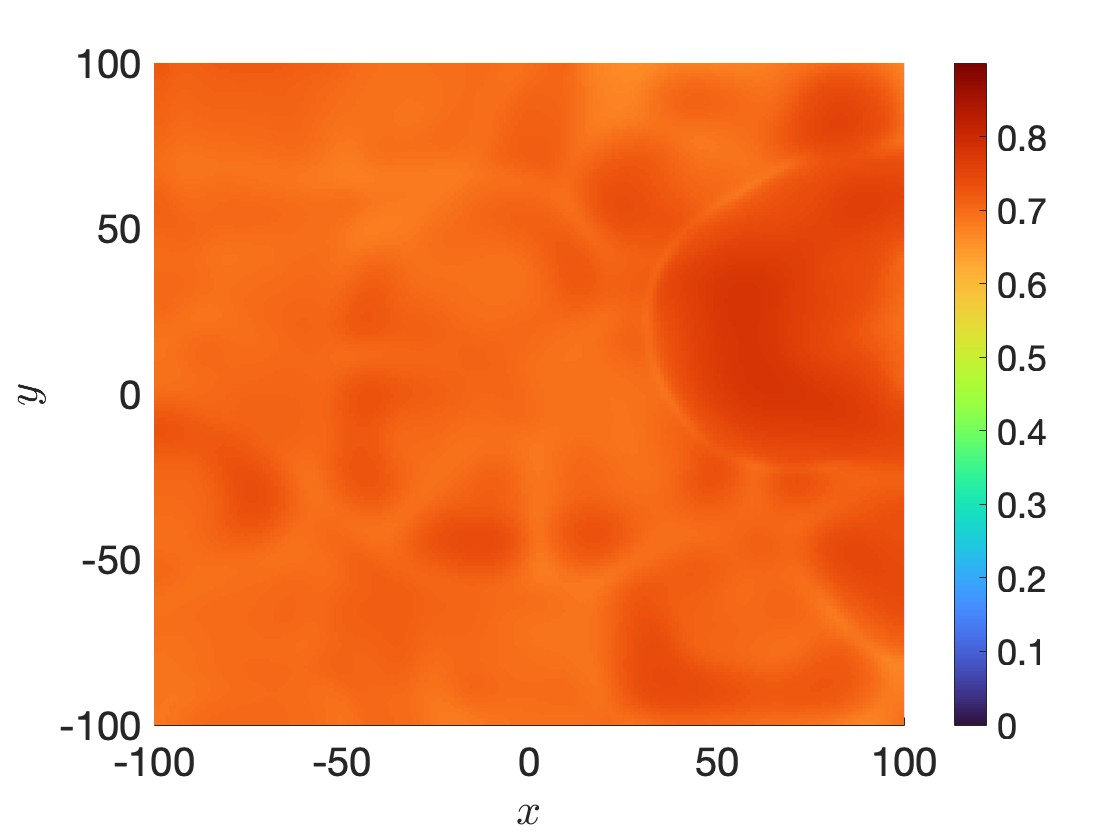}       \caption*{Predators - \(T_0 = 0\)}   \label{fig: colormap_Z_T0=0_t=201.jpg}
    \end{subfigure}   
    \begin{subfigure}[b]{0.24\textwidth}
         \centering        \includegraphics[width=\textwidth]{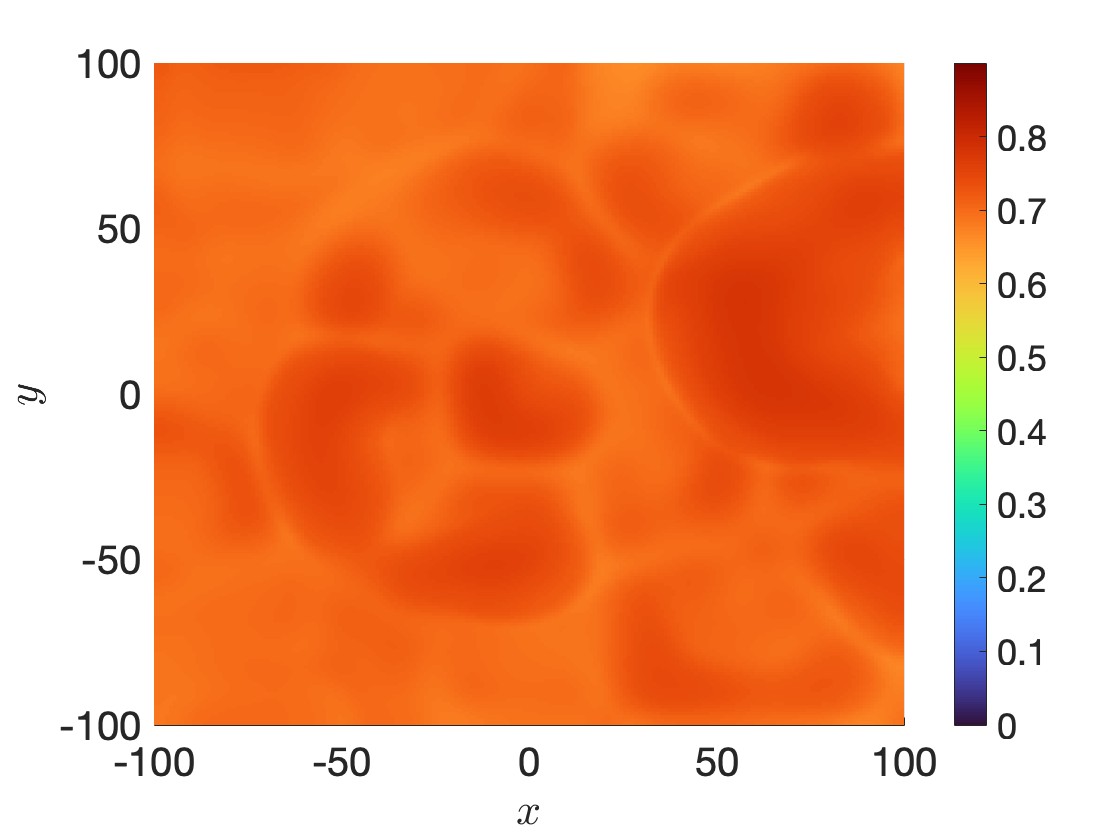}       \caption*{Predators - \(T_0 = 6\)}   \label{fig: colormap_Z_T0=6_t=201.jpg}
    \end{subfigure}
    \begin{subfigure}[b]{0.24\textwidth}
         \centering        \includegraphics[width=\textwidth]{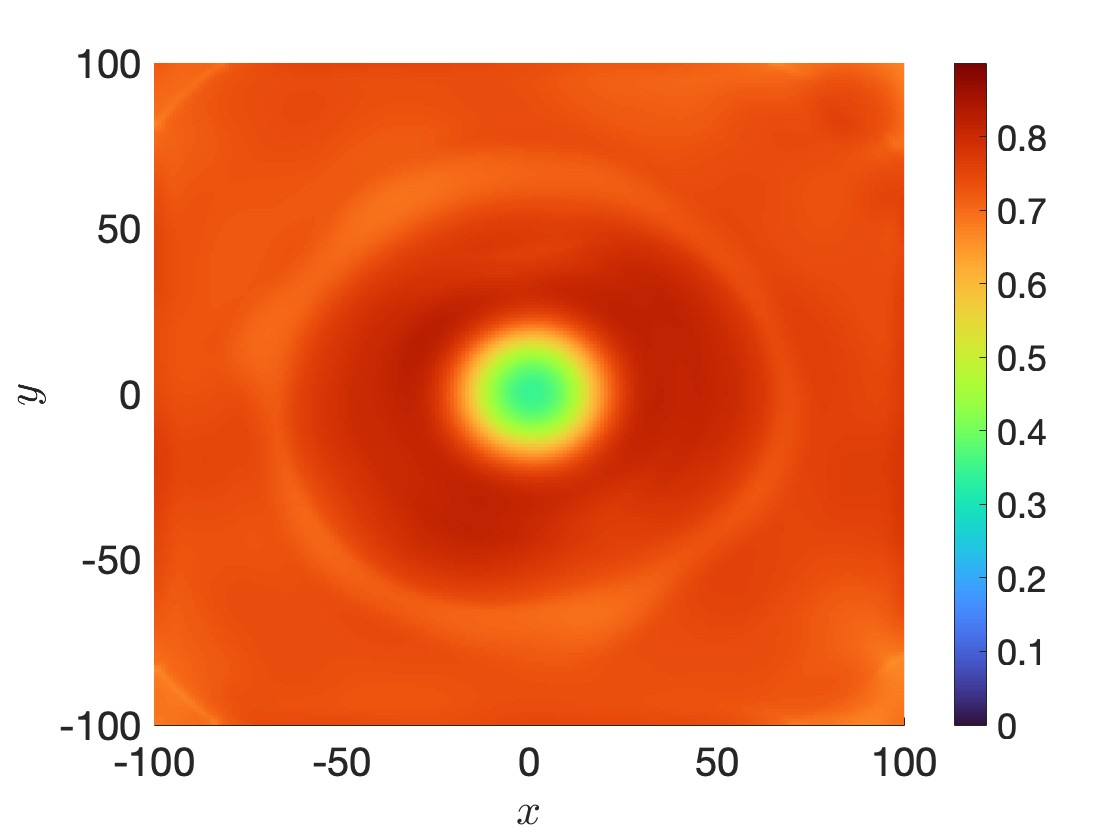}       \caption*{Predators - \(T_0 = 30\)}   \label{fig: colormap_Z_T0=30_t=201.jpg}
           \end{subfigure} 
    \begin{subfigure}[b]{0.24\textwidth}
         \centering        \includegraphics[width=\textwidth]{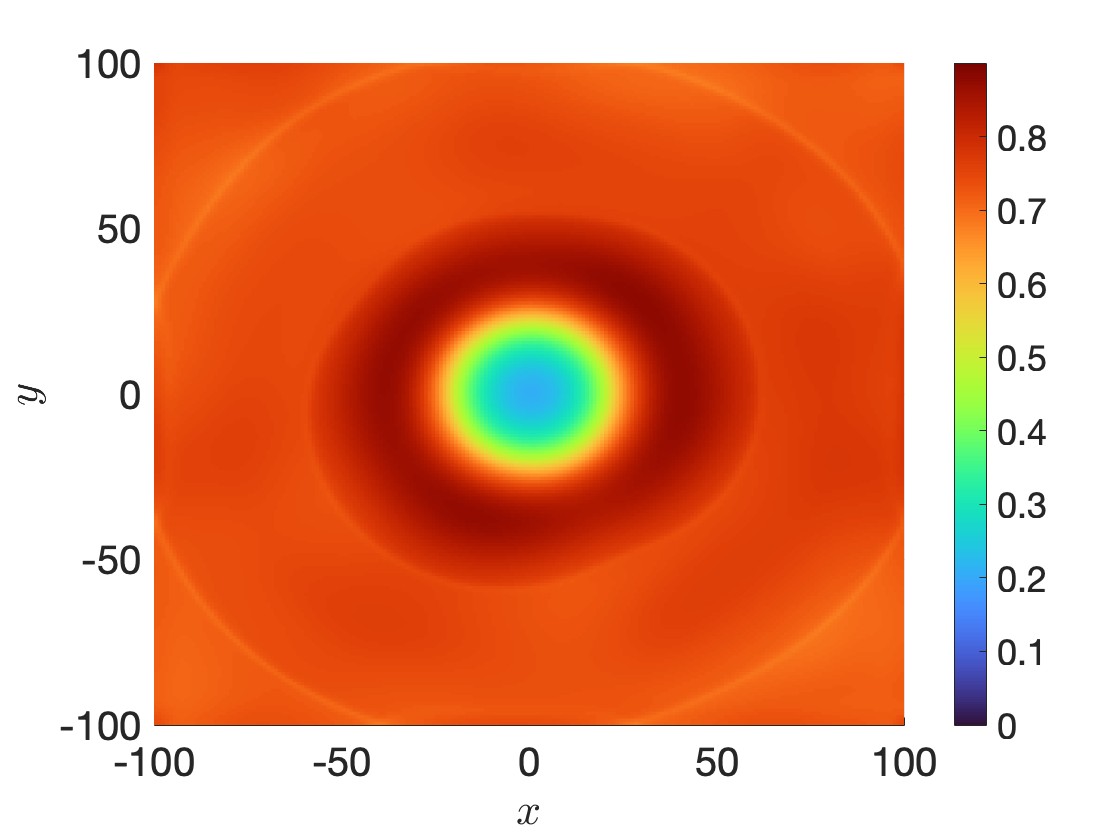}       \caption*{Predators - \(T_0 = 100\)}   \label{fig: colormap_Z_T0=100_t=201.jpg}       
     \end{subfigure}
     \caption{\footnotesize Spatial distribution of populations in a 2D space under radially symmetric toxin concentration: The first row represents prey, the second row bold grazers, the third row shy grazers, and the fourth row predators. The first column corresponds to \(T_0 = 0\), the second to \(T_0 = 6\), the third to \(T_0 = 30\), and the fourth to \(T_0 = 100\).}
     \label{fig: 2D_exponential_decay}
\end{figure}

In two-dimensional cases, when the pollution source is centrally located, the distribution is approximately radially symmetric, with the concentration nearly uniform at any given distance from the source in all directions. To model this, we extend the toxin distribution function from the one-dimensional case and consider a radially symmetric toxin concentration. The highest level, \(T_0\), is at the origin, and the concentration decays exponentially with distance as follows
\[
   T(x, y) = T_0 e^{-\frac{x^2 + y^2}{15^2}}.
\]
Similar population structures emerge; moreover, more complex and intriguing patterns across different toxin levels are found, as shown in Figures \ref{fig: 2D_exponential_decay} at \( t = 200 \). Population densities, ranging from low to high, are depicted by a color gradient from blue to red. 
The population demonstrates periodic patterns. Moreover, we found that the spatial patterns for different species vary significantly, and increasing toxin levels lead to heightened aggregation behavior. 

In the absence of toxins, the spatial distribution of species is highly random. After 200 days, all species coexist, but the densities of the grazer and producer show an opposite trend. The two grazer species exhibit similar spatial distributions, though bold grazers have higher population densities compared to shy grazers.

At low contaminant levels (\( T_0 = 6 \)), population densities still exhibit a certain extent of randomness, but clear differences emerge between the bold and shy species. Shy grazers tend to cluster in the pollution center, while bold grazers are more dispersed around the perimeter of the polluted region.

At moderate contaminant levels (\( T_0 = 30 \)), the population distribution becomes more organized and radially symmetric. The population moves in periodic circular patterns, spreading outward from the center over time. Aggregation behavior appears: bold grazers aggregate farther from the pollution center, while shy grazers concentrate in the highly polluted central region. Within this polluted center, prey and shy grazers reach high population densities, while predator densities remain low.

When the contaminant levels are high (\( T_0 = 100 \)), the aggregation becomes more intense, with the areas of aggregation forming tighter loops. No grazers survive in the toxin's center. Shy grazers aggregate near, but not within, the central region, whereas bold grazers cluster farther from the center. Compared to \( T_0 = 30 \), the toxin's range of influence is broader, for example, predators occupy larger areas where they can only maintain low population densities.

\subsubsection{Influence spatial distribution of toxin  \label{sec: Influence spatial distribution of toxin}}

\begin{figure}[h!]
    \centering
    \begin{subfigure}[t]{0.24\textwidth}
        \centering
        \includegraphics[width=\textwidth]{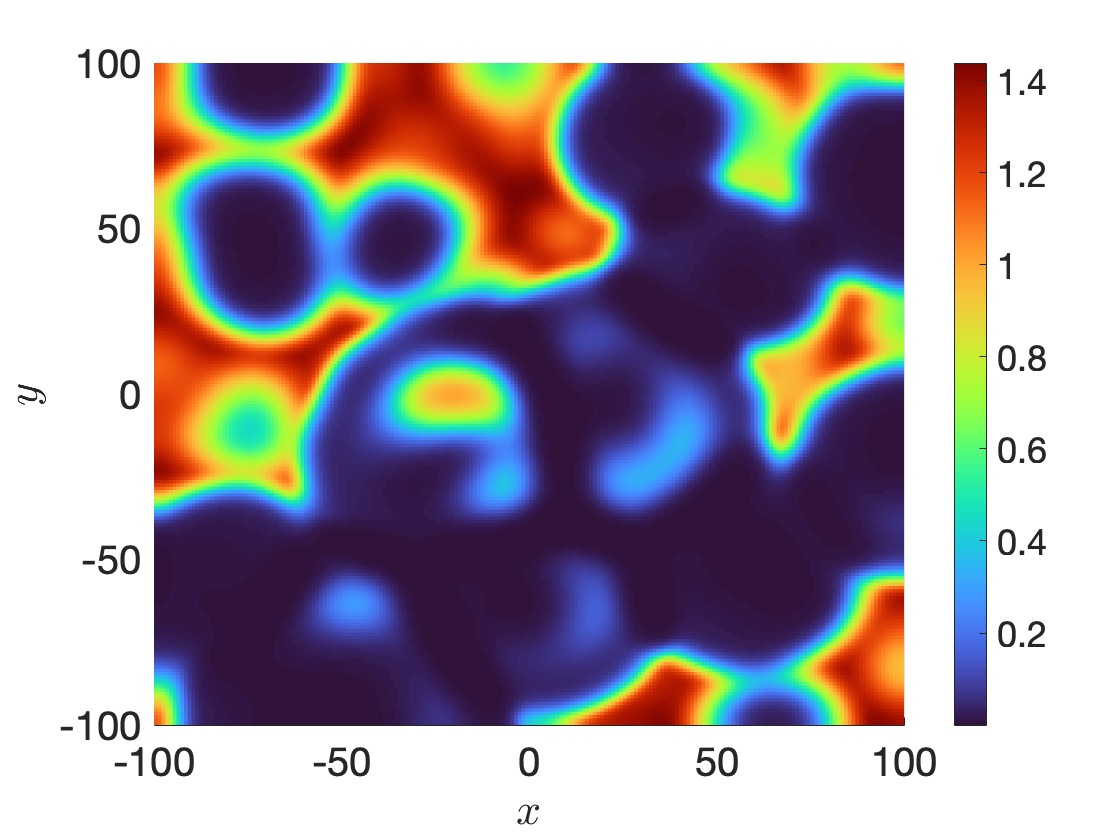}
        \caption*{Prey - Constant Toxin}
        \label{fig: constantT_X_T0=10_t=201.jpg}
    \end{subfigure}
    \hspace{0.01\textwidth}
    \begin{subfigure}[t]{0.24\textwidth}
         \centering
         \includegraphics[width=\textwidth]{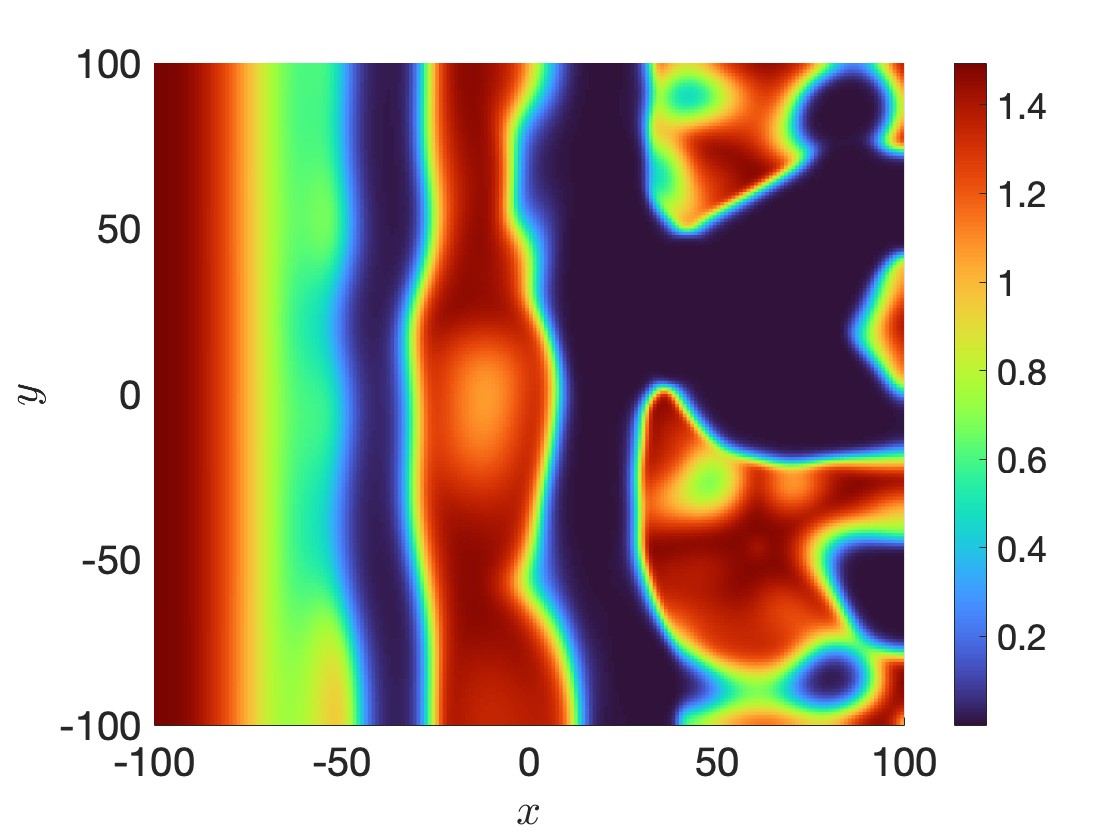}
         \caption*{Prey - Linear Gradient}
         \label{fig: Line_X_T0=30_t=201.jpg}
    \end{subfigure}   
    \begin{subfigure}[t]{0.24\textwidth}
         \centering
         \includegraphics[width=\textwidth]{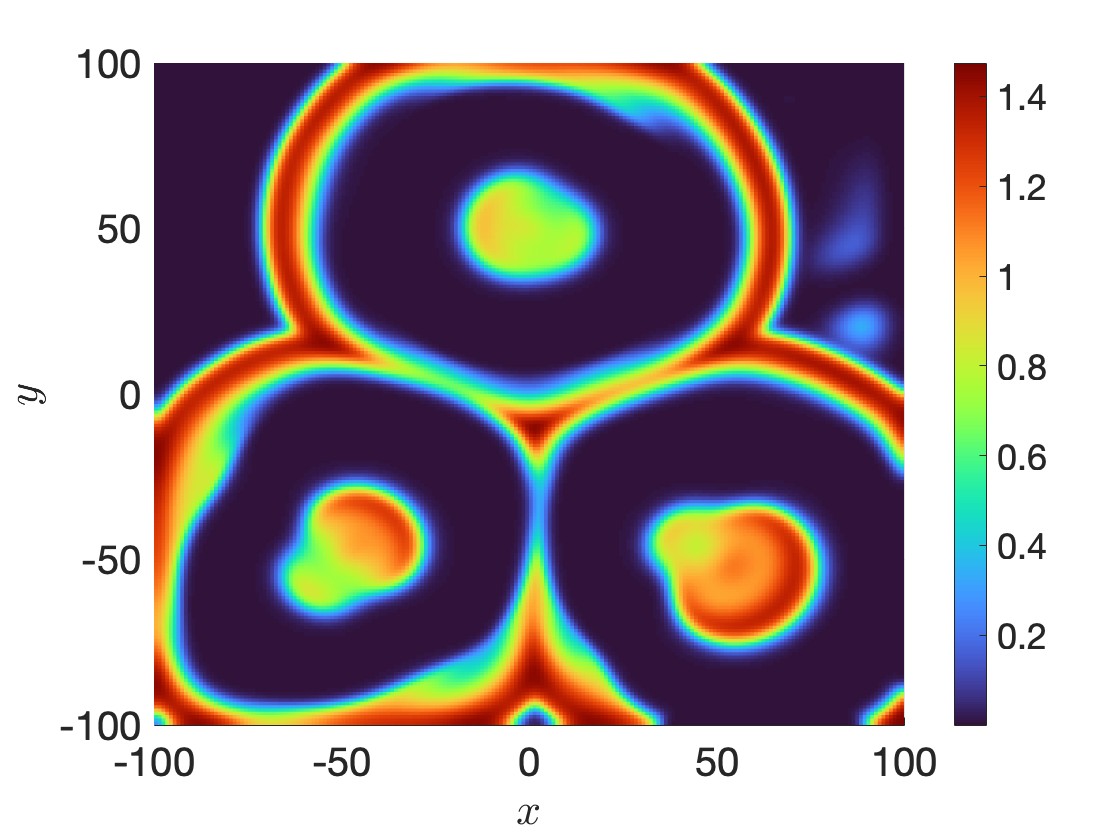}
         \caption*{Prey - Multi-Peak}
         \label{fig: ThreeCircles_X_T0=30_t=201.jpg}
    \end{subfigure}
    \begin{subfigure}[t]{0.24\textwidth}
         \centering
         \includegraphics[width=\textwidth]{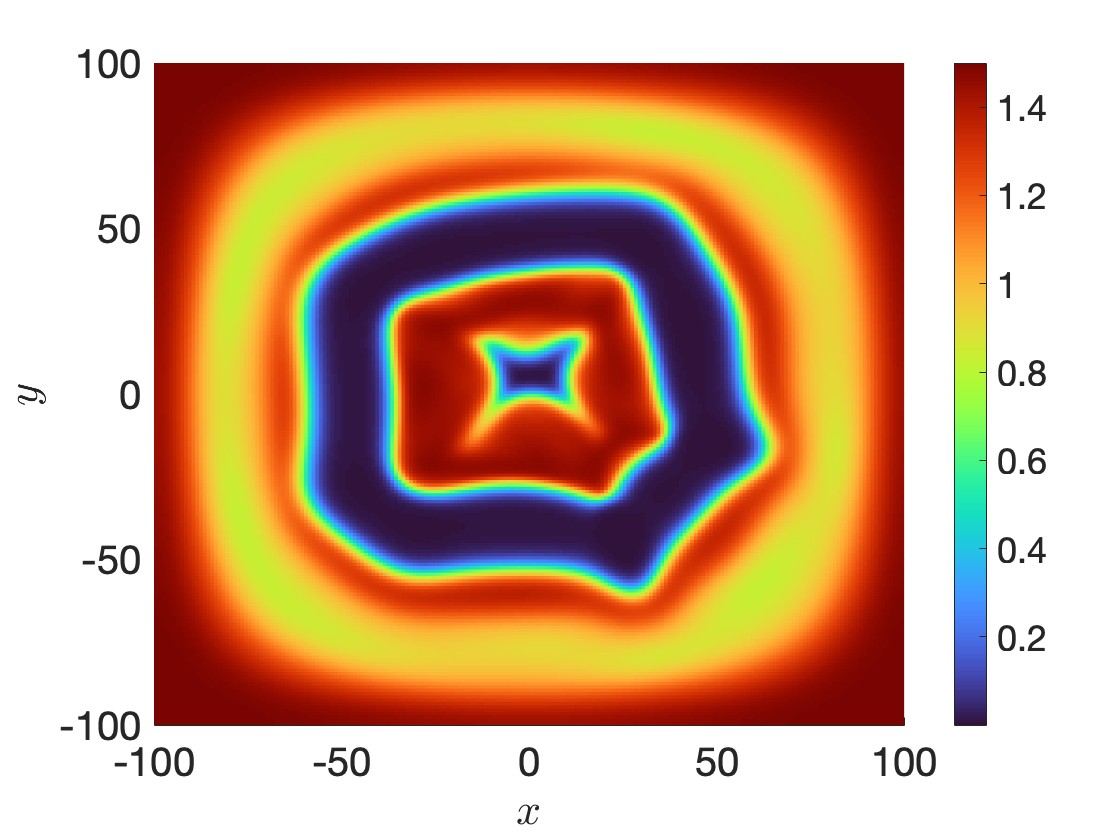}
         \caption*{Prey - Rectangular}
         \label{fig: EdgeCircles_X_T0=30_t=201.jpg}
    \end{subfigure} 
    
    \begin{subfigure}[t]{0.24\textwidth}
        \centering
        \includegraphics[width=\textwidth]{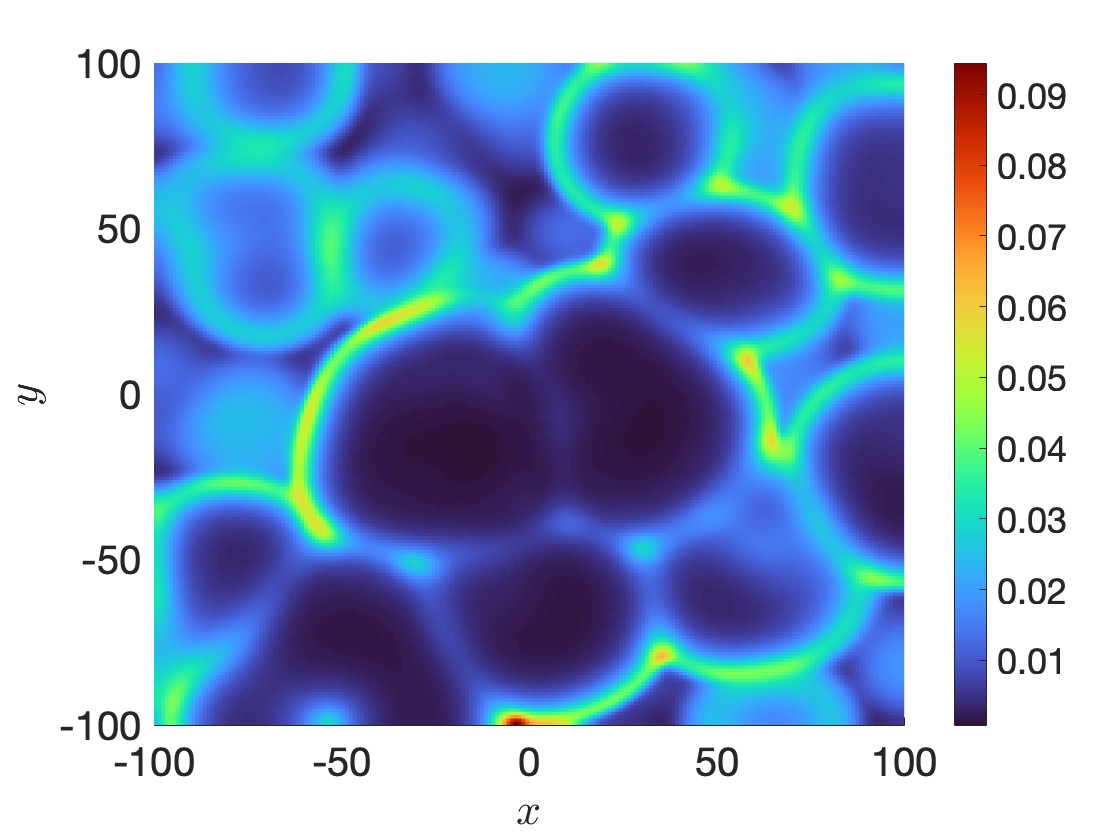}
        \caption*{Bold Grazers - Constant Toxin}
        \label{fig: constantT_Y1_T0=10_t=201.jpg}
    \end{subfigure}
    \hspace{0.01\textwidth}
    \begin{subfigure}[t]{0.24\textwidth}
         \centering
         \includegraphics[width=\textwidth]{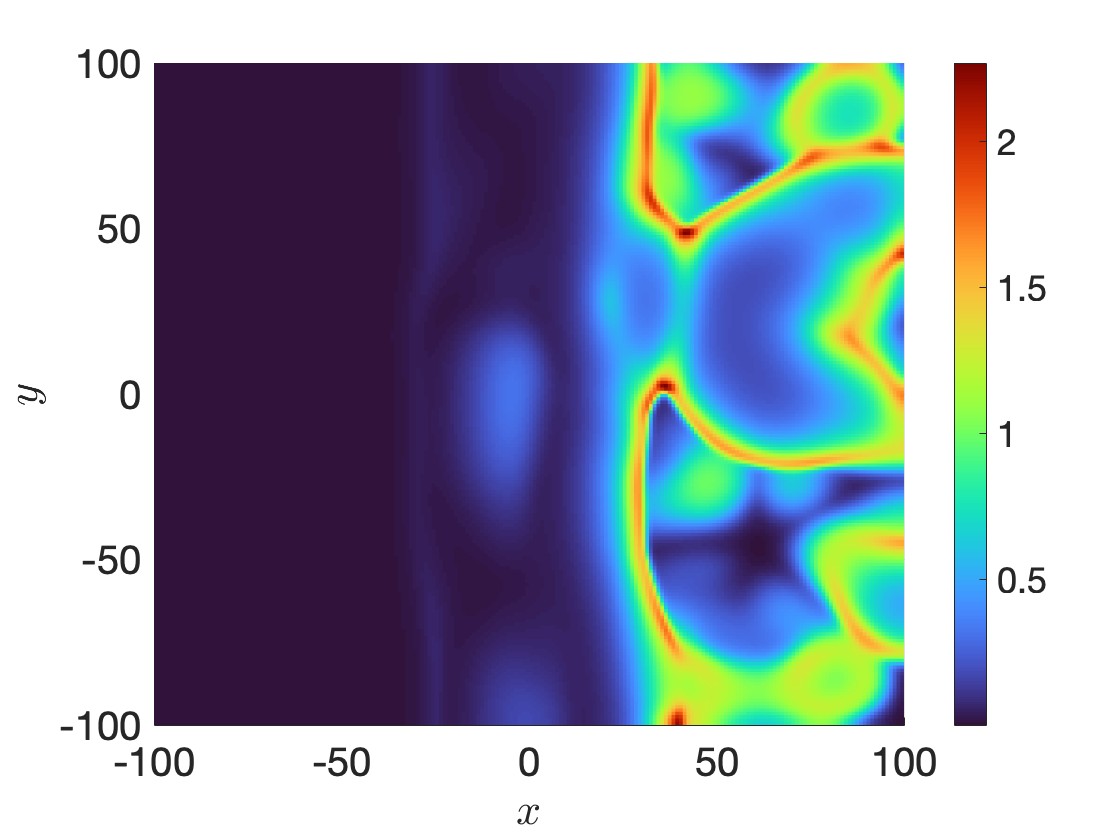}
         \caption*{Bold Grazers - Linear Gradient}
         \label{fig: Line_Y1_T0=30_t=201.jpg}
     \end{subfigure}   
    \begin{subfigure}[t]{0.24\textwidth}
         \centering
         \includegraphics[width=\textwidth]{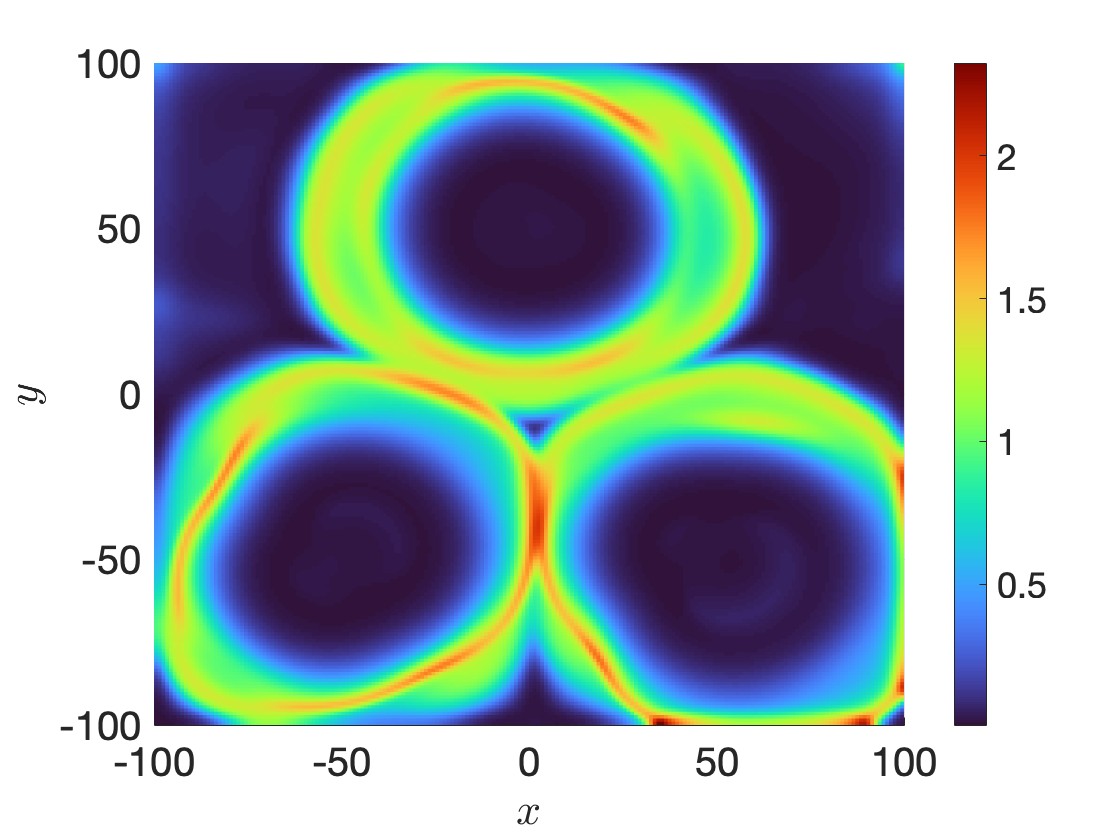}
         \caption*{Bold Grazers - Multi-Peak}
         \label{fig: ThreeCircles_Y1_T0=30_t=201.jpg}
    \end{subfigure}
    \begin{subfigure}[t]{0.24\textwidth}
         \centering
         \includegraphics[width=\textwidth]{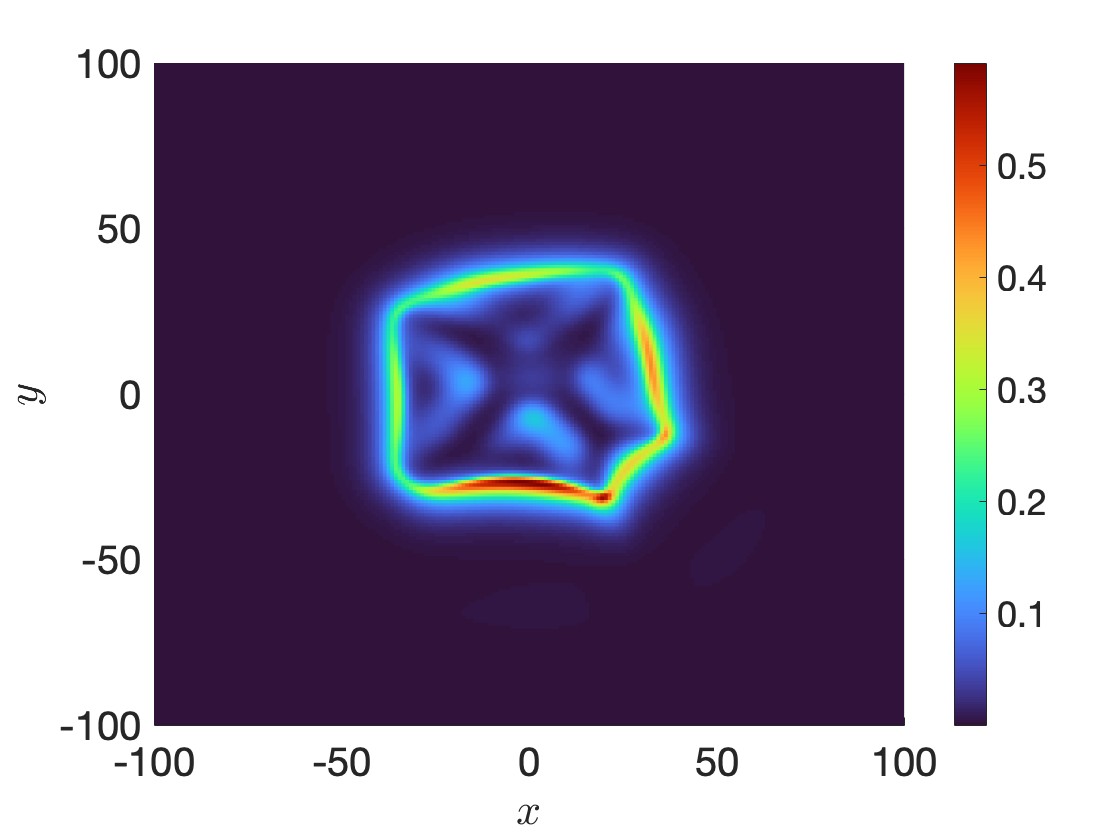}
         \caption*{Bold Grazers - Rectangular}
         \label{fig: EdgeCircles_Y1_T0=30_t=201.jpg}
    \end{subfigure}
    
    \begin{subfigure}[t]{0.24\textwidth}
        \centering
        \includegraphics[width=\textwidth]{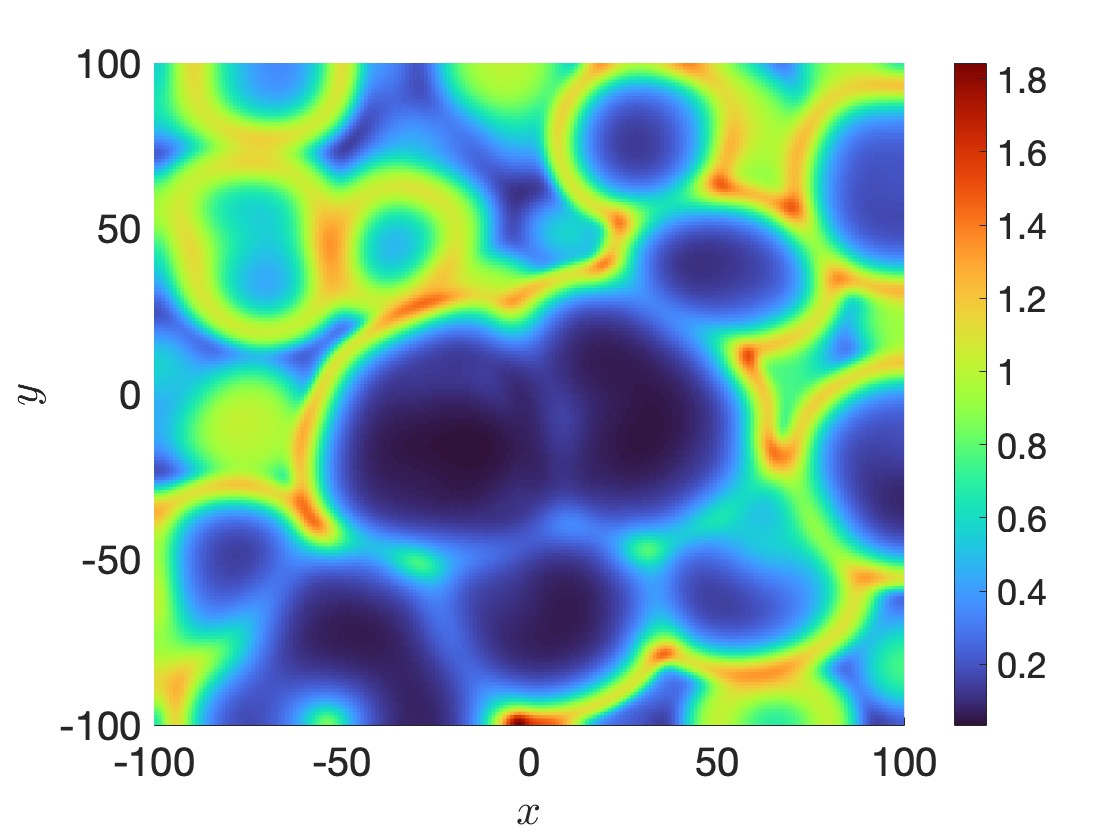}
        \caption*{Shy Grazers - Constant Toxin}
        \label{fig: constantT_Y2_T0=10_t=201.jpg}
    \end{subfigure}
    \hspace{0.01\textwidth}
    \begin{subfigure}[t]{0.24\textwidth}
         \centering
         \includegraphics[width=\textwidth]{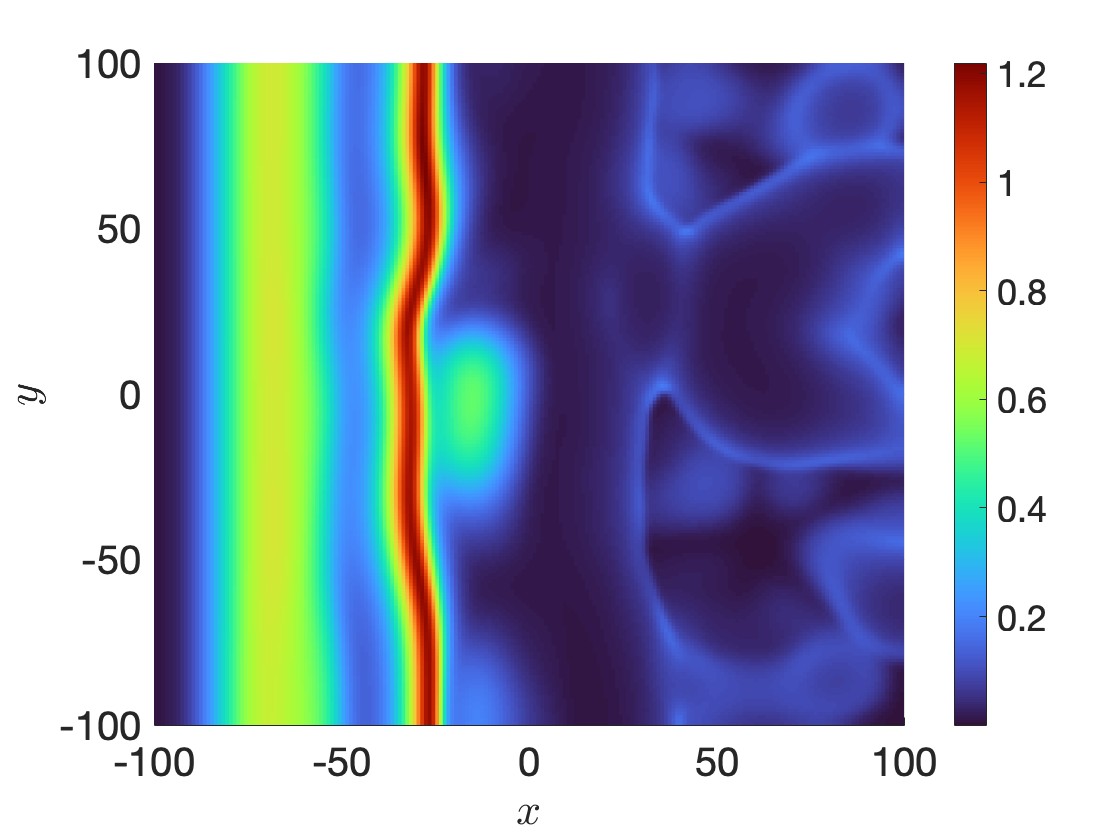}
         \caption*{Shy Grazers - Linear Gradient}
         \label{fig: Line_Y2_T0=30_t=201.jpg}
     \end{subfigure}   
    \begin{subfigure}[t]{0.24\textwidth}
         \centering
         \includegraphics[width=\textwidth]{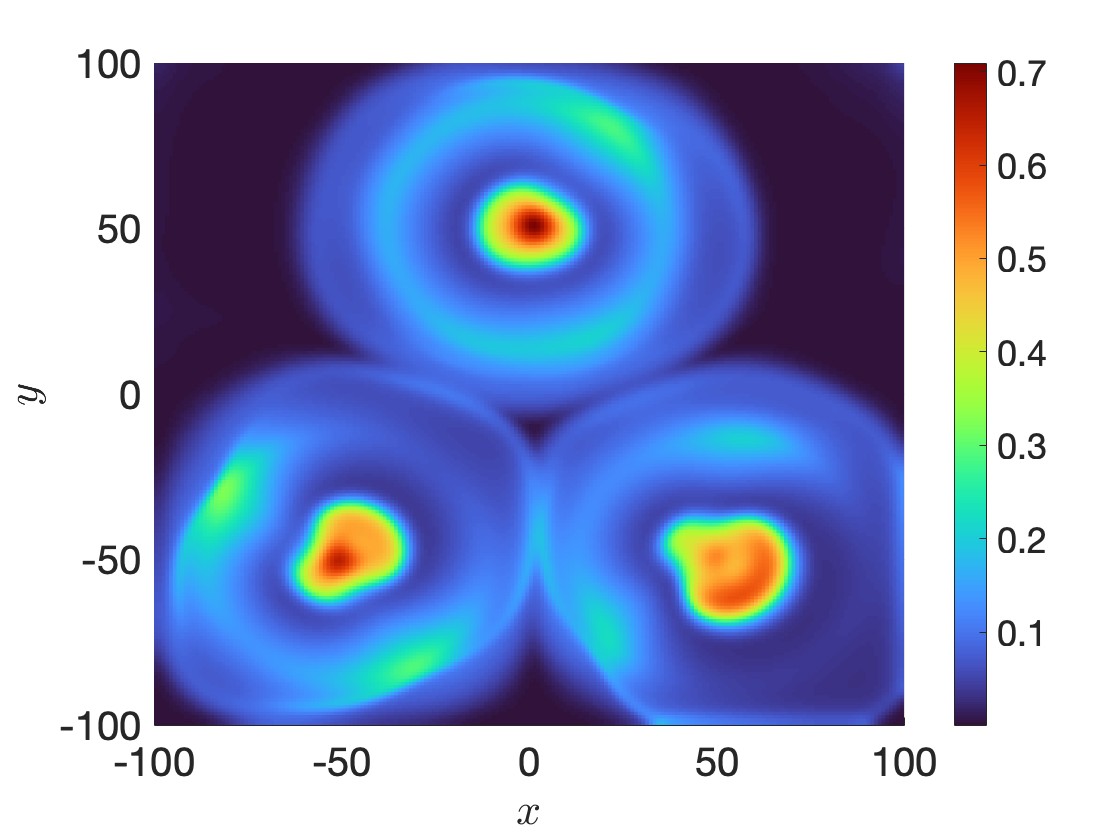}
         \caption*{Shy Grazers - Multi-Peak}
         \label{fig: ThreeCircles_Y2_T0=30_t=201.jpg}
    \end{subfigure}
    \begin{subfigure}[t]{0.24\textwidth}
         \centering
         \includegraphics[width=\textwidth]{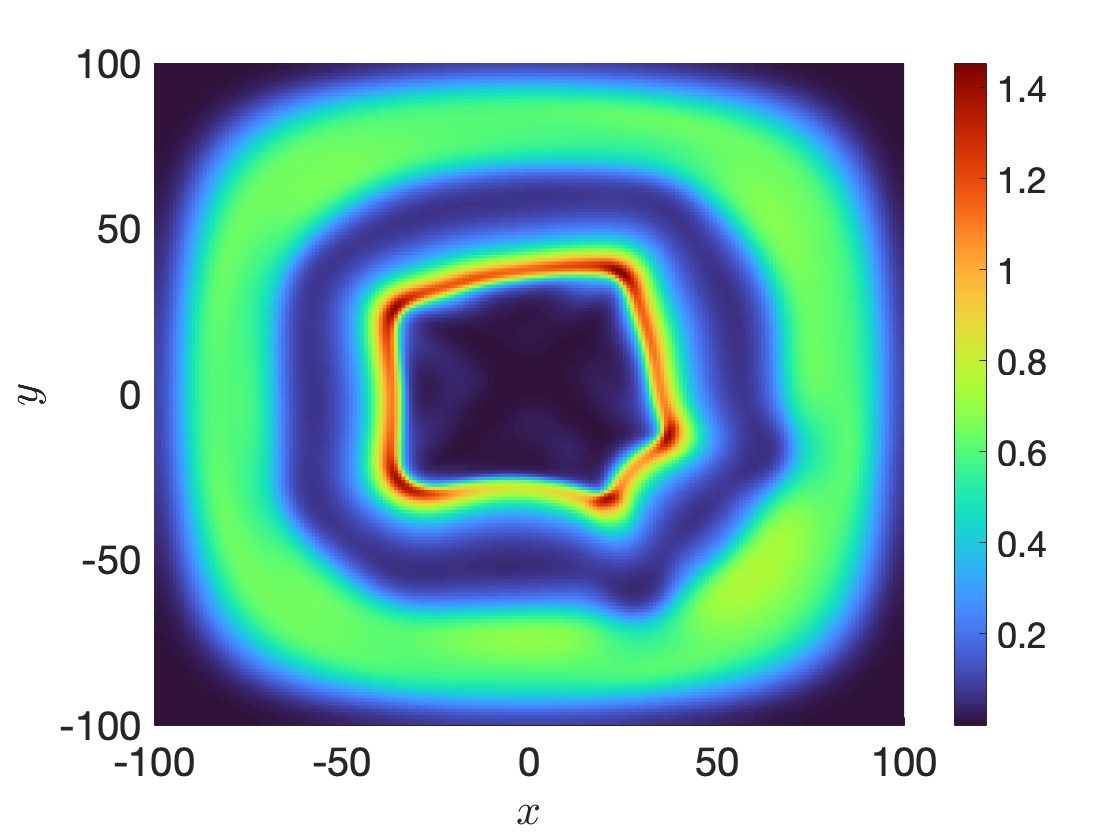}
         \caption*{Shy Grazers - Rectangular}
         \label{fig: EdgeCircles_Y2_T0=30_t=201.jpg}
    \end{subfigure}
    
    \begin{subfigure}[t]{0.24\textwidth}
        \centering
        \includegraphics[width=\textwidth]{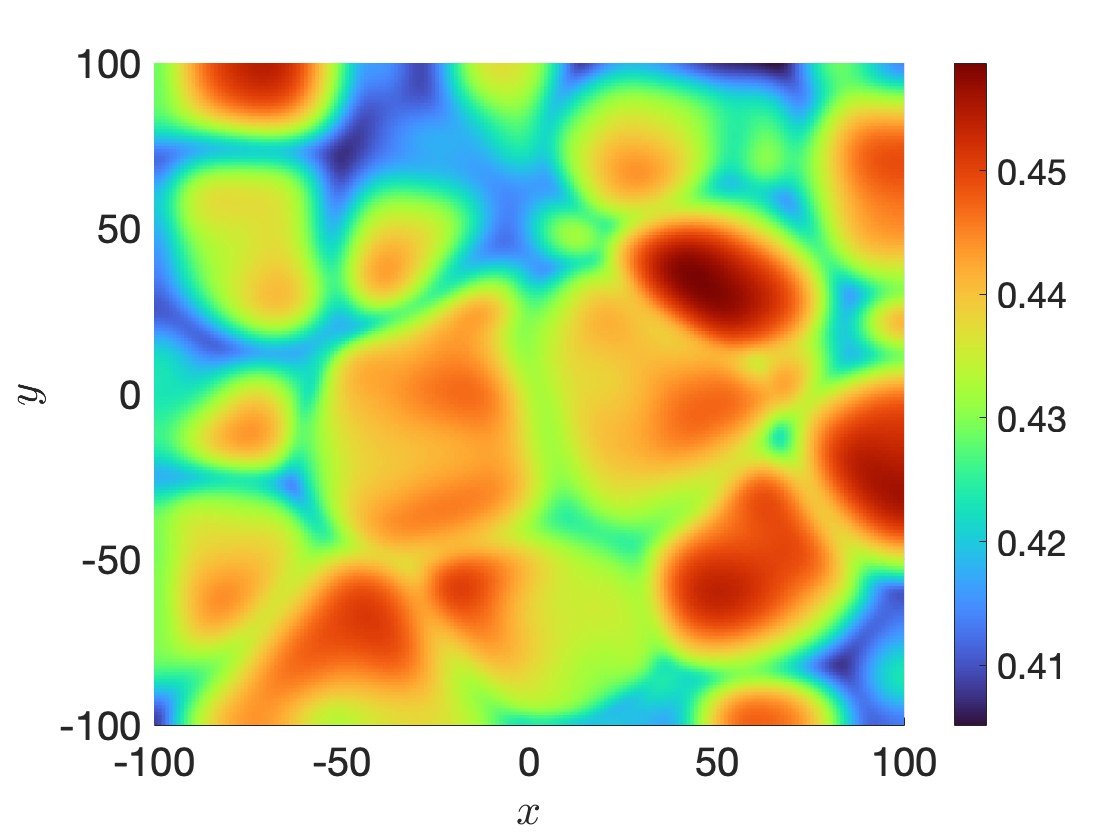}
        \caption*{Predators - Constant Toxin}
        \label{fig: constantT_Z_T0=10_t=201.jpg}
    \end{subfigure}
    \hspace{0.01\textwidth}
    \begin{subfigure}[t]{0.24\textwidth}
         \centering
         \includegraphics[width=\textwidth]{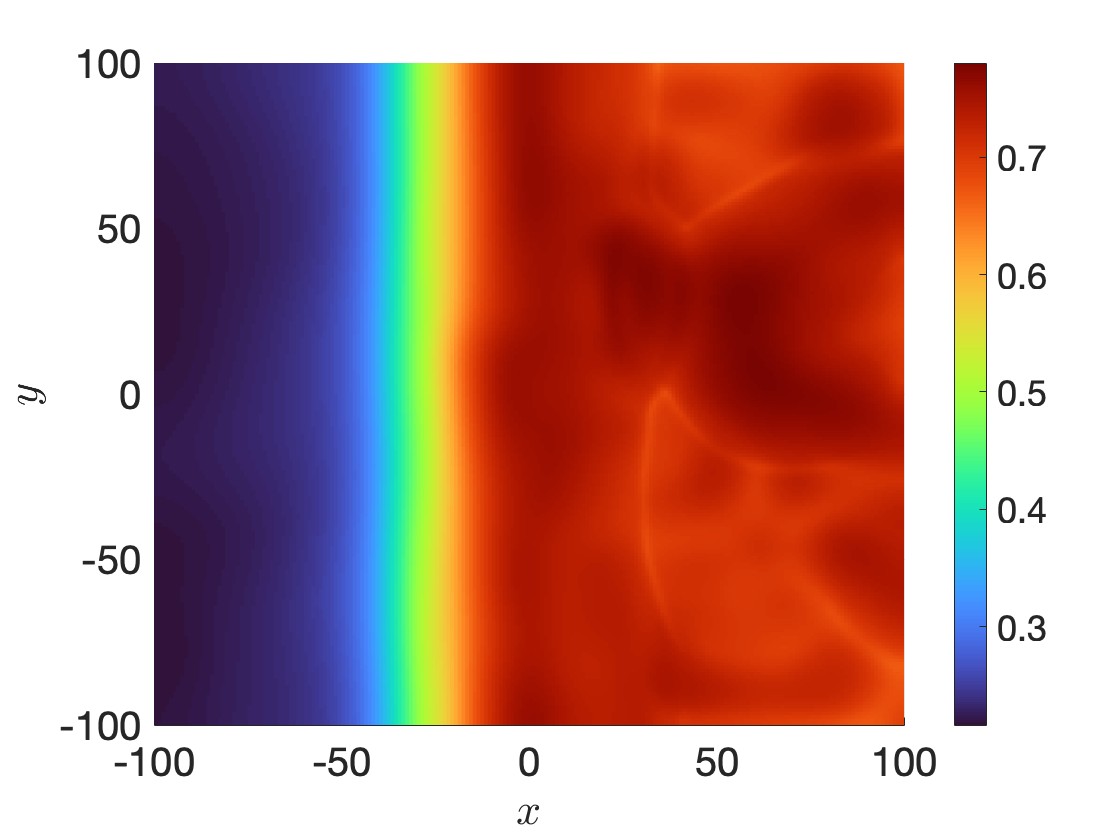}
         \caption*{Predators - Linear Gradient}
         \label{fig: Line_Z_T0=30_t=201.jpg}
     \end{subfigure}   
    \begin{subfigure}[t]{0.24\textwidth}
         \centering
         \includegraphics[width=\textwidth]{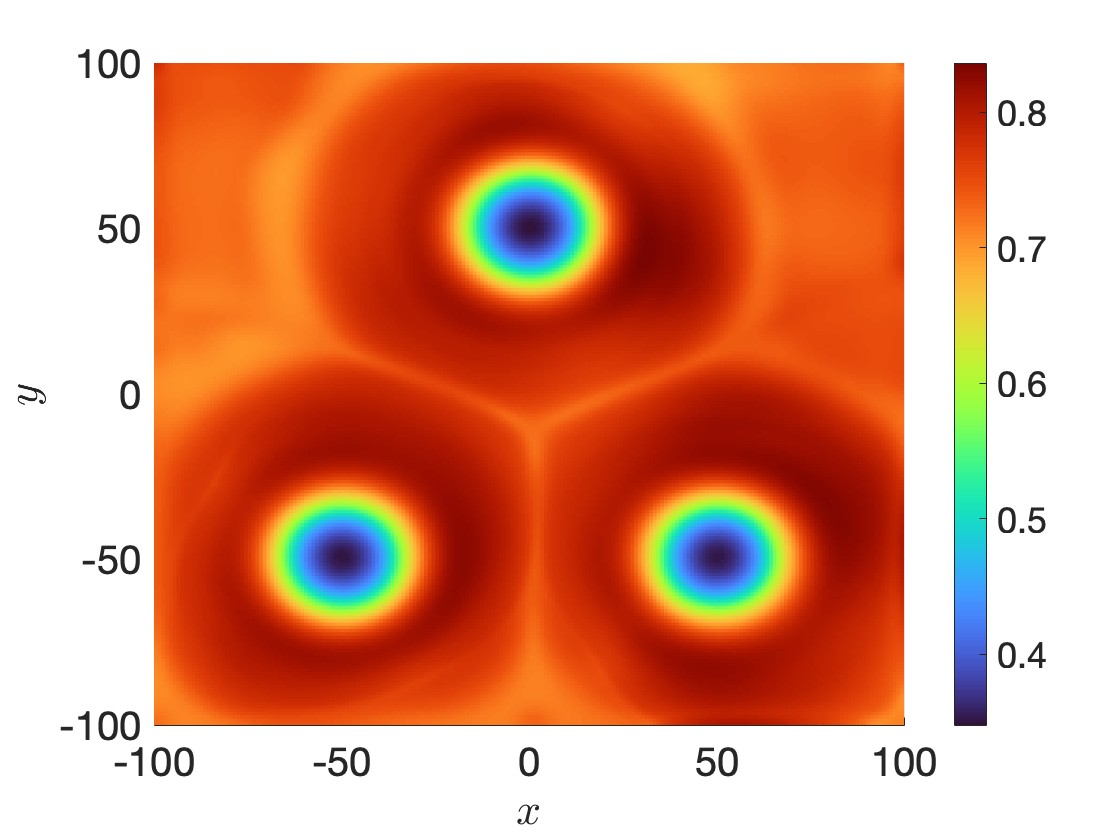}
         \caption*{Predators - Multi-Peak}
         \label{fig: ThreeCircles_Z_T0=30_t=201.jpg}
    \end{subfigure}
    \begin{subfigure}[t]{0.24\textwidth}
         \centering
         \includegraphics[width=\textwidth]{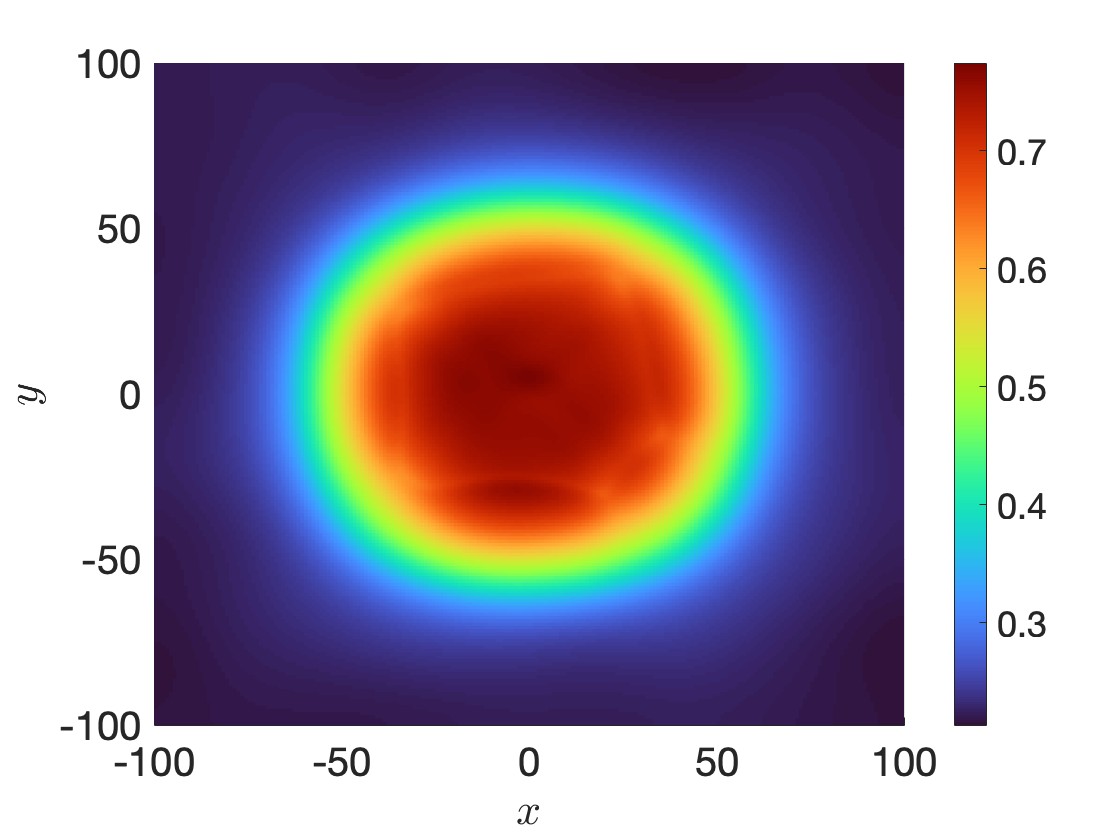}
         \caption*{Predators - Rectangular}
         \label{fig: EdgeCircles_Z_T0=30_t=201.jpg}
    \end{subfigure}
    
\caption{\footnotesize Spatial distribution patterns of populations in a 2D space with different toxin distributions. The first row represents prey, the second row bold grazers, the third row shy grazers, and the fourth row predators. The first column corresponds to a constant toxin concentration with \(T_0 = 10\), the second to a linear gradient toxin distribution with \(T_0 = 30\), the third to a multi-peak toxin distribution with \(T_0 = 30\), and the fourth to a rectangular toxin distribution with \(T_0 = 30\).}
\label{fig: Multiple_toxin_distribution_2D_T0=30}
\end{figure}
To better simulate real-world scenarios, we explored various types of toxin distributions, as illustrated in Figure \ref{fig: Multiple_toxin_distribution_2D_T0=30}. These include:

\begin{itemize}
    \item[1.] Constant concentration
    
    In certain localized regions within large aquatic environments, toxin levels can remain relatively uniform. This scenario often occurs in small areas with limited water flow or diffusion, where pollution levels are stable. In such cases, we model the toxin concentration as constant, \(T(x) = T_0\), reflecting a homogeneous distribution of toxins over a small region.
    
    \item[2.] Linear gradient concentration
    
    In large water bodies, such as oceans, pollution often concentrates near the edge exposed to human activity. For example, factories along the shoreline may discharge industrial waste directly into the water, or microplastics can accumulate along the coast due to littering. In a localized area near the shore, the toxin concentration is highest at the edge and gradually diminishes as it spreads farther into the water. This scenario can be modeled using a linear gradient, where the toxin level peaks at the left edge (\(T_0\)) and decreases to zero at \(x = 0\):
   \[
   T(x) = \max\left\{ -T_0 \frac{x}{100}, 0 \right\}.
   \]

\item[3.] Multi-peak concentration

In larger environments, pollution often originates from multiple sources, such as industrial zones or densely populated areas with several waste discharge points. Toxins spread from these multiple locations simultaneously, creating complex pollution patterns. To capture this, we use a multi-peak distribution, with three peaks located at (-50, -50), (0, 50), and (50, -50), each decaying exponentially:
   \[
   T(x, y) = T_0 \max\left\{e^{-\frac{x^2 + (y-50)^2}{15^2}}, e^{-\frac{(x+50)^2 + (y+50)^2}{15^2}}, e^{-\frac{(x-50)^2 + (y+50)^2}{15^2}}\right\}.
   \]

\item[4.] Rectangular distribution

In small lakes or enclosed water bodies, pollutants often accumulate near the edges and gradually diffuse toward the center. This can occur when pollution is introduced from surrounding land or from various points along the shore. As a result, the highest concentrations are found along the edges, with the toxin levels decaying as they move inward. We model this scenario with a rectangular distribution, where the toxin concentration is highest along the edges and decreases toward the center:
   \[
   T(x, y) = T_0 \left( 1 - \left(1 - \left(\frac{x}{100}\right)^2 \right)\left( 1 - \left(\frac{y}{100}\right)^2 \right)\right).
   \]

\end{itemize}

The results clearly demonstrate that population distribution is strongly influenced by the spatial distribution of toxins. For a homogeneously distributed toxin concentration at a relatively low level (\(T_0 = 10\)), shy individuals dominate and persist at much higher densities compared to bold grazers. However, when the toxin concentration increases significantly, all grazers eventually go extinct over time. For heterogeneous toxin distributions with \(T_0 = 30\), the population exhibits more complex patterns and moves periodically, such as spreading strips and loops. Species tend to move along toxin concentration gradients. Bold grazers prefer to aggregate in toxin-free or very low-toxin areas, while shy grazers tend to cluster in regions with low to moderate toxin levels. The population structure varies significantly depending on the spatial distribution of toxins. If the low-toxin region is larger, shy grazers dominate a greater area, whereas bold grazers become more dominant in toxin-free or very low-toxin regions. In areas where the toxin decays quickly, shy grazers tend to be dominant. Conversely, in areas with highly concentrated toxin levels, neither grazers nor predators can survive, such as the left edge in the linear gradient distribution and the corners of the rectangular distribution, which aligns with the results in Lemma \ref{lemma: boundedness of integral of Z}.

\section{Discussion}
\label{sec: Discussion}

Pollution is widely recognized as a significant threat to aquatic ecosystems, and animals exhibit varying behavioral responses influenced by their personality traits. However, the population dynamics of animals with different personalities in spatially heterogeneous polluted environments remain poorly understood. In this study, we develop a prey-taxis model that incorporates spatially dependent predation functions and taxis coefficients to investigate these dynamics.

Mathematically, we focused on the global existence of classical solutions for the general model \eqref{eq: general model} with sublinear predation functions and bounded taxis functions. We thoroughly studied the $L^1$ boundedness of the solution. Using a bootstrap strategy, we gradually enhanced the smoothness of the solution and established global boundedness.

The model was then applied to examine the population dynamics of an ecosystem exposed to a spatially heterogeneous polluted environment. Based on experimental data and relevant mathematical biology background, we derived specific functional forms. We compared the behavioral responses of grazers with different personality traits (bold vs. shy) under various environmental conditions. Specifically, we analyzed the aggregation behavior, population structure, and movement traces of animals in a one-dimensional space and further explored the spatial patterns in a two-dimensional space.

Although bold grazers achieve greater predation success, they also face higher predation risks \citep{wolf2007life,reale2010personality,church2019ideal}. Our findings indicate that, despite this trade-off, bold grazers are still more competitive in toxin-free areas. In this case, over time, ecosystems tend to evolve toward a predominance of bold personalities in such environments.

Interestingly, in regions with low to moderate toxin levels, the dominant personality shifts, with shy grazers becoming more competitive. Bold grazers tend to follow toxin gradients, occupying toxin-free or very low-toxin areas, while shy grazers are more likely to remain in low to moderately polluted regions. One possible reason is that in toxic environments, species experience higher metabolic rates and increased energetic needs \citep{mckenzie2007complex}, which enhance foraging behavior and activity. For bold grazers, their naturally high movement activity already demands significant energy, and the additional strain from toxin exposure becomes overwhelming. In contrast, shy grazers, being more reserved, can benefit from a moderate increase in activity without the same energetic cost. Another potential explanation is that bold grazers may accumulate more toxins \citep{chen2022fish}, leading to greater negative effects and higher death rates compared to shy grazers. As bold grazers decline, shy grazers gain access to more resources, improving their chances of growth. This suggests that pollutants can sometimes provide positive feedback, depending on toxin levels and personality traits. However, in highly polluted areas, no population can survive. 

This feedback also alters population structure and aggregation behavior. The proportion of bold personality is lower near polluted center, however, shy grazers dominate a larger area near polluted center as pollution levels rise. The spatial distribution pattern of population is highly influenced by toxin distribution. In a totally toxin-free environment, the spatial distribution of species exhibits high randomness. With the presence of toxin, the population moves along the gradient of toxin and shows periodic pattern. Moreover, As the concentration of toxin increases, the aggregation behavior becomes more pronounced for all prey, grazers, and predators.

The degree of personality difference also plays a significant role. When the difference between bold and shy grazers is small, shy grazers exhibit behaviors more similar to bold grazers and can occupy a larger area. However, when the discrepancy is large, shy grazers may become too timid to survive anywhere. There may be more interesting outcomes, and we call for further experimental research to explore the influence of personality discrepancies.

The dynamics of the entire grazer population, however, are different. The population as a whole shows some resilience to pollutants. Near the pollution source, when the maximum toxin concentration is relatively low, the total grazer population density remains stable. Interestingly, as the toxin concentration surpasses a certain threshold, the population increases, possibly due to adaptive strategies or positive feedback from pollutant-induced higher activity in shy grazers \citep{brodin2013dilute,montiglio2014contaminants}. This positive feedback can sometimes not only mitigate the negative effects of pollutants but also offer advantages. However, as toxin levels continue to rise, the overall population eventually declines, leading to extinction when the toxin concentration becomes too high. Ultimately, the environment becomes unsustainable for the ecosystem, driving all grazers to extinction.

This study examines the interaction between two distinct personality types, bold and shy. As personality traits often lie on a continuum, future research could explore treating personality as a continuous variable or function, which may provide a more nuanced understanding of behavioral dynamics. The model introduced here is general enough to be adapted to model various heterogeneous signals or sources beyond toxins. This versatility makes it applicable to a wide range of real-world scenarios. To further improve the model’s predictive accuracy, incorporating more precise data on toxin distribution and concentration will be crucial. This enhancement will lead to a more comprehensive understanding of population dynamics influenced by pollution or other environmental factors. From a mathematical perspective, future research could explore the stability and instability of solutions to the general model, potentially providing deeper insights into its long-term behavior and the emergence of pattern formation.

\section*{Acknowledgments}
The research of Hao Wang was partially supported by the Natural Sciences
and Engineering Research Council of Canada (Individual Discovery Grant RGPIN-2020-03911 and Discovery
Accelerator Supplement Award RGPAS-2020-00090) and the Canada Research Chairs program (Tier 1 Canada
Research Chair Award).

\section*{Data Availability}
All data we used in this paper came from published papers in the literature.

\section*{Conflict of Interest}
The authors declare that they have no known conflict of interest.

\begin{appendix}
\section{Appendix}
\subsection{Local existence and positivity}
\label{app: proof for local existence}
We rewrite the general model \eqref{eq: general model} in a more concise form
\begin{equation}
    \label{eq: concise form}
    \frac{\partial U}{\partial t}= \nabla \left (\mathbb A(x, U) \nabla U \right ) + \Phi (x, U), \quad \text{in } \Omega \times (0, \infty),
\end{equation}
where \( U=(X,Y_1,Y_2,Z) \) is the solution vector, and the matrix \(\mathbb A\) is
\begin{equation}
    \label{eq: matrix mathbb A}
    \mathbb A(x, U) = 
    \begin{pmatrix}
        \delta_X & 0 & 0 & 0 \\
        -\alpha_1(x) h(Y_1) & \delta_{Y_1} & 0 & 0 \\
        -\alpha_2(x) h(Y_2) & 0 & \delta_{Y_2} & 0 \\
        0 & -\gamma_1(x) h(Z) & -\gamma_2(x) h(Z) & \delta_Z
    \end{pmatrix},
\end{equation}
and the function \(\Phi(x, U)\) is given by
\begin{equation}
    \label{eq: vector Phi}
    \Phi(x, U) = 
    \begin{pmatrix}
        r(X) - f_1(X, Y_1, x) - f_2(X, Y_2, x) - d_1(x) X \\
        e_1 f_1(X, Y_1, x) - g_1(Y_1, Z, x) - d_{21}(x) Y_1 \\
        e_1 f_2(X, Y_2, x) - g_2(Y_2, Z, x) - d_{22}(x) Y_2 \\
        e_2 g_1(Y_1, Z, x) + e_2 g_2(Y_2, x) - d_3(x) Z
    \end{pmatrix}.
\end{equation}

\begin{proof}[\rm \bf Proof of Theorem \ref{th: local existence}]
\label{proof: th: local existence}
    Fix a number $p > n$ and define
    \begin{equation}
        \label{eq: V_eps}
        V_\epsilon = \{ v \in W^{1, p} (\Omega; \mathbb{R}^4): v(x) \in G_\epsilon = (-\epsilon, \infty)^4, \ \forall x \in \overline{\Omega} \}
    \end{equation}
    for some $\epsilon > 0$. Observe that all eigenvalues of $\mathbb A$ have positive real parts, and assumption \eqref{eq: assumption initial} implies that the initial values belong to $V_{\epsilon}$. By the theorem on local existence of quasilinear parabolic equations (\citep{amann1990dynamic}, p. 17), we obtain
    \begin{equation}
        \label{eq: local smoothness 1}
        U \in C \big( [0, T_{\max}); V_{\epsilon} \big) \cap C^\infty \big( \overline{\Omega} \times (0, T_{\max}); \mathbb R^4 \big),
    \end{equation}
    where $T_{\max} \in (0, \infty]$ is the lifespan defined by
    \begin{equation*}
        T_{\max} := \sup \{ T > 0: U(\cdot, t) \in V_{\epsilon}, \ \forall t \in [0, T] \}.
    \end{equation*}
    By Sobolev embedding and the fact that $U(\cdot, t) \in C(\overline{\Omega}; G_\epsilon)$ for all $t \geq 0$, we obtain
    \begin{equation}
        \label{eq: local smoothness 2}
        U \in C \big( [0, T_{\max}), C(\overline{\Omega}; G_\epsilon) \big) = C \big( \overline{\Omega} \times [0, T_{\max}); G_\epsilon \big).
    \end{equation}
    Moreover, $U = 0$ is a supersolution of \eqref{eq: general model} and the initial value is not identically 0 by assumption \eqref{eq: assumption initial}. Using comparison principle for parabolic equation with Neumann boundary condition, we have $U(t, x) \in \mathbb{R}^4_{+}$ for all $t > 0$ and $x \in \Omega$. This fact combined with equations \eqref{eq: local smoothness 1} and \eqref{eq: local smoothness 2} imply equation \eqref{eq: smoothness}. Since the solution $U$ takes value away from the boundary of $G_\epsilon$ and the matrix $\mathbb{A}$ in equation \eqref{eq: matrix mathbb A} is a lower triangular matrix, we obtain the blow-up criterion \eqref{eq: blow up condition simplified} by (Theorem 3, \citep{amann1989dynamic}).
\end{proof}

\subsection{$L^1$ estimates}
\label{App: proof for L1 L2 estimates}

\begin{proof}[\rm\bf Proof of Lemma \ref{lem: X bounded}]
\label{proof: lem: X bounded}
    The first equation of \eqref{eq: general model} and assumptions \eqref{eq: assumption r}, \eqref{eq: assumption f} imply
    \begin{equation}
        \label{eq: proof of X bounded}
        \frac{\partial X}{\partial t} - \delta_X \Delta X \leq r(X) \leq 0
    \end{equation}
    wherever $X(x, t) \geq K$. It follows that $K$ is a subsolution of the first equation in \eqref{eq: general model}. Using comparison principle for parabolic equation with Neumann boundary condition, we obtain the bound \eqref{eq: X bounded}.
\end{proof}

\begin{proof}[\rm\bf Proof of Lemma \ref{lemma: boundedness of integral of Y_1}]
\label{proof: lemma: boundedness of integral of Y_1}
    We multiply the first equation in \eqref{eq: general model} by $e_1$, then add the result to the second equation in \eqref{eq: general model}. An integration by parts yields
    \begin{equation*}
        \frac{\mathrm{d}}{\mathrm{d}t} \int_\Omega \left(e_1X+Y_1\right) + \int_\Omega d_{21}(x) Y_1 
        = \int_\Omega \Big( e_1 r(X) - e_1 f_2(X,Y_2,x) - g_1(Y_1,Z,x) - e_1 d_1(x) X \Big).
    \end{equation*}
    It follows from nonnegativity of $X$, $Y_1$ and $Y_2$ that 
    \begin{equation}
    \label{eq: proof 1 of boundedness of integral of Y_1 }
        \frac{\mathrm{d}}{\mathrm{d}t} \int_\Omega \left(e_1X+Y_1\right)  + \int_\Omega d_{21m}(x) (e_1 X + Y_1)  \leq e_1 \int_\Omega (r(X) + \left(d_{21m}-d_{21}(x)\right) X).
    \end{equation}
    By the assumption on $d_{21}$ and Lemma \ref{lem: X bounded}, one obtains
    \begin{equation}
    \label{eq: proof 2 of boundedness of integral of Y_1 }
        \begin{aligned}
            \frac{\mathrm{d}}{\mathrm{d}t} \int_\Omega (e_1 X + Y_1) + d_{21m} \int_\Omega \left(e_1X + Y_1\right) 
            &\leq e_1 \left( \overline{r(X)} + d_{21m} \overline{X}\right) |\Omega|,
        \end{aligned}
    \end{equation}
    where 
    \begin{equation}
        \label{eq: r(X) bar}
        \overline{r(X)}=\max\left\{r(X): X \in [0, \overline{X}] \right\}.
    \end{equation}
    Applying Gr\"onwall's inequality to \eqref{eq: proof 2 of boundedness of integral of Y_1 }, we get
    \begin{equation*}
        \int_\Omega \left(e_1 X + Y_1 \right)
        \leq \max \left\{ e_1 \|X_0\|_{L^1} + \|Y_{10}\|_{L^1}, \ e_1 |\Omega| \left(   \frac{\overline{r(X)}}{d_{21m}} +  \overline{X}\right)  \right\} : = C_1,
    \end{equation*}
    which implies \eqref{eq: lemma of boundedness of integral of Y_1}.

    Moreover, an integration by parts with respect to the second equation in \eqref{eq: general model} yields
    \begin{equation*}
        \frac{\mathrm{d} }{\mathrm{d}t} \int_\Omega Y_1  
        \leq \int_\Omega \left(e_1 f_1(X,Y_1,x) - d_{21m} Y_1\right) \leq \int_\Omega \left(e_1 C_{f,1} - d_{21m} \right)Y_1,
    \end{equation*}
    where the assumption on $f$ \eqref{eq: assumption f} has been used.
It follows that
    \begin{equation*}
        \int_\Omega Y_1 \leq e^{\left(e_1 C_{f,1} - d_{21m} \right) t} \int_\Omega Y_{10}.
    \end{equation*}
    If $e_1 C_{f,1} < d_{21m}$, by nonnegativity of $Y_1$, we have
    \begin{equation*}
        \lim_{t\to \infty} \int_\Omega Y_1 =0.
    \end{equation*}
\end{proof}

\begin{proof}[\rm\bf Proof of Lemma \ref{lemma: boundedness of integral of Y_2}]
    \label{proof: lemma: boundedness of integral of Y_2}
By similar argument to Lemma \ref{lemma: boundedness of integral of Y_1}, we can prove \eqref{eq: lemma of boundedness of integral of Y_2} with
    \begin{equation*}
        C_2 := \max \left\{ e_1 \|X_0\|_{L^1} + \|Y_{20}\|_{L^1}, \ \ e_1 |\Omega| \left(   \frac{\overline{r(X)}}{d_{22m}} +  \overline{X}\right) \right\}.
    \end{equation*} 
We omit detailed proof here.
\end{proof}

\begin{proof}[\rm\bf Proof of Lemma \ref{lemma: boundedness of integral of Z}]
\label{proof: lemma: boundedness of integral of Z}
We multiply the first equation in \eqref{eq: general model} by $e_1e_2$, the second and third equations by $e_2$, then add the results to the fourth equation. An integration by parts yields
    \begin{equation}
        \label{eq: proof 1 of boundedness of integral of Z}
        \begin{aligned}
        & \frac{\mathrm{d}}{\mathrm{d}t}\int_\Omega \left(e_1 e_2 X + e_2 Y_1 + e_2 Y_2 + Z \right)  +  \int_\Omega \Big( e_1e_2 d_1(x) X + e_2 d_{21}(x) Y_1  + e_2 d_{22}(x) Y_2  + d_3(x) Z \Big)\\
        &= e_1 e_2 \int_\Omega r(X).
        \end{aligned}
    \end{equation}
    Denote $d_m = \min\{ d_{21m}, d_{22m}, d_{3m} \}$. Adding $e_1 e_2 d \int_\Omega X $to both sides of \eqref{eq: proof 1 of boundedness of integral of Z}, along with Lemma \ref{lem: X bounded}, one obtains the inequality
    \begin{equation}
    \label{eq: proof 2 of boundedness of integral of Z }
        \begin{aligned}
            & \quad \ \frac{\mathrm{d}}{\mathrm{d}t} \int_\Omega (e_1 e_2 X + e_2 Y_1 + e_2 Y_2 + Z )  + d_m \int_\Omega \left(e_1 e_2 X + e_2 Y_1 + e_2 Y_2 + Z \right)  \\
            & \leq e_1 e_2 \int_\Omega r(X) + e_1 e_2 d_m \int_\Omega X  \\
            & \leq e_1 e_2 \left( \overline{r(X)} + d_m \overline{X} \right) |\Omega|,
        \end{aligned}
    \end{equation}
    where $\overline{r(X)}$ is defined in \eqref{eq: r(X) bar}. Applying Gr\"onwall's inequality to \eqref{eq: proof 2 of boundedness of integral of Z }, we get
    \begin{equation}
    \label{eq: L1 sum upper bound}
    \begin{aligned}
        &\quad \ \int_\Omega \left(e_1 e_2 X + e_2 Y_1 + e_2 Y_2 + Z \right)  \\
        & \leq \max \left\{e_{1} e_{2} \|X_0 \|_{L^1} + e_{2} \| Y_{10}\|_{L^1} + e_{2} \| Y_{20}\|_{L^1} + \| Z_0 \|_{L^1}, \ e_1 e_2 \left( \frac{\overline{r(X)}}{d_m} +  \overline{X}\right) |\Omega| \right\} := C_3,
        \end{aligned}
    \end{equation}
    which gives \eqref{eq: lemma of boundedness of integral of Z}. Moreover, from \eqref{eq: proof 2 of boundedness of integral of Z }, we can obtain
    \begin{equation}
    \label{eq: L1 sum limsup upper bound}
        \limsup_{t\to \infty}\int_\Omega \left(e_1 e_2 X + e_2 Y_1 + e_2 Y_2 + Z \right) 
        \leq e_1 e_2 \left( \frac{\overline{r(X)}}{d_m} +  \overline{X}\right) |\Omega|.
    \end{equation}
    

    On the other hand, an integration by parts with respect to the forth equation in \eqref{eq: general model} yields
    \begin{equation*}
        \frac{\mathrm{d}}{\mathrm{d}t} \int_\Omega Z \leq \int_\Omega \left( e_2\left(g_1(Y_1,Z,x) + g_2(Y_2,Z,x)\right) - d_3(x) Z\right) \leq \int_\Omega \left(e_2\left( C_{g,1} + C_{g,2}\right) - d_{3m} \right) Z.
    \end{equation*}
    If $e_2\left( C_{g,1} + C_{g,2}\right) < d_{3m}$, by nonnegativity of $Z$, we have
    \begin{equation*}
        \lim\limits_{t \to \infty} \int_\Omega Z = 0.
    \end{equation*}
\end{proof}

\subsection{Global existence}
\label{App: proof for global existence}
In this section, we prove global existence of the solution to the system \eqref{eq: general model}-\eqref{eq: bc ic}. Moreover, we restrict ourselves to $n \leq 3$ in our analysis, since for general dimension $n$, we can iterate the proof below and gradually increase the integrability and smoothness of the solution. 

In order to obtain further estimates, we have to require that the matrix $\mathbb{A}$ defined in equation \eqref{eq: matrix mathbb A} is positive definite. Below we provide a sufficient condition for it.

\begin{lemma}
    \label{lem: A positive definite}
    If 
    \begin{equation}
        \label{eq: condition A positive definite}
        \begin{aligned}
            & \quad \frac{\alpha_{1M}^2}{\delta_{Y_1}} + \frac{\alpha_{2M}^2}{\delta_{Y_2}} < 4 \frac{\delta_X}{C_h^2}, \\
            & \quad \left( \frac{\alpha_{1M} \gamma_{1M} }{\delta_{Y_1}} + \frac{\alpha_{2M} \gamma_{2M}}{\delta_{Y_2}} \right)^2  < \left( 4 \frac{\delta_X}{C_h^2} - \frac{\alpha_{1M}^2}{\delta_{Y_1}} - \frac{\alpha_{2M}^2}{\delta_{Y_2}} \right) \cdot \left( 4 \frac{\delta_Z}{C_h^2} -  \frac{\gamma_{1M}^2} {\delta_{Y_1}} - \frac{\gamma_{2M}^2} {\delta_{Y_2}} \right),
        \end{aligned}
    \end{equation}
    then the matrix $\mathbb{A} = \mathbb{A} (x, U)$ defined in equation \eqref{eq: matrix mathbb A} is uniformly positive definite in the sense of
    \begin{equation}
    \label{eq: A positive definite}
        \xi^T \mathbb A(x, U) \xi \geq \lambda |\xi|^2, \quad \forall x \in \Omega, \ U \in \mathbb R_{\geq 0}^4, \ \xi \in \mathbb R^4,
    \end{equation}
    for some $\lambda > 0$.
\end{lemma}
\begin{proof}
\label{proof: lem: A positive definite}
    First we prove that $\mathbb{A}$ is positive definite (not necessarily uniformly) under condition \eqref{eq: condition A positive definite}. This is true if and only if $\mathbb{B} = \mathbb{A} + \mathbb{A}^T$ is positive definite. It is equivalent to require that the determinants of the following submatrices of $\mathbb{B}$ are positive:
    \begin{equation*}
        \begin{aligned}
            & \mathbb{B}_1 = 2\delta_{Y_1}, \\
            & \mathbb{B}_2 = \text{diag} (2\delta_{Y_1}, 2\delta_{Y_2}), \\
            & \mathbb{B}_3 = 
            \begin{pmatrix}
                2\delta_X & -\alpha_1(x) h(Y_1) & -\alpha_2(x) h(Y_2) \\
                -\alpha_1(x) h(Y_1) & 2\delta_{Y_1} & 0 \\
                -\alpha_2(x) h(Y_2) & 0 & 2\delta_{Y_2}
            \end{pmatrix}, \\
            & \mathbb{B}_4 = \mathbb{B}.
        \end{aligned}
    \end{equation*}
    It is clear that $\mathbb{B}_1$ and $\mathbb{B}_2$ have positive determinants. By direct calculation, $\mathbb{B}_3$ and $\mathbb{B}_4$ have positive determinant if and only if
    \begin{equation}
        \label{eq: proof of A positive definite}
        \begin{aligned}
            & \frac{b_1^2}{\delta_{Y_1}} + \frac{b_2^2}{\delta_{Y_2}} < 4 \delta_X, \\
            & \left( \frac{b_1 b_3}{\delta_{Y_1}} + \frac{b_2 b_4}{\delta_{Y_2}} \right)^2 < \left( 4 \delta_X - \frac{b_1^2}{\delta_{Y_1}} - \frac{b_2^2}{\delta_{Y_2}} \right) \cdot \left( 4 \delta_Z - \frac{b_3^2}{\delta_{Y_1}} - \frac{b_4^2} {\delta_{Y_2}} \right),
        \end{aligned}
    \end{equation}
    where
    \begin{align*}
        b_1 = \alpha_1(x) h(Y_1), \quad b_2 = \alpha_2(x) h(Y_2), 
        \quad b_3 = \gamma_1(x) h(Z), \quad b_4 = \gamma_2(x) h(Z).
    \end{align*}
    By taking the minimum over the left hand side and taking the maximum over the right hand side of \eqref{eq: proof of A positive definite}, we obtain \eqref{eq: condition A positive definite}.

    Note that, if equation \eqref{eq: condition A positive definite} holds, then there exists $\lambda > 0$ so that it is still true when $\delta_X$, $\delta_{Y_1}$, $\delta_{Y_2}$ and $\delta_Z$ are replaced by $\delta_X - \lambda$, $\delta_{Y_1} - \lambda$, $\delta_{Y_2} - \lambda$ and $\delta_Z - \lambda$ respectively. The modified inequality shows that $\mathbb{A} - \lambda I_4$ is positive definite. This implies \eqref{eq: A positive definite}.
\end{proof}

As we will see in subsequent proofs, condition \eqref{eq: condition A positive definite} only guarantees time-dependent estimates of the solution. To make the estimates uniform in time, we require that the following quantity
\begin{equation}
\label{eq: U L^2 boundedness def M}
\begin{aligned}
    M = \max \left\{ C_{f, 1} - d_{21m}, \ C_{f, 2} - d_{22m}, \ C_{g, 1} + C_{g, 2} - d_{3m} \right\}
\end{aligned}
\end{equation}
is negative.
    
\begin{lemma}
    \label{lemma: L^2 boundedness}
    Suppose condition \eqref{eq: condition A positive definite} holds, then the solution to equation \eqref{eq: general model}-\eqref{eq: bc ic} satisfies
    \begin{equation}
        \label{eq: U L^2 bounded}
        \left\|U\right\|_{L^2} \leq C_4 \big( ||U_0||_{L^2}, ||X_0||_{L^\infty}, t \big), \quad \forall t \in [0, T_{\max}).
    \end{equation}
    and
    \begin{equation}
        \label{eq: nabla U L2 L2 bounded}
        \| \nabla U \|_{L^2(\Omega) L^2(0, t)} \leq C_5 \big( ||U_0||_{L^2}, ||X_0||_{L^\infty}, t \big), \quad \forall t \in [0, T_{\max}).
    \end{equation}
\end{lemma} 
\begin{proof}
\label{proof: lemma: L^2 boundedness}
    Multiplying equation \eqref{eq: concise form} by $U$ and integrating by parts, we obtain
    \begin{equation*}
        \frac{1}{2}\frac{\mathrm{d}}{\mathrm{d}t}\int_\Omega |U|^2  = - \int_\Omega \nabla U ^T \mathbb A \nabla U  + \int_\Omega \Phi(x,U)U ,
    \end{equation*}
    where $\mathbb{A}$ and $\Phi$ are defined in \eqref{eq: matrix mathbb A} and \eqref{eq: vector Phi} respectively. Condition \eqref{eq: condition A positive definite} implies that there exists a positive number $\lambda > 0$ such that 
    \begin{equation*}
        \nabla U ^T \mathbb A \nabla U \geq \lambda |\nabla U|^2.
    \end{equation*}
By direct calculation we also have
\begin{equation}
\label{eq: Phi U bounded}
    \Phi(x,U) U \leq \overline{r(X)}\overline X + M |U|^2,
\end{equation}
where $M$ is defined in \eqref{eq: U L^2 boundedness def M}. It follows that
\begin{equation}
\label{eq: nabla bounded lemma equation1}
    \frac{1}{2}\frac{\mathrm{d}}{\mathrm{d}t} \int_\Omega |U|^2  + \lambda \int_\Omega |\nabla U|^2  \leq |\Omega| \overline{r(X)} \overline X + M \int_\Omega |U|^2 .
\end{equation}
By Gr\"onwall's inequality, we obtain that
\begin{equation}
\label{eq: nabla bounded lemma equation2}
    \left\|U\right\|^2_{L^2}
    \leq e^{2Mt} \left\|U_0 \right\|^2_{L^2} + \frac{e^{2Mt} - 1}{M}| \Omega| \overline{r(X)} \overline X =: C_{4,t}^2.
\end{equation}
Integrate \eqref{eq: nabla bounded lemma equation1} from $0$ to $t$, and we get
\begin{equation*}
    \int_\Omega |U(t)|^2  - \int_\Omega |U(0)|^2  + 2 \lambda \int_0^t \left(\int_\Omega |\nabla U|^2  \right) \mathrm{d}s \leq 2t |\Omega| \overline{r(X)}\overline{X} + 2 M \int_0^t \left( \int_\Omega |U|^2  \right) \mathrm{d} s.
\end{equation*}
Together with equation \eqref{eq: nabla bounded lemma equation2}, we obtain
\begin{equation*}
    \| \nabla U \|_{L^2(\Omega) L^2(0, t)}^2 \leq  \frac{|\Omega|\overline{r(X)}\overline X}{2 \lambda} \frac{ e^{2Mt} - 1 }{M} + \frac{1}{2 \lambda} e^{2Mt} ||U_0||_{L^2}^2 =: C_{5,t}^2.
\end{equation*}
\end{proof}

\begin{lemma}
    Assume that condition \eqref{eq: condition A positive definite} holds, then the solution for equation \eqref{eq: general model}-\eqref{eq: bc ic} satisfies
    \begin{equation}
    \label{eq: nabla X L2}
        ||\nabla X||_{L^2} \leq C_6 \big( ||U_0||_{L^\infty}, ||\nabla X_0||_{L^2}, t \big), \quad \forall t \in [0, T_{\max}).
    \end{equation}
\end{lemma}
\begin{proof}
    Multiplying the first equation by $-\Delta X$ and integrating by parts, we obtain
    \begin{equation*}
        \frac{\mathrm{d}}{\mathrm{d}t} \int_\Omega |\nabla X|^2 + \delta_X \int_\Omega |\Delta X|^2 = - \int_\Omega r(X) \Delta X + \int_\Omega f_1(X,Y_1,x) \Delta X + \int_\Omega f_2(X,Y_2,x) \Delta X + \int_\Omega d_1(x) \Delta X.
    \end{equation*}
    By Assumption \eqref{eq: assumption f} and Young's inequality, we have
    \begin{equation*}
    \begin{aligned}
        &\quad \ \frac{\mathrm{d}}{\mathrm{d}t} \int_\Omega |\nabla X|^2  + \delta_X \int_\Omega |\Delta X|^2  \\
        &\leq \overline{r(X)}\int_\Omega |\Delta X| + C_{f,1} \int_\Omega Y_1 |\Delta X| + C_{f,2} \int_\Omega Y_2 |\Delta X| + d_{1M} \int_\Omega |\Delta X| \\
        &\leq \frac{2\overline{r(X)} |\Omega|}{\delta_X} + \frac{2C_{f,1}}{\delta_X}\int_\Omega |Y_1|^2  + \frac{2C_{f,2}}{\delta_X} \int_\Omega |Y_2|^2 + \frac{2d_{1M} |\Omega|}{\delta_X} + \frac{1}{2} \delta_X \int_\Omega |\Delta X|^2 .
    \end{aligned}
    \end{equation*}
    It follows that
    \begin{equation}
    \label{eq: nabla X 1 lemma nabla X}
        \frac{\mathrm{d}}{\mathrm{d}t} \int_\Omega |\nabla X|^2 + \frac{1}{2} \delta_X \int_\Omega |\Delta X|^2 
        \leq \frac{2 (\overline{r(X)} + d_{1M}) |\Omega|}{\delta_X} + \frac{2 C_{f,1}}{\delta_X} \int_\Omega |Y_1|^2 + \frac{2 C_{f,2}}{\delta_X} \int_\Omega |Y_2|^2.
    \end{equation}
    Moreover, by Young's inequality, 
    \begin{equation}
    \label{eq: nabla X 2 lemma nabla X}
        \int_\Omega |\nabla X|^2 = \int_\Omega \nabla X \cdot \nabla X  = - \int_\Omega X \Delta X 
        \leq \frac{1}{2 \delta_X} \int_\Omega |X|^2  + \frac{1}{2} \delta_X \int_\Omega |\Delta X|^2.
    \end{equation}
    Substituting \eqref{eq: nabla X 2 lemma nabla X} into \eqref{eq: nabla X 1 lemma nabla X}, we obtain
    \begin{equation*}
    \begin{aligned}
        &\quad \ \frac{\mathrm{d}}{\mathrm{d}t} \int_\Omega |\nabla X|^2  + \int_\Omega |\nabla X|^2  \\
        &\leq \frac{2 (\overline{r(X)} + d_{1M}) |\Omega|}{\delta_X} + \frac{1}{2\delta_X} \int_\Omega |X|^2 + \frac{2C_{f,1}}{\delta_X}\int_\Omega |Y_1|^2  + \frac{2C_{f,2}}{\delta_X} \int_\Omega |Y_2|^2  \\
        &\leq \frac{2 (\overline{r(X)} + d_{1M}) |\Omega|}{\delta_X} + \frac{\max\left\{ 1, 4C_{f,1}, 4C_{f,2} \right\}}{2\delta_X} \int_\Omega |U|^2 .
    \end{aligned}
    \end{equation*}
    Then Gr\"onwall's inequality and estimate \eqref{eq: U L^2 bounded} on $\| U \|_{L^2}$ yield
    \begin{equation*}
    \begin{aligned}
        ||\nabla X||_{L^2}^2 
        &\leq ||\nabla X_0||_{L^2}^2 + \frac{\max\left\{ 1, 4C_{f,1}, 4C_{f,2} \right\}}{2\delta_X} \int_0^t e^{-(t-s)} \left( C_{4, s}^2 + \frac{2 (\overline{r(X)} + d_{1M}) |\Omega|}{\delta_X} \right) \mathrm{d}s \\
        &\leq ||\nabla X_0||_{L^2}^2 + \frac{\max\left\{ 1, 4C_{f,1}, 4C_{f,2} \right\}}{2\delta_X} 
        \left( ||C_{4, s}||_{L^\infty (0, t)}^2 + \frac{2 (\overline{r(X)} + d_{1M}) |\Omega|}{\delta_X} \right) \\
        &=: C_{6,t}^2.
    \end{aligned}
    \end{equation*}
\end{proof}

 The following $L^p$-$L^q$ estimate (cf. \citep{winkler2010LpLq}) for the heat semigroup is fundamental in our subsequent analysis.

\begin{lemma}
    \label{lem: L^p-L^q estimate}
    Let $\left\{ e^{td\Delta} \right\}_{t\geq 0} $ be the Newmann heat semigroup in $\Omega$, and let $\lambda_1>0$ denote the first nonzero eigenvalue of $-\Delta$ in $\Omega$ under Neumann boundary conditions, where $d$ is a positive constant. Then for all $t>0$, there exist some constant $C$ depending only on $\Omega$ such that 
    \begin{itemize}
        \item[(i)] If $2\leq p < \infty$, then
        \begin{equation}
        \label{eq: lemma Lp Lq estimate 1}
            ||\nabla e^{td\Delta}z||_{L^p} \leq C \left(1+t^{-\frac{n}{2}\left(\frac{1}{q}-\frac{1}{p}\right)} \right) e^{-d\lambda_1 t}||\nabla z||_{L^q}
        \end{equation}
        for all $z \in W^{1,q}(\Omega)$.
        \item[(ii)] If $1 \leq q \leq p \leq \infty$, then
        \begin{equation}
        \label{eq: lemma Lp Lq estimate 2}
            ||\nabla e^{td\Delta}z||_{L^p} \leq C \left(1+t^{ - \frac{1}{2} -\frac{n}{2}\left(\frac{1}{q}-\frac{1}{p}\right)} \right) e^{-d\lambda_1 t}||z||_{L^q}
        \end{equation}
        for all $z \in L^q(\Omega)$.
        \item[(iii)] If $1\leq q\leq p\leq \infty$, then 
        \begin{equation}
        \label{eq: lemma Lp Lq estimate 3}
            ||e^{td\Delta}z||_{L^p} \leq C \left(1+t^{ -\frac{n}{2}\left(\frac{1}{q}-\frac{1}{p}\right)} \right) e^{-d\lambda_1 t}
            ||z||_{L^q}
        \end{equation}
        for all $z \in L^q(\Omega)$.
        \item[(iv)] If $1\leq q\leq p\leq \infty$, then 
        \begin{equation}
        \label{eq: lemma Lp Lq estimate 4}
            ||e^{td\Delta}\nabla \cdot z||_{L^p} \leq C \left(1+t^{- \frac{1}{2} -\frac{n}{2}\left(\frac{1}{q}-\frac{1}{p}\right)} \right) e^{-d\lambda_1 t}||z||_{L^q}
        \end{equation}
        for all $z \in \left(C_0^\infty (\Omega)\right)^n $.
    \end{itemize}
\end{lemma}

\begin{lemma}
    \label{lem: Young's convolution inequality on T}
    Define an operator $\mathcal{T}_{\rho, \lambda}$ by
    \begin{equation}
        \label{eq: operator T_rho,lambda}
        \mathcal{T}_{\rho, \lambda} \phi (t) = \int_0^t (1 + (t - s)^{-\rho}) e^{-\lambda (t - s)} \phi(s) \mathrm d s,
    \end{equation}
    where $\rho \in (0, 1)$ and $\lambda > 0$. Suppose
    \begin{equation}
        \label{eq: prop condition on p q}
        1 \leq p \leq q \leq \infty
        \quad \text{and } \quad
        \frac{1}{p} + \rho < \frac{1}{q} + 1,
    \end{equation}
    then for any $t \in [0, \infty]$, $\mathcal{T}_{\rho, \lambda}$ is a bounded operator from $L^{p} (0, t)$ to $L^{q} (0, t)$, and the bound is uniform in $t$.
\end{lemma}

\begin{proof}
    Fix $t \geq 0$ and define $\widetilde{\phi} (s) = \phi (s) \boldone_{(0, t)} (s)$, $s \in \mathbb R$. Besides, we let $\widetilde{\psi}(s) = (1 + s^{-\rho}) e^{-\lambda s} \boldone_{(0, t)} (s)$, then $\mathcal{T}_{\rho, \lambda} \phi = \widetilde{\phi} * \widetilde{\psi}$ on $(0, t)$. Let $r$ be a real number such that
    \begin{equation*}
        \frac{1}{p} + \frac{1}{r} = \frac{1}{q} + 1,
    \end{equation*}
    then from \eqref{eq: prop condition on p q} we know that $1 \leq r < \frac{1}{\rho}$. It follows that
    \begin{equation*}
        \| \widetilde{\psi} \|_{L^r (\mathbb R)}
        \leq \| e^{-\lambda s} \|_{L^r (0, t)} + \| s^{-\rho} e^{-\lambda s} \|_{L^r (0, t)}
        \leq (r \lambda)^{-\frac{1}{r}} + (r \lambda)^{\rho - \frac{1}{r}} \Gamma(1 - \rho r)^\frac{1}{r}
        =: c_r < \infty.
    \end{equation*}
  By Young's convolution inequality, we get
    \begin{equation}
        \label{eq: Lp Lq boundedness of T_rho lambda}
        \| \mathcal{T}_{\rho, \lambda} \phi \|_{L^q (0, t)}
        \leq \| \widetilde{\phi} * \widetilde{\psi} \|_{L^q (\mathbb R)}
        \leq \| \widetilde{\phi} \|_{L^p (\mathbb R)} \| \widetilde{\psi} \|_{L^r (\mathbb R)}
        \leq c_r \| \phi \|_{L^p (0, t)}.
    \end{equation}  
\end{proof}

\begin{lemma}
    \label{lem: Ln+1 Boundedness of Y_1}
    Assume condition \eqref{eq: condition A positive definite} holds, then the solution for equation \eqref{eq: general model}-\eqref{eq: bc ic} satisfies
    \begin{equation}
        \label{eq: Ln+1 Boundedness of Y_1}
        ||Y_1||_{L^{n+1}} \leq C_7 \big( ||U_0||_{L^\infty}, ||\nabla X_0||_{L^2}, t \big), \quad \forall t \in [0, T_{\max}).
    \end{equation}
\end{lemma}

\begin{proof}
    We apply the variation of constants formula to the second equation in \eqref{eq: general model} and obtain
    \begin{equation}
    \begin{aligned}
        \label{eq: variation of formula Y_1}
        Y_1(t) &= e^{(\delta_{Y_1} \Delta -d_{21m}) t}Y_{10} + \int_0^t e^{(\delta_{Y_1} \Delta - d_{21m}) (t-s)} \\
        & \quad \, \Big( e_1 f_1(X,Y_1,x) - g_1(Y_1,Z,x) - (d_{21}(x) - d_{21m})Y - \nabla \left( \alpha_1(x) h(Y_1)\nabla X \right) \Big) \, \mathrm{d}s \\
        & \leq e^{(\delta_{Y_1}\Delta - d_{21m}) t} Y_{10} + \int_0^t e^{(\delta_{Y_1} \Delta - d_{21m}) (t-s)} \left(e_1 f_1(X,Y_1,x) -\nabla (\alpha_1(x) h(Y_1) \nabla X \right)) \, \mathrm{d}s.
    \end{aligned}
    \end{equation}
    Applying \eqref{eq: lemma Lp Lq estimate 3} and \eqref{eq: lemma Lp Lq estimate 4} and setting
    \begin{equation}
        \label{eq: rho}
        \rho = \frac{n}{2} \left(\frac{1}{2} - \frac{1}{n+1}\right) \in \Big[ 0, \frac{1}{2} \Big),
    \end{equation}
    we derive 
    \begin{equation*}
    \begin{aligned}
        ||Y_1(t)||_{L^{n+1}} 
        & \leq \left\| e^{(\delta_{Y_1} \Delta -d_{21m}) t}Y_{10} \right\|_{L^{n+1}} + e_1 \int_0^t \left\| e^{(\delta_{Y_1}\Delta - d_{21m}) (t-s)} f_1(X,Y_1,x) \right\|_{L^{n+1}} \mathrm{d}s \\
        & \quad + \int_0^t \left\| e^{(\delta_{Y_1} \Delta - d_{21m}) (t-s)} \nabla (\alpha_1(x) h(Y_1)\nabla X) \right\|_{L^{n+1}} \mathrm{d}s\\
        & \leq c_1 e^{-(\delta_{Y_1} \lambda_1 + d_{21m}) t} ||Y_{10} ||_{L^{n+1}} + c_1 e_1 \mathcal{T}_{\rho,\delta_{Y_1} \lambda_1 + d_{21m}} ||f_1(X,Y_1,x)||_{L^2}  \\
        & \quad + c_1 \mathcal{T}_{ \frac{1}{2} + \rho,\delta_{Y_1} \lambda_1 + d_{21m}} ||\alpha_1(x) h(Y_1)\nabla X||_{L^2} 
    \end{aligned}
    \end{equation*}
    for some constant $c_1 > 0$ and $\mathcal{T}$ is defined in \eqref{eq: operator T_rho,lambda}. By assumptions \eqref{eq: assumption f}, \eqref{eq: assumption h} and estimates \eqref{eq: U L^2 bounded} on $\| U \|_{L^2}$ and \eqref{eq: nabla X L2} on $\| \nabla X \|_{L^2}$, we get
\begin{equation}
\label{eq: f1 L2 t_bounded}
    ||f_1(X, Y_1, x)||_{L^2} \leq C_{f, 1} C_{4, s}
\end{equation}
and
\begin{equation}
\label{eq: alpha_1(x)h(Y_1)nabla X L2 t_bounded}
    ||\alpha_1(x) h(Y_1) \nabla X||_{L^2} \leq \alpha_{1M} C_h C_{6, s}.
\end{equation}
Using Lemma \ref{lem: Young's convolution inequality on T} with $p = q = \infty$, we obtain
\begin{equation*}
    \mathcal{T}_{\rho,\delta_{Y_1} \lambda_1 + d_{21m}} ||f_1(X,Y_1,x)||_{L^2}
    \leq c_2 C_{f, 1} \| C_{4, s} \|_{L^\infty (0, t)}
\end{equation*}
and
\begin{equation*}
    \mathcal{T}_{ \frac{1}{2} + \rho,\delta_{Y_1} \lambda_1 + d_{21m}} ||\alpha_1(x) h(Y_1)\nabla X||_{L^2}
    \leq c_2 \alpha_{1M} C_h \| C_{6, s} \|_{L^\infty (0, t)}
\end{equation*}
for some constant $c_2 > 0$. Therefore, 
\begin{equation*}
    ||Y_1(t)||_{L^{n+1}} 
    \leq c_1 ||Y_{10}||_{L^{n+1}} + c_1 c_2 e_1 C_{f, 1} \| C_{4, s} \|_{L^\infty (0, t)} + c_1 c_2 \alpha_{1M} C_h \| C_{6, s} \|_{L^\infty (0, t)} =: C_{7,t}.
\end{equation*}
\end{proof}

By a similar argument as for Lemma \ref{lem: Ln+1 Boundedness of Y_1}, we can prove the following lemma.
\begin{lemma}
    \label{lemma: Ln+1 Boundedness of $Y_2$}
    Assume condition \eqref{eq: condition A positive definite} holds, then the solution for equation \eqref{eq: general model}-\eqref{eq: bc ic} satisfies
    \begin{equation*}
        ||Y_2||_{L^{n+1}} \leq C_8 \big( ||U_0||_{L^\infty}, ||\nabla X_0||_{L^2}, t \big), \quad \forall t \in [0, T_{\max}).
    \end{equation*}
\end{lemma}

\begin{lemma}
    \label{lemma: L^n+1 Boundedness of nabla X}
    Assume condition \eqref{eq: condition A positive definite} holds, then the solution for equation \eqref{eq: general model}-\eqref{eq: bc ic} satisfies
    \begin{equation}
        \label{eq: L^n+1 Boundedness of nabla X}
        ||\nabla X||_{L^{n+1}} \leq C_9 \big( ||U_0||_{L^\infty}, ||\nabla X_0||_{L^{n + 1}}, t \big), \quad \forall t \in [0, T_{\max}).
    \end{equation}
\end{lemma}
 
\begin{proof}
    Apply the variation of constants formula to the first equation in \eqref{eq: general model} and obtain
    \begin{equation}
        \label{eq: variation of constants X}
        \begin{aligned}
            X(t) & = e^{\delta_X t \Delta} X_0 + \int_0^t e^{\delta_X (t-s)\Delta}\left(r(X) - f_1(X,Y_1,x) - f_2(X,Y_2,x) - d_1(x) X \right) \mathrm{d}s.
        \end{aligned}
    \end{equation}
    Taking gradient at each side, we get
    \begin{equation*}
        \begin{aligned}
            \nabla X(t) & = \nabla e^{\delta_X t \Delta} X_0 + \int_0^t \nabla e^{\delta_X (t-s)\Delta} \left(r(X) - f_1(X,Y_1,x) - f_2(X,Y_2,x) - d_1(x) X \right) \mathrm{d}s.
        \end{aligned}
    \end{equation*}
    Applying \eqref{eq: lemma Lp Lq estimate 1} and \eqref{eq: lemma Lp Lq estimate 2}, and together with nonnegativity of $d_1$ we get
    \begin{equation*}
    \begin{aligned}
        ||\nabla X||_{L^{n+1}} & \leq c_1 e^{-\delta_X \lambda_1 t}||\nabla X_0||_{L^{n+1}} \\
        & \quad  + c_1 \mathcal{T}_{\frac{1}{2} + \rho, \delta_X \lambda_1} 
        \Big( || r(X)||_{L^2} + || f_1(X,Y_1,x) ||_{L^2} + || f_2(X,Y_2,x) ||_{L^2} + || d_1(x) X ||_{L^2} \Big)
        \end{aligned}
    \end{equation*}
    for $\rho$ in equation \eqref{eq: rho} and for some $c_1 > 0$. By assumptions \eqref{eq: assumption r}, \eqref{eq: assumption f}, we obtain
    \begin{equation}
        \begin{aligned}
        \label{eq: f_i L2 t_bounded}
            || f_i(X,Y_i) ||_{L^2} \leq C_{f,i} C_{4,s}, \quad i=1,2.
        \end{aligned}
    \end{equation}
    Together with the estimate \eqref{eq: U L^2 bounded} on $\| U \|_{L^2}$, we have
    \begin{equation*}
        \begin{aligned}
            ||\nabla X||_{L^{n+1}} 
            & \leq c_1 e^{-\delta_X \lambda_1 t} ||\nabla X_0||_{L^{n+1}} + c_1 \mathcal{T}_{\frac{1}{2} + \rho, \delta_X \lambda_1 } \left( \overline{r(X)} |\Omega|^{\frac{1}{2}} + (C_{f,1} + C_{f,2} + d_{1M} ) C_{4,s} \right)  \\
            & \leq c_1 ||\nabla X_0||_{L^{n+1}} + c_1 c_2 \left( \overline{r(X)} |\Omega|^{\frac{1}{2}} + (C_{f,1} + C_{f,2} + d_{1M} ) \| C_{4,s} \|_{L^\infty (0, t)} \right) \\
            & =: C_{9,t}.
        \end{aligned}
    \end{equation*}
\end{proof}

To summarize, Lemma \ref{lem: Ln+1 Boundedness of Y_1} (resp. Lemma \ref{lemma: Ln+1 Boundedness of $Y_2$}) provides an estimate of $||Y_1||_{L^{n + 1}}$ (resp. $||Y_2||_{L^{n + 1}}$) via the estimate of $||\nabla X||_{L^2}$ and $||Y_1||_{L^2}$ (resp. $||Y_2||_{L^2}$). Similarly, Lemma \ref{lemma: L^n+1 Boundedness of nabla X} provides an estimate of $||\nabla X||_{L^{n + 1}}$ via the estimate of $||Y_1||_{L^2}$ and $||Y_2||_{L^2}$. We can verify that the exact same proofs can be conducted by replacing the index pair $(2, n + 1)$ by $(n + 1, \infty)$. Thus, we deduce the following $L^\infty$ estimate.

\begin{lemma}
    \label{lem: L^inf boundedness of Y1 Y2 nabla X}
    Assume condition \eqref{eq: condition A positive definite} holds, then the solution for equation \eqref{eq: general model}-\eqref{eq: bc ic} satisfies
    \begin{equation}
        \label{eq: L^inf boundedness of Y1 Y2 nabla X}
        ||Y_1||_{L^\infty}, \ ||Y_2||_{L^\infty}, \ ||\nabla X||_{L^\infty} \leq C_{10} \big( ||U_0||_{L^\infty}, ||\nabla X_0||_{L^\infty}, t \big), \quad \forall t \in [0, T_{\max}).
    \end{equation}
\end{lemma}

\begin{lemma}
    \label{lem: Delta X L2 Lp bounded}
    Assume condition \eqref{eq: condition A positive definite} holds, then for all $p \in [2, \infty)$, the solution for equation \eqref{eq: general model}-\eqref{eq: bc ic} satisfies
    \begin{equation}
        \label{eq: lemma Delta X L2 Lp+infty bounded}
        ||\Delta X||_{L^2(\Omega) L^{p} (0, t)} \leq C_{11} \big( ||U_0||_{L^\infty}, ||\nabla X_0||_{L^\infty}, ||\Delta X_0||_{L^2}, t \big), \quad \forall t \in [0, T_{\max}).
    \end{equation}
\end{lemma}

\begin{proof}
We take $\Delta$ to both sides of equation \eqref{eq: variation of constants X}. Note that $\Delta$ commutes with $e^{\delta_X t \Delta}$, so we get
\begin{align}
\label{eq: Delta X formula}
    \Delta X(t) = e^{\delta_X t \Delta} \Delta X_0 + \int_0^t e^{\delta_X (t-s)\Delta} \Delta (r(X) - f_1(X,Y_1,x) - f_2(X,Y_2,x) -d_1(x) X ) \mathrm{d}s.
\end{align}

Let $p=q=2$ in \eqref{eq: lemma Lp Lq estimate 3} and \eqref{eq: lemma Lp Lq estimate 4}, then there exists a constant $c_1$ such that
\begin{equation*}
\begin{aligned}
    \left\| \Delta X \right\|_{L^2} 
    & \leq c_1 e^{-\delta_X \lambda_1 t} \left\| \Delta X_0 \right\|_{L^2}\\
    & \quad + c_1  \mathcal{T}_{\frac{1}{2} \delta_X \lambda_1 } \left(\| \nabla (r(X) - f_1(X,Y_1,x) - f_2(X,Y_2,x) - d_1(x) X ) \|_{L^2} \right) \\
    & = c_1 e^{-\delta_X \lambda_1 t} \left\| \Delta X_0 \right\|_{L^2} + c_1 \mathcal{T}_{\frac{1}{2}, \delta_X \lambda_1} \big( \| \nabla (r(X) - f_1(X,Y_1,x) - f_2(X,Y_2,x) - d_1(x) X ) \|_{L^2} \big).
\end{aligned}
\end{equation*}
We have
\begin{equation}
\begin{aligned}
\label{eq: proof nabla (r-f1-f2) L2}
    & \quad \ \| \nabla (r(X) - f_1(X,Y_1,x) - f_2(X,Y_2,x) - d_1(x) X) \|_{L^2} \\
    & \leq \left( \| r'(X) \|_{L^\infty} + \left\| \frac{\partial f_1}{\partial X} \right\|_{L^\infty} + \left\| \frac{\partial f_2}{\partial X} \right\|_{L^\infty} + ||d_1||_{L^\infty} \right) \| \nabla X \|_{L^2} 
     + \left\| \frac{\partial f_1}{\partial Y_1} \right\|_{L^\infty} \| \nabla Y_1 \|_{L^2} \\
     & \quad + \left\|  \frac{\partial f_2}{\partial Y_2} \right\|_{L^\infty} + \| \nabla Y_2 \|_{L^2} + \| \nabla_x f_1 \|_{L^2} + \| \nabla_x f_2 \|_{L^2} + \left\| \nabla d_1\right\|_{L^\infty} \left\| X\right\|_{L^2}.
\end{aligned}
\end{equation}
The $L^\infty$ estimates on $X$ \eqref{eq: X bounded}, $Y_1$ and $Y_2$ \eqref{eq: L^inf boundedness of Y1 Y2 nabla X} imply that the derivatives of $r$, $f_1$ and $f_2$ are bounded by a constant depending on $t$. Together with the smoothness of $d_1$ over $\overline{\Omega}$, we can find another constant $c_{2, t}$ depending on $t$ such that
\begin{equation}
    \label{eq: proof Delta X L2 Lp bounded}
    \begin{aligned}
        \| \Delta X \|_{L^2} 
        \leq c_1 e^{-\delta_X \lambda_1 t} \| \Delta X_0 \|_{L^2} + c_1 \mathcal{T}_{\frac{1}{2}, \delta_X \lambda_1} \big( c_{2, t} \left( \| \nabla X \|_{L^2} + \| \nabla Y_1 \|_{L^2} + \| \nabla Y_2 \|_{L^2} + 1 \right) \big).
    \end{aligned}
\end{equation}
For the first term, there exists a constant $c_3$ independent of $t$ such that
\begin{equation}
    \label{eq: Delta X L2 Lp+inf item 1}
    \begin{aligned}
       \left\| e^{-\delta_X \lambda_1 t } || \Delta X_0||_{L^2} \right\|_{L^{p} (0,t)}
       \leq \left\| e^{-\delta_X \lambda_1 t } \right\|_{L^{p} (0,t)} ||\Delta X_0||_{L^2} \leq c_3 ||\Delta X_0||_{L^2}.
    \end{aligned}
\end{equation} 
By Lemma \ref{lem: Young's convolution inequality on T}
and estimate \eqref{eq: nabla U L2 L2 bounded} on $||\nabla U||_{L^2(\Omega) L^2(0, t)}$, for any $p \in [2, \infty)$, there exists a constant $c_4$ such that
\begin{equation*}
    \begin{aligned}
        & \quad \ \left\| \mathcal{T}_{\frac{1}{2}, \delta_X \lambda_1} \big( c_{2, t} \left( \| \nabla X \|_{L^2} + \| \nabla Y_1 \|_{L^2} + \| \nabla Y_2 \|_{L^2} + 1 \right) \big) \right\|_{L^{p} (0,t)} \\
        & \leq c_4 \left\| c_{2, s} \left( \| \nabla X \|_{L^2} + \| \nabla Y_1 \|_{L^2} + \| \nabla Y_2 \|_{L^2} + 1 \right) \right\|_{L^{2} (0,t)} \\
        & \leq c_4 ||c_{2, s}||_{L^\infty(0, t)} \left( \| \nabla X \|_{L^2(\Omega)L^2(0,t)} + \| \nabla Y_1 \|_{L^2(\Omega)L^2(0,t)} + \| \nabla Y_2 \|_{L^2(\Omega)L^2(0,t)} + \|1\|_{L^2(0,t)} \right) \\
        & \leq c_4 ||c_{2, s}||_{L^\infty(0, t)} \left( C_{5, t} + \sqrt{t} \right).
    \end{aligned}
\end{equation*}
Therefore, taking $L^{p}$ norm on \eqref{eq: proof Delta X L2 Lp bounded} with respect to time, we obtain
\begin{align*}
    \quad \ \| \Delta X \|_{L^2(\Omega) L^{p} (0,t)} 
    \leq c_1 c_3 \| \Delta X_0 \|_{L^2} + c_1 c_4 ||c_{2, s}||_{L^\infty(0, t)} \left( C_{5, t} + \sqrt{t} \right) =: C_{11,t}. 
\end{align*}
\end{proof}

\begin{lemma}
    \label{lem: nabla Y_1 L2 Lp bounded}
    Assume condition \eqref{eq: condition A positive definite} holds, then for all $p \in [2, \infty)$, the solution for equation \eqref{eq: general model}-\eqref{eq: bc ic} satisfies
    \begin{equation}
        \label{eq: lemma nabla Y_1 L2 Lp+infty bounded}
        ||\nabla Y_1||_{L^2(\Omega) L^{p}(0, t)} \leq C_{12} \big( ||U_0||_{L^\infty}, ||\nabla X_0||_{L^\infty}, ||\nabla Y_{10}||_{L^2}, ||\Delta X_0||_{L^2}, t \big), \quad \forall t \in [0, T_{\max}).
    \end{equation}
\end{lemma}

\begin{proof}
Taking gradient on equation \eqref{eq: variation of formula Y_1}, we get
\begin{equation}
\label{eq: proof1: variation of formula Y_1}
    \begin{aligned}
        \nabla Y_1(t) 
        &= \nabla e^{(\delta_{Y_1}\Delta -d_{21m}) t} Y_{10} + \int_0^t \nabla e^{(\delta_{Y_1}\Delta - d_{21m}) (t-s)} \\
        & \quad \, \left(e_1f_1(X,Y_1,x) - g_1(Y_1,Z,x) - (d_{21}(x) - d_{21m}) Y -\nabla (\alpha_1(x) h(Y_1)\nabla X\right)) \mathrm{d}s.
    \end{aligned}
\end{equation}
By \eqref{eq: lemma Lp Lq estimate 1} and \eqref{eq: lemma Lp Lq estimate 2}, there exists $c_1$ such that
\begin{equation}
    \begin{aligned}
    \label{eq: proof nabla Y_1 L2 Lp bounded}
        \left\|\nabla Y_1(t)\right\|_{L^{2}} 
        & \leq \left\| \nabla e^{(\delta_{Y_1}\Delta -d_{21m})t}Y_{10} \right\|_{L^{2}} \\
        & \quad + \int_0^t \left\| \nabla e^{(\delta_{Y_1}\Delta - d_{21m}) (t-s)} \left(e_1f_1(X,Y_1,x) - g_1(Y_1,Z,x) - (d_{21}(x) - d_{21m}) Y \right)\right\|_{L^{2}}\mathrm{d}s \\
        & \quad + \int_0^t \left\| \nabla e^{(\delta_{Y_1}\Delta - d_{21m}) (t-s)} \nabla (\alpha_1(x) h(Y_1)\nabla X) \right\|_{L^{2}}\mathrm{d}s \\
        & \leq c_1 e^{-(\delta_{Y_1}\lambda_1 + d_{21m}) t } || \nabla Y_{10}||_{L^{2}} \\
        & \quad + c_1 \mathcal{T}_{\frac{1}{2},\delta_{Y_1} \lambda_1 + d_{21m}} \Big( e_1 \left\| f_1(X,Y_1,x) \right\|_{L^2} + \left\| g_1(Y_1,Z,x) \right\|_{L^2} +  \left\| (d_{21}(x) - d_{21m}) Y \right\|_{L^2} \! \Big) \\
        & \quad + c_1 \mathcal{T}_{\frac{1}{2},\delta_{Y_1} \lambda_1 + d_{21m}} \big(\left\| \nabla (\alpha_1(x) h(Y_1)\nabla X) \right\|_{L^2} \big),
    \end{aligned}
\end{equation}
where we used the definition \eqref{eq: operator T_rho,lambda} of operator $\mathcal{T}$. For the first term, using a similar argument as in \eqref{eq: Delta X L2 Lp+inf item 1}, we have 
\begin{equation}
\label{eq: proof nabla Y_1 L2 Lp+infty item 1}
    \begin{aligned}
       \left\| e^{-(\delta_{Y_1}\lambda_1 + d_{21m}) t } || \nabla Y_{10}||_{L^{2}}\right\|_{L^{p}(0,t)} \leq c_2 || \nabla Y_{10}||_{L^{2}(0,t)},
    \end{aligned}
\end{equation}
for some constant $c_2$.
By Assumptions \eqref{eq: assumption f}, \eqref{eq: assumption g} and $L^2$ boundedness of $U$ \eqref{eq: U L^2 bounded}, we obtain
     \begin{equation}
     \label{eq: f_L2 g_L2 bounded}
        \begin{aligned}
            || f_i(X,Y_i) ||_{L^2} \leq C_{f,i} C_{4,s}, \quad 
            || g_i(Y_i,Z) ||_{L^2} \leq C_{g,i} C_{4,s}, \quad i=1,2.
        \end{aligned}
    \end{equation}
It follows from Lemma \ref{lem: Young's convolution inequality on T} that 
\begin{equation}
\label{eq: proof nabla Y_1 L2 Lp+infty item 2}
    \begin{aligned}
        & \quad \ \bigg \| \mathcal{T}_{\frac{1}{2},\delta_{Y_1} \lambda_1 + d_{21m}} \Big( e_1 \left\| f_1(X,Y_1,x) \right\|_{L^2} + \left\| g_1(Y_1,Z,x) \right\|_{L^2} + \left\| (d_{21}(x) - d_{21m}) Y \right\|_{L^2} \! \Big)  \bigg\|_{L^{p}(0,t)}\\
        & \leq \left( e_1C_{f,1} + C_{g,1} + d_{21M} - d_{21m} \right) \left\| \mathcal{T}_{\frac{1}{2},\delta_{Y_1} \lambda_1 + d_{21m}} (C_{4,s}) \right\|_{L^{p}(0,t)} \\
        & \leq c_3 \left( e_1 C_{f,1} + C_{g,1} + d_{21M} - d_{21m} \right) \|C_{4,s} \|_{L^\infty(0,t)}
    \end{aligned}
\end{equation}
for some constant $c_3$. By the estimates of $||\nabla U ||_{L^2(\Omega) L^2(0,t)}$ \eqref{eq: nabla U L2 L2 bounded}, $||Y_1||_{L^\infty}$, $||\nabla X||_{L^\infty}$  \eqref{eq: L^inf boundedness of Y1 Y2 nabla X}, and $||\Delta X||_{L^2(\Omega)L^{p}(0, t)}$ \eqref{eq: lemma Delta X L2 Lp+infty bounded}, there exists a constant $c_{4,t}$ such that
\begin{equation}
    \label{eq: nabla (alpha_1 h nabla X) L2 L2 infty bounded}
    \begin{aligned}
        &\quad \, \left\|\nabla ( \alpha_1(x) h(Y_1)\nabla X) \right\|_{L^2(\Omega) L^{2}(0,t)} \\
        & \leq \left\| \nabla \alpha_1(x) h(Y_1)\nabla X \right\|_{L^2(\Omega) L^{2}(0,t)} 
        + \left\| \alpha_1(x) h'(Y_1) \nabla Y_1 \nabla X \right\|_{L^2(\Omega) L^{2}(0,t)}
        + \left\| \alpha_1(x) h(Y_1) \Delta X \right\|_{L^2(\Omega) L^{2}(0,t)} \\
        & \leq c_{4,t} \left( C_h ||\nabla X||_{L^2(\Omega) L^2(0,t)} + ||\nabla X||_{L^\infty} ||\nabla Y_1||_{L^2(\Omega)L^2(0,t)} + C_h ||\Delta X||_{L^2(\Omega)L^{2}(0,t)} \right) \\
        & \leq c_{4,t} \left( C_h C_{5,t} + C_{10,t} C_{5,t} + C_h C_{11,t} \right).
    \end{aligned}
\end{equation}
For any $p \in [2, \infty)$, by Lemma \ref{lem: Young's convolution inequality on T}, there exists a constant $c_5$ such that
\begin{equation}
\label{eq: proof nabla Y_1 L2 Lp+infty item 3}
    \begin{aligned}
        & \quad \ \left\| \mathcal{T}_{\frac{1}{2},\delta_{Y_1} \lambda_1 + d_{21m}} \left\| \nabla (\alpha_1(x) h(Y_1)\nabla X) \right\|_{L^2}  \mathrm{d}s \right\|_{L^{p}(0,t)} \\
        & \leq c_5 \left\| \nabla (\alpha_1(x) h(Y_1) \nabla X ) \right \|_{L^2(\Omega) L^{2}(0,t)} \\
        & \leq c_5 c_{4,t} \left( C_h C_{5,t} + C_{10,t} C_{5,t} + C_h C_{11,t} \right).
    \end{aligned}
\end{equation}
Along with above inequalities \eqref{eq: proof nabla Y_1 L2 Lp+infty item 1}, \eqref{eq: proof nabla Y_1 L2 Lp+infty item 2}, \eqref{eq: proof nabla Y_1 L2 Lp+infty item 3}, after taking $L^{p}$ norm on \eqref{eq: proof nabla Y_1 L2 Lp bounded} with respect to time, we obtain
\begin{equation}
    \begin{aligned}
    \label{eq: proof nabla Y_1 L2 Lp t_bounded}
        & \quad \ \left\|\nabla Y_1(t)\right\|_{L^{2}(\Omega)L^{p}(0,t)} \\
        & \leq c_1 c_2 || \nabla Y_{10}||_{L^{2}} + c_1 c_3\left( e_1C_{f,1} + C_{g,1} + d_{21M} - d_{21m}\right) \| C_{4,s} \|_{L^\infty(0,t)} \\
        & \quad + c_1 c_5 c_{4,t} \left( C_h C_{5,t} + C_{10,t} C_{5,t} + C_h C_{11,t} \right) \\
        & =: C_{12,t}.
    \end{aligned}
\end{equation}
\end{proof}

\begin{lemma}
    \label{lem: nabla Y_1 Ln+1 bounded}
    Assume condition \eqref{eq: condition A positive definite} holds, then the solution for equation \eqref{eq: general model}-\eqref{eq: bc ic} satisfies
    \begin{equation}
        \label{eq: nabla Y_1 Ln+1 bounded}
        ||\nabla Y_1||_{L^{n+1}} \leq C_{13} \big( ||U_0||_{L^\infty}, ||\nabla X_0||_{L^\infty}, ||\nabla Y_{10}||_{L^{n+1}}, ||\Delta X_0||_{L^2}, t \big),
        \quad \forall t \in [0, T_{\max}).
    \end{equation}
\end{lemma}

\begin{proof}
We prove by a similar argument as in Lemma \ref{lem: nabla Y_1 L2 Lp bounded}.
    Applying \eqref{eq: lemma Lp Lq estimate 1} and \eqref{eq: lemma Lp Lq estimate 2} on \eqref{eq: proof1: variation of formula Y_1}, we obtain that
    \begin{equation*}
    \begin{aligned}
        & \quad \left\|\nabla Y_1(t)\right\|_{L^{n+1}}\\ 
        & \leq c_1 e^{-(\delta_{Y_1} \lambda_1 + d_{21m}) t} || \nabla Y_{10} ||_{L^{n+1}} \\
        & \quad + c_1 \mathcal{T}_{\frac{1}{2} + \rho,\delta_{Y_1} \lambda_1 + d_{21m}} \Big( e_1 \left\| f_1(X,Y_1,x) \right\|_{L^2} + \left\| g_1(Y_1,Z,x) \right\|_{L^2} + \left\|(d_{21}(x) - d_{21m}) Y \right\|_{L^2} \! \Big) \\
        & \quad + c_1 \mathcal{T}_{\frac{1}{2} + \rho,\delta_{Y_1} \lambda_1 + d_{21m}} \big( \left\| \nabla (\alpha_1(x) h(Y_1)\nabla X) \right\|_{L^2} \big)
    \end{aligned}
\end{equation*}
for $\rho$ in equation \eqref{eq: rho} and for some constant $c_1 > 0$. Now we use equation \eqref{eq: f_L2 g_L2 bounded} and Lemma \ref{lem: Young's convolution inequality on T} with $p = q = \infty$ to deduce
\begin{equation*}
    \begin{aligned}
        &\quad \ \mathcal{T}_{\frac{1}{2} + \rho,\delta_{Y_1} \lambda_1 + d_{21m}} \Big( e_1 \left\| e_1f_1(X,Y_1,x) \right\|_{L^2} + \left\| g_1(Y_1,Z,x) \right\|_{L^2}  + \left\|(d_{21}(x) - d_{21m}) Y \right\|_{L^2} \! \Big) \\
        &\leq (e_1C_{f,1} + C_{g,1} + d_{21M} - d_{21m} ) \mathcal{T}_{\frac{1}{2} + \rho,\delta_{Y_1} \lambda_1 + d_{21m}} \big(\|C_{4,s}\|_{L^\infty(0,t)}\big) \\ 
        &\leq c_2 (e_1 C_{f,1} + C_{g,1} + d_{21M} - d_{21m} ) \|C_{4,s}\|_{L^\infty(0,t)}.
    \end{aligned}
\end{equation*}
for some $c_2 > 0$. Take $q = \infty$ and a sufficiently large $p$ in Lemma \ref{lem: Young's convolution inequality on T}, we obtain 
\begin{equation*}
    \begin{aligned}
        \mathcal{T}_{\frac{1}{2} + \rho, \delta_{Y_1} \lambda_1 + d_{21m}} \big( \left\| \nabla (\alpha_1(x) h(Y_1)\nabla X) \right\|_{L^2} \big)
        \leq c_3 \left\| \nabla (\alpha_1(x) h(Y_1)\nabla X ) \right \|_{L^2(\Omega) L^{p}(0,t)}
    \end{aligned}
\end{equation*}
for some constant $c_3>0$. By the estimates of $\|\nabla X\|_{L^2}$ \eqref{eq: nabla X L2}, $\left\| \nabla X \right\|_{L^\infty}$ \eqref{eq: L^inf boundedness of Y1 Y2 nabla X}, $||\nabla Y_1||_{L^2(\Omega)L^{p}(0,t)}$ \eqref{eq: lemma nabla Y_1 L2 Lp+infty bounded} and $||\Delta X||_{L^2(\Omega)L^{p}(0,t)}$ \eqref{eq: lemma Delta X L2 Lp+infty bounded}, there exists a constant $c_{4,t}$ such that
\begin{equation}
    \label{eq: nabla (alpha_1 h nabla X) L2 Lp infty bounded}
    \begin{aligned}
        &\quad \ \left\|\nabla ( \alpha_1(x) h(Y_1)\nabla X) \right\|_{L^2(\Omega) L^{p}(0,t)} \\
         & \leq c_{4,t} \left( C_h||\nabla X||_{L^2(\Omega) L^\infty(0,t)} + ||\nabla X||_{L^\infty}  ||\nabla Y_1||_{L^2(\Omega)L^{p}(0,t)} + C_h||\Delta X||_{L^2(\Omega)L^{p}(0,t)} \right) \\
         & \leq c_{4,t} \left( C_h \| C_{6,s}\|_{L^\infty (0,t)} + C_{10,t} C_{12,t} + C_h C_{11,t} \right).
    \end{aligned}
\end{equation}
It follows that
\begin{equation*}
    \begin{aligned}
        \left\|\nabla Y_1(t)\right\|_{L^{n+1}} 
        & \leq c_1 e^{-(\delta_{Y_1} \lambda_1 + d_{21m}) t} || \nabla Y_{10} ||_{L^{n+1}}  \\
        & \quad + c_1 c_2 (e_1C_{f,1} + C_{g,1} + d_{21M} - d_{21m} ) \|C_{4,s}\|_{L^\infty(0,t)}\\
        & \quad + c_1 c_3  c_{4,t} \left( C_h \| C_{6,s}\|_{L^\infty (0,t)} + C_{10,t} C_{12,t} + C_h C_{11,t} \right) \\
        & =: C_{13,t}.
    \end{aligned}
\end{equation*}
\end{proof}

By a similar argument as for the estimate of $Y_1$, we can demonstrate the following lemma for $Y_2$.
\begin{lemma}
    \label{lem: nabla Y_2 Ln+1 bounded}
    Assume condition \eqref{eq: condition A positive definite} holds, then the solution for equation \eqref{eq: general model}-\eqref{eq: bc ic} satisfies
    \begin{equation}
        \label{eq: nabla Y_2 Ln+1 bounded}
        ||\nabla Y_2||_{L^{n+1}} \leq C_{14} \big( ||U_0||_{L^\infty}, ||\nabla X_0||_{L^\infty}, ||\nabla Y_{20}||_{L^{n+1}}, ||\Delta X_0||_{L^2}, t \big), \quad \forall t \in [0, T_{\max}).
    \end{equation}
\end{lemma}

\begin{lemma}
    \label{lem: Z L_infnity bounded}
    Assume condition \eqref{eq: condition A positive definite} holds, then the solution for equation \eqref{eq: general model}-\eqref{eq: bc ic} satisfies
    \begin{equation}
        \label{eq: Z L_infnity bounded}
        ||Z||_{L^{\infty}} \leq C_{15} \big( ||U_0||_{W^{1, \infty}}, ||\Delta X_0||_{L^2}, t \big), \quad \forall t \in [0, T_{\max}).
    \end{equation}
\end{lemma}
\begin{proof}
    Applying the variation of constants to the fourth equation of \eqref{eq: general model}, we obtain
    \begin{equation}
    \label{eq: variation of formula Z}
        \begin{aligned}
            Z(t) = & \ e^{(\delta_Z \Delta - d_{3m})t} Z_0 + \int_0^t e^{(\delta_Z \Delta - d_{3m}) (t - s)} \big( e_2\left(g_1(Y_1,Z,x) + g_2(Y_2,Z,x) \right) \\
            & - (d_3(x) - d_{3m}) Z - \nabla (\gamma_1(x)h(Z)\nabla Y_1) - \nabla (\gamma_2(x)h(Z)\nabla Y_2) \big) \mathrm{d}s.
        \end{aligned}
    \end{equation}
    It follows that
    \begin{equation}
        \label{eq: proof 0 Z L_infnity bounded}
        \begin{aligned}
            ||Z||_{L^{\infty}} \leq & \left\| e^{(\delta_Z \Delta - d_{3m})t} Z_0 \right\|_{L^{\infty}} \\
            & + \int_0^t \left\| e^{(\delta_Z \Delta - d_3) (t - s)} \left( e_2g_1(Y_1,Z,x) + e_2g_2(Y_2,Z,x) + (d_{3}(x) - d_{3m})Z \right) \right\|_{L^{\infty}} \mathrm{d}s \\
            & + \int_0^t \left\| e^{(\delta_Z \Delta - d_3) (t - s)} \nabla (\gamma_1(x)h(Z)\nabla Y_1) \right\|_{L^{\infty}} \mathrm{d}s \\
            & + \int_0^t \left\| e^{(\delta_Z \Delta - d_3) (t - s)}  \nabla (\gamma_2(x)h(Z)\nabla Y_2) \right\|_{L^{\infty}}\mathrm{d}s.
        \end{aligned}
    \end{equation}
    First, from equation \eqref{eq: lemma Lp Lq estimate 1}, we obtain 
    \begin{equation}
        \label{eq: proof 1 Z L_infnity bounded}
        \left\| e^{(\delta_Z \Delta - d_3)t} Z_0 \right\|_{L^{\infty}} 
        \leq c_1 e^{-( \delta_Z \lambda_1 + d_3) t} ||Z_0||_{L^\infty}
    \end{equation}
    for some $c_1 > 0$. Referring to \eqref{eq: lemma Lp Lq estimate 3}, with $p=\infty$ and $q=2$, and utilizing the estimate of $||g_i||_{L^2}$ as in \eqref{eq: f_L2 g_L2 bounded}, we obtain
    \begin{equation}
        \label{eq: proof 2 Z L_infnity bounded}
        \begin{aligned}
            & \quad \ \left\| e^{(\delta_Z \Delta - d_3) (t - s)} \left(e_2g_1(Y_1,Z,x) + e_2g_2(Y_2,Z,x) + (d_3(x) - d_{3m}) Z \right) \right\|_{L^{\infty}} \\
            & \leq c_2 \left( 1 + (t-s)^{-n / 4}\right) e^{-( \delta_Z \lambda_1 + d_3) (t-s)} \\
            & \qquad \cdot \left( e_2\left\|g_1(Y_1,Z,x)\right\|_{L^2} + e_2||g_2(Y_2,Z,x)||_{L^2} + \left\| (d_3(x) - d_{3m}) Z \right\|_{L^2} \right) \\
            & \leq c_2 \left( 1 + (t-s)^{-n / 4}\right) e^{-( \delta_Z \lambda_1 + d_3) (t-s)} \left( e_2 C_{g,1} + e_2 C_{g,2} + d_{3M} - d_{3m} \right) C_{4,s}
        \end{aligned}
    \end{equation}
    for some $c_2 > 0$. Take $p= \infty$ and $q=n+1$ in \eqref{eq: lemma Lp Lq estimate 4}, then there exists a constant $c_3 >0 $ such that
\begin{equation*}
    \begin{aligned}
        & \quad \ \left\| e^{(\delta_Z \Delta - d_3) (t - s)} \nabla (\gamma_i(x) h(Z) \nabla Y_i) \right\|_{L^{\infty}} \\
        & \leq c_3 \left( 1 + (t-s) ^{-\rho'}\right) e^{- (\delta_Z \lambda_1 + d_3) (t-s)} ||\gamma_i(x) h(Z) \nabla Y_i||_{L^{n+1}}
    \end{aligned}
\end{equation*}
for $\rho' = \frac{2n + 1}{2(n+1)} < 1$ and $i = 1, 2$. Moreover, by the estimates on $Y_1$ \eqref{eq: nabla Y_1 Ln+1 bounded}, $h$ \eqref{eq: assumption h} and $\gamma_1$, we have 
\begin{equation*}
    ||\gamma_1(x) h(Z) \nabla Y_1||_{L^{n+1}} \leq \gamma_{1M} C_h C_{13,s}.
\end{equation*}
Therefore,
\begin{equation}
    \label{eq: proof 3 Z L_infnity bounded}
    \begin{aligned}
        \int_0^t \left\| e^{(\delta_Z \Delta - d_3) (t - s)} \nabla (\gamma_1(x) h(Z) \nabla Y_1) \right\|_{L^{\infty}} \mathrm{d}s 
        \leq c_3 \gamma_{1M} C_h \mathcal{T}_{\rho', \delta_Z \lambda_1 + d_{3m}} (C_{13, t}).
    \end{aligned}
\end{equation}
Similarly, for $Y_2$, we derive from \eqref{eq: nabla Y_2 Ln+1 bounded} that
\begin{equation}
    \label{eq: proof 4 Z L_infnity bounded}
    \begin{aligned}
        \int_0^t \left\| e^{(\delta_Z \Delta - d_3) (t - s)} \nabla (\gamma_2(x) h(Z) \nabla Y_2) \right\|_{L^{\infty}} \mathrm{d}s
        \leq c_3 \gamma_{2M} C_h \mathcal{T}_{\rho', \delta_Z \lambda_1 + d_{3m}} (C_{14, t}).
    \end{aligned}
    \end{equation}
    Finally, combining equations \eqref{eq: proof 0 Z L_infnity bounded}-\eqref{eq: proof 4 Z L_infnity bounded}, we conclude that
    \begin{equation*}
        \begin{aligned}
            ||Z||_{L^\infty} &\leq c_1 e^{-( \delta_Z \lambda_1 + d_3) t} ||Z_0||_{L^\infty} \\
            & \quad + c_2 \left( e_2 C_{g,1} + e_2 C_{g,2} + d_{3M} - d_{3m} \right) \mathcal{T}_{\frac{n}{4}, \delta_Z \lambda_1 + d_{3m}} \left( C_{4,s} \right) \\
            & \quad + c_3 C_h  \mathcal{T}_{\rho', \delta_Z \lambda_1 + d_3} (\gamma_{1M} C_{13,s} + \gamma_{2M} C_{14,s}) =: C_{15,t}.
        \end{aligned}
    \end{equation*}
\end{proof}

\begin{proof}[\rm\bf{Proof of Theorem \ref{th: global existence}}]
    Lemmas \ref{lem: X bounded}, \ref{lem: L^inf boundedness of Y1 Y2 nabla X} and \ref{lem: Z L_infnity bounded} guarantee that the solution to equation \eqref{eq: general model}-\eqref{eq: bc ic} does not blow up under $L^\infty$ norm in finite time. The global existence of the solution is a direct consequence of Theorem \ref{th: local existence}.
\end{proof}

\end{appendix}

\bibliographystyle{apalike}
\bibliography{reference}
\end{document}